\documentclass[12pt, a4paper, makeidx]{memoir}

\usepackage{graphicx}
\usepackage{epsfig}
\usepackage{amsmath}
\usepackage{amssymb}
\usepackage{amsthm}
\usepackage{wasysym}
\usepackage{booktabs}
\usepackage{stmaryrd}
\usepackage{url}
\usepackage{longtable}
\usepackage[figuresright]{rotating}

\usepackage{algorithmic}
\usepackage{algorithm}

\usepackage{polski}
\usepackage[cp1250]{inputenc}
\selecthyphenation{polish}
\usepackage[tableposition=top,font={small,rm},labelfont=bf,format=hang]{caption}
\DeclareCaptionLabelSeparator{kropka}{.\hspace{1ex}}
\makeindex

\usepackage{citeref}
\renewcommand{\bibitempages}[1]{\newblock {\scriptsize [\mbox{cytowanie na str.\ }#1]}}

\usepackage{color}

\definecolor{greenyellow}   {cmyk}{0.15, 0   , 0.69, 0   }
\definecolor{yellow}        {cmyk}{0   , 0   , 1   , 0   }
\definecolor{goldenrod}     {cmyk}{0   , 0.10, 0.84, 0   }
\definecolor{dandelion}     {cmyk}{0   , 0.29, 0.84, 0   }
\definecolor{apricot}       {cmyk}{0   , 0.32, 0.52, 0   }
\definecolor{peach}         {cmyk}{0   , 0.50, 0.70, 0   }
\definecolor{melon}         {cmyk}{0   , 0.46, 0.50, 0   }
\definecolor{yelloworange}  {cmyk}{0   , 0.42, 1   , 0   }
\definecolor{orange}        {cmyk}{0   , 0.61, 0.87, 0   }
\definecolor{burntorange}   {cmyk}{0   , 0.51, 1   , 0   }
\definecolor{bittersweet}   {cmyk}{0   , 0.75, 1   , 0.24}
\definecolor{redorange}     {cmyk}{0   , 0.77, 0.87, 0   }
\definecolor{mahogany}      {cmyk}{0   , 0.85, 0.87, 0.35}
\definecolor{maroon}        {cmyk}{0   , 0.87, 0.68, 0.32}
\definecolor{brickred}      {cmyk}{0   , 0.89, 0.94, 0.28}
\definecolor{red}           {cmyk}{0   , 1   , 1   , 0   }
\definecolor{orangered}     {cmyk}{0   , 1   , 0.50, 0   }
\definecolor{rubinered}     {cmyk}{0   , 1   , 0.13, 0   }
\definecolor{wildstrawberry}{cmyk}{0   , 0.96, 0.39, 0   }
\definecolor{salmon}        {cmyk}{0   , 0.53, 0.38, 0   }
\definecolor{carnationpink} {cmyk}{0   , 0.63, 0   , 0   }
\definecolor{magenta}       {cmyk}{0   , 1   , 0   , 0   }
\definecolor{violetred}     {cmyk}{0   , 0.81, 0   , 0   }
\definecolor{rhodamine}     {cmyk}{0   , 0.82, 0   , 0   }
\definecolor{mulberry}      {cmyk}{0.34, 0.90, 0   , 0.02}
\definecolor{redviolet}     {cmyk}{0.07, 0.90, 0   , 0.34}
\definecolor{fuchsia}       {cmyk}{0.47, 0.91, 0   , 0.08}
\definecolor{lavender}      {cmyk}{0   , 0.48, 0   , 0   }
\definecolor{thistle}       {cmyk}{0.12, 0.59, 0   , 0   }
\definecolor{orchid}        {cmyk}{0.32, 0.64, 0   , 0   }
\definecolor{darkorchid}    {cmyk}{0.40, 0.80, 0.20, 0   }
\definecolor{purple}        {cmyk}{0.45, 0.86, 0   , 0   }
\definecolor{plum}          {cmyk}{0.50, 1   , 0   , 0   }
\definecolor{violet}        {cmyk}{0.79, 0.88, 0   , 0   }
\definecolor{royalpurple}   {cmyk}{0.75, 0.90, 0   , 0   }
\definecolor{blueviolet}    {cmyk}{0.86, 0.91, 0   , 0.04}
\definecolor{periwinkle}    {cmyk}{0.57, 0.55, 0   , 0   }
\definecolor{cadetblue}     {cmyk}{0.62, 0.57, 0.23, 0   }
\definecolor{cornflowerblue}{cmyk}{0.65, 0.13, 0   , 0   }
\definecolor{midnightblue}  {cmyk}{0.98, 0.13, 0   , 0.43}
\definecolor{navyblue}      {cmyk}{0.94, 0.54, 0   , 0   }
\definecolor{royalblue}     {cmyk}{1   , 0.50, 0   , 0   }
\definecolor{blue}          {cmyk}{1   , 1   , 0   , 0   }
\definecolor{cerulean}      {cmyk}{0.94, 0.11, 0   , 0   }
\definecolor{cyan}          {cmyk}{1   , 0   , 0   , 0   }
\definecolor{processblue}   {cmyk}{0.96, 0   , 0   , 0   }
\definecolor{skyblue}       {cmyk}{0.62, 0   , 0.12, 0   }
\definecolor{turquoise}     {cmyk}{0.85, 0   , 0.20, 0   }
\definecolor{tealblue}      {cmyk}{0.86, 0   , 0.34, 0.02}
\definecolor{aquamarine}    {cmyk}{0.82, 0   , 0.30, 0   }
\definecolor{bluegreen}     {cmyk}{0.85, 0   , 0.33, 0   }
\definecolor{emerald}       {cmyk}{1   , 0   , 0.50, 0   }
\definecolor{junglegreen}   {cmyk}{0.99, 0   , 0.52, 0   }
\definecolor{seagreen}      {cmyk}{0.69, 0   , 0.50, 0   }
\definecolor{green}         {cmyk}{1   , 0   , 1   , 0   }
\definecolor{forestgreen}   {cmyk}{0.91, 0   , 0.88, 0.12}
\definecolor{pinegreen}     {cmyk}{0.92, 0   , 0.59, 0.25}
\definecolor{limegreen}     {cmyk}{0.50, 0   , 1   , 0   }
\definecolor{yellowgreen}   {cmyk}{0.44, 0   , 0.74, 0   }
\definecolor{springgreen}   {cmyk}{0.26, 0   , 0.76, 0   }
\definecolor{olivegreen}    {cmyk}{0.64, 0   , 0.95, 0.40}
\definecolor{rawsienna}     {cmyk}{0   , 0.72, 1   , 0.45}
\definecolor{sepia}         {cmyk}{0   , 0.83, 1   , 0.70}
\definecolor{brown}         {cmyk}{0   , 0.81, 1   , 0.60}
\definecolor{tan}           {cmyk}{0.14, 0.42, 0.56, 0   }
\definecolor{gray}          {cmyk}{0   , 0   , 0   , 0.50}
\definecolor{black}         {cmyk}{0   , 0   , 0   , 1   }
\definecolor{white}         {cmyk}{0   , 0   , 0   , 0   }

\ifpdf
        \usepackage[plainpages=false,pdfpagelabels,bookmarksnumbered,%
        colorlinks=true,%
        linkcolor=sepia,%
        citecolor=sepia,%
        filecolor=maroon,%
        pagecolor=red,%
        urlcolor=sepia,%
        pdftex,%
        unicode]{hyperref}
    \input supp-mis.tex
    \input supp-pdf.tex
    \usepackage{thumbpdf}
\else
    \usepackage{hyperref}
\fi

\usepackage{memhfixc}

\settypeblocksize{*}{32pc}{1.618}

\setlrmarginsandblock{1.5in}{1.0in}{*}
\setulmarginsandblock{1.0in}{1.0in}{*}

\setheadfoot{\onelineskip}{2\onelineskip}
\setheaderspaces{*}{2\onelineskip}{*}

\def\baselinestretch{1.5}

\checkandfixthelayout

\makechapterstyle{mychapterstyle}{%
    \renewcommand{\chapnamefont}{\LARGE\sffamily\bfseries}%
    \renewcommand{\chapnumfont}{\LARGE\sffamily\bfseries}%
    \renewcommand{\chaptitlefont}{\Huge\sffamily\bfseries}%
    \renewcommand{\printchaptertitle}[1]{%
        \chaptitlefont\hrule height 0.5pt \vspace{1em}%
        {##1}\vspace{1em}\hrule height 0.5pt%
        }%
    \renewcommand{\printchapternum}{%
        \chapnumfont\thechapter%
        }%
}

\chapterstyle{mychapterstyle}

\setsecheadstyle{\Large\sffamily\bfseries}
\setsubsecheadstyle{\large\sffamily\bfseries}
\setsubsubsecheadstyle{\normalfont\sffamily\bfseries}
\setparaheadstyle{\normalfont\sffamily}

\makeevenhead{headings}{\thepage}{}{\small\slshape\leftmark}
\makeoddhead{headings}{\small\slshape\rightmark}{}{\thepage}

\settocdepth{subsection}

\setsecnumdepth{subsection}
\maxsecnumdepth{subsection}
\settocdepth{subsection}
\maxtocdepth{subsection}

\usepackage{fourier}
\usepackage[scaled=.92]{helvet}
\usepackage[T1]{fontenc}
\usepackage{color}
\definecolor{ChapGrey}{rgb}{0.6,0.6,0.6}
\newcommand{\LargeFont}{
\usefont{\encodingdefault}{\rmdefault}{b}{n}%
\fontsize{60}{80}\selectfont\color{ChapGrey}}
\makeatletter
\makechapterstyle{GreyNum}{%
\renewcommand{\chapnamefont}{\large\sffamily\bfseries\itshape}
\renewcommand{\chapnumfont}{\LargeFont}
\renewcommand{\chaptitlefont}{\Huge\sffamily\bfseries\itshape}
\setlength{\beforechapskip}{0pt}
\setlength{\midchapskip}{40pt}
\setlength{\afterchapskip}{60pt}
\renewcommand\chapterheadstart{\vspace*{\beforechapskip}}
\renewcommand\printchaptername{%
\begin{tabular}{@{}c@{}}
\chapnamefont \@chapapp\\}
\renewcommand\chapternamenum{\noalign{\vskip 2ex}}
\renewcommand\printchapternum{\chapnumfont\thechapter\par}
\renewcommand\afterchapternum{%
\end{tabular}
\par\nobreak\vskip\midchapskip}
\renewcommand\printchapternonum{}
\renewcommand\printchaptertitle[1]{%
{\chaptitlefont{##1}\par}}
\renewcommand\afterchaptertitle{\par\nobreak\vskip \afterchapskip}
}
\makeatother
\chapterstyle{GreyNum}

\renewcommand*{\cftchaptername}{Rozdział\space}
\renewcommand*{\cftchapteraftersnum}{.}

\setsecnumformat{\csname the#1\endcsname.\quad}

\begin{document}

\frontmatter


\pagestyle{empty}

\noindent

\vfill

\begin{center}
    \large
\hfill Rozprawa doktorska\\
\end{center}

\vfill
\begin{center}
    \LARGE\bfseries
\hfill   Modelowanie numeryczne transportu\\
\hfill   płynów przez ośrodki porowate\\
\end{center}

\hfill\line(1,0){200}

\begin{center}
    \Huge\bfseries
\hfill    Maciej Matyka
\end{center}

\vfill\vfill\vfill
\begin{center}
    \large
\hfill    Promotor: dr hab. Zbigniew Koza
\end{center}


\begin{center}
    \large
\hfill    Uniwersytet Wrocławski\\
\hfill    Wydział Fizyki i Astronomii\\
\end{center}

\vfill
\begin{center}
\large
    Wrocław, 2008
\end{center}

\clearpage



\begin{abstract}
The aim of the thesis is to present and analyze two particular problems of transport in porous media flow. The first of them is related to the process of saturation of porous building materials. Recently, M. K\"untz and P. Laval\'ee, using a computer model of this process, have concluded that the anomalous diffusion assumption is correct. In this thesis I present an alternative explanation of this results without any refer to anomalous diffusion. The second part of the thesis covers the numerical analysis of the tortuosity of the flow -- one of a very interesting physical macroscopic variables characterizing transport in porous media.

-----

Celem niniejszej rozprawy jest przedstawienie oraz analiza dwóch szczegółowych problemów związanych z transportem płynów w ośrodku porowatym. Pierwszy z nich związany jest z zagadnieniem zwilżania materiałów budowlanych. Niedawno M. K\"untz i P. Laval\'ee opracowali model komputerowy tego procesu i stwierdzili, że mamy w nim do czynienia z dyfuzją anomalną. W niniejszej rozprawie przedstawiam alternatywne wyjaśnienie tych wyników bez odwoływania się do dyfuzji anomalnej. Drugie zagadnienie dotyczy szczegółowej analizy numerycznej krętości przepływu -- jednej z ciekawszych makroskopowych wielkości fizycznych charakteryzujących transport w ośrodkach porowatych.
\end{abstract}

\clearpage

\pagenumbering{roman}
\pagestyle{ruled}
\nouppercaseheads
\tableofcontents*


\mainmatter
\newcommand\vect[1]{\mathbf{#1}}
\newcommand\bm[1]{\mathbf{#1}}
\renewcommand\v[1]{\vect{#1}}
\renewcommand\bf[1]{\textbf{#1}}
\newcommand\tc[2]{\textcolor{#1}{#2}}
\newcommand\zk[1]{#1}
\newcommand\qu[1]{\tc{red}{(#1)}}
\newcommand\qqqq{[\qu{????}]}
\newcommand\qc{[\cq]}
\newcommand\kb{k_\mathrm{B}}
\newcommand{\vf}{\bar v_\mathrm{f}}

\hyphenation{Bol-tzman-na}
\hyphenation{je-dno-re-la-ksa-cy-jnym}


\renewcommand\leq{\leqslant}
\renewcommand\geq{\geqslant}
\renewcommand\le{<}
\renewcommand\ge{>}


\newcommand{\lb}{\linebreak}
\newcommand{\nlb}{\nolinebreak}
\newcommand{\nbl}{\nlb}
\chapter{Wstęp}

\section{Wprowadzenie}
\label{sec:chapter1:Wprowadzenie}

Transport ciepła, cieczy i gazów przez ośrodki porowate jest zjawiskiem wszechobecnym, a badanie jego właściwości ma ogromne znaczenie dla wielu gałęzi nauki i \nlb techniki. Do dziedzin, w których zjawisko to ma fundamentalne znaczenie, zaliczyć można przemysł wydobywczy gazu ziemnego i ropy naftowej \cite{Bear72, Li05, Taylor98}, hydrogeologię (budowa zbiorników wodnych, tam, podziemnych ujęć wody) i rolnictwo (odprowadzanie nadmiaru wody deszczowej i utrzymywanie odpowiedniej wilgotności gleby). Zjawisko to wykorzystuje się też w przemyśle petrochemicznym (transport ciepła oraz znaczenie polimerów w procesach wydobycia surowców) \cite{Taylor98,Scott05}, przy przetwarzaniu źródeł energii (poszukiwanie materiałów porowatych magazynujących energię w kolektorach słonecznych) \cite{Sopian07}, w ochronie przeciwpożarowej (transport przez materiały izolujące od wysokich temperatur) \cite{Figueiredo04,Ma04}, w farmacji (dostarczanie leków do organizmu) \cite{Lemaire03,Khanafer06}, w ochronie środowiska (filtracja cieczy w zbiornikach wodnych oraz odzyskiwanie związków żelaza z odpadów przemysłowych) \cite{Mohamed08,Hemond85,Dorofeev02}, diagnostyce medycznej (elektrody jonoselektywne użyteczne w pomiarach stężeń jonów w warunkach klinicznych) \cite{Conroy00} i technologii nuklearnej (wzbogacanie uranu) \cite{Cardew99}. Transport w ośrodkach porowatych leży też u podstaw działania reaktorów katalitycznych, membran \cite{Cardew99} i większości typów filtrów. Stanowi też punkt wyjścia dla teorii perkolacji \cite{Havlin02,Stauffer96}. Porowatość i przepuszczalność dla ciepła, gazów i wilgoci to podstawowe parametry niemal wszystkich materiałów budowlanych oraz materiałów codziennego użytku (np. tekstyliów). Jednym z ciekawszych przykładów układu o strukturze i cechach ośrodka porowatego przepuszczalnego zarówno dla gazów (powietrze), jak i płynów (krew) są płuca, które ze względu na swoją rolę w procesie oddychania stały się przedmiotem intensywnych badań doświadczalnych i teoretycznych \cite{Tseng98, Butler02}.

Wobec tak wielkiego znaczenia dla różnych gałęzi przemysłu i nauki, zagadnienie transportu w ośrodkach porowatych jest problemem interdyscyplinarnym podejmowanym zarówno przez badaczy nauk podstawowych \cite{Johnson82,Andrade99,Wyllie50,Johnson86}, jak i przyrodniczych \cite{Zacarias05,Benhamou04}, technicznych \cite{Sopian07,Ghandy04,Pel95}, a nawet medycznych \cite{Rusakov98,Nicholson98,Starly07}. Mimo, że w \nlb niemal niezliczonej liczbie prac zbadano już całe spektrum różnych aspektów tego zjawiska i że poświęcono mu wiele monografii \cite{Bear72,Dullien79,Nield98,Ingham98,Ingham02,Ingham05,Torquato01}, wciąż kryje ono przed nami wiele tajemnic.

Celem niniejszej rozprawy jest analiza dwóch szczegółowych problemów związanych z transportem płynów w ośrodku porowatym. Pierwszy z nich związany jest z zagadnieniem zwilżania materiałów budowlanych. Niektóre wyniki doświadczalne i teoretyczne sugerują, że zjawisko to wiąże się z tzw. dyfuzją anomalną. Niedawno M. \nlb K\"untz i P. Laval\'ee opracowali prosty model komputerowy tego procesu i \nlb stwierdzili, że rzeczywiście mamy w nim do czynienia z dyfuzją anomalną. W niniejszej rozprawie przedstawię alternatywne wyjaśnienie wyników uzyskanych w modelu K\"untza i Laval\'ee'go, nie wymagające odwoływania się do dyfuzji anomalnej.

Drugie zagadnienie dotyczy szczegółowej analizy numerycznej krętości przepływu -- jednej z ciekawszych makroskopowych wielkości fizycznych charakteryzujących transport w ośrodkach porowatych \cite{Clennell97,Lacks08}.

\section{Dyfuzja normalna i anomalna}\label{sec:dyfuzja}

W roku 1827 brytyjski botanik Robert Brown zauważył, że pyłki kwiatów znajdujące się na powierzchni cieczy poruszają się ruchem nieuporządkowanym. Zjawisko to pozostawało długo niewyjaśnione i dopiero Albert Einstein (1905) i Marian Smoluchowski (1906) opisali ten ruch jako wynik mikroskopijnych zderzeń pyłków z cząsteczkami cieczy. Odkrycie Einsteina i Smoluchowskiego ma szereg konsekwencji dla współczesnej fizyki, szczególnie dla teorii kinetycznej oraz fizyki statystycznej. To dzięki nim wiadomo, że przesunięcie średniokwadratowe cząsteczek w transporcie dyfuzyjnym jest proporcjonalny do pierwiastka z czasu. Prawo to można zapisać przy pomocy relacji:
\begin{equation}\label{eq:ralpha}
\sqrt{\langle r^2\rangle}\propto t^{\alpha},
\end{equation}
gdzie $\langle r^2\rangle$ jest średnią odległością przebytą w czasie $t$, a wykładnik $\alpha$ w przypadku dyfuzji normalnej przyjmuje wartość $1/2$. Wielkości makroskopowe takie jak szerokość lub pozycja profili temperatury (lub koncentracji) w transporcie dyfuzyjnym ciepła (lub materii) również skalują się w czasie zgodnie z równaniem (\ref{eq:ralpha}), tzn. opisane są jako funkcja jednej bezwymiarowej wielkości $x/\sqrt{\mathrm{Dt}}$. Okazuje się jednak, że istnieją przypadki, gdy prawo to nie jest spełnione i $\alpha\neq1/2$. Mówimy wtedy o dyfuzji anomalnej.

Istnieje szereg doniesień eksperymentalnych i teoretycznych na temat dyfuzji anomalnej pojedynczych molekuł oraz anomalnej dyfuzji chemicznej \cite{Beijeren85,Kutner99,Kosztolowicz05}. Dotyczą one m.in. zwilżania materiałów budowlanych \cite{Ghandy04,Pel95,Kuntz01,Lockington03}, dyfuzji siarczanu miedzi w cieczy po dejonizacji \cite{Carey95,Kuntz04}, transportu w układach polimerowych \cite{Rehage70,Snively99}, dyfuzji przez membrany syntetyczne \cite{Fowlkes06} oraz dyfuzji w żywych układach biologicznych \cite{Amblard98,Caspi00}. W pewnych warunkach anomalną dyfuzję zaobserwować można również w dyfuzji powierzchniowej \cite{Naumovec99}, a nawet w ruchu cząsteczek w zanieczyszczonej zawiesinie plazmowej \cite{Liu08}. Fizyczne mechanizmy dyfuzji anomalnej \lb w tych układach są różne i nie zawsze kompletnie poznane. Dyfuzję anomalną pojedynczych molekuł tłumaczy się między innymi tym, że odległości pokonywane przez molekuły w kolejnych krokach czasowych lub czasy pomiędzy kolejnymi skokami nie spełniają założeń centralnego twierdzenia granicznego i nie mogą być opisane przy pomocy rozkładu normalnego \cite{Metzler00, Metzler04}.

Oprócz pojedynczych molekuł, istnieje szereg układów, w których wielkości makroskopowe nie spełniają skalowania klasycznego (\ref{eq:ralpha}), a przykładem takich zjawisk są procesy zwilżania \cite{Ghandy04}. Jedną z hipotez tłumaczących dyfuzję anomalną w tego typu układach jest wysunięta przez K\"{u}ntza i Lavall\'{e}e'go zależność współczynnika dyfuzji $D$ od koncentracji dyfundującej substancji \cite{Kuntz03}. Zasugerowali oni, że dla \lb $D$ rosnącego z koncentracją $c$ wykładnik $\alpha$ jest większy niż $1/2$ i mamy do czynienia z procesami przebiegającymi szybciej od klasycznej dyfuzji (\emph{ang. superdiffusion}). Z \nlb drugiej strony dla $D$ malejącego z $c$ mamy do czynienia ze spowolnieniem tego procesu \cite{Kuntz04}. Hipoteza ta oparta została na wynikach symulacji transportu płynu w \nlb modelu ośrodka porowatego z losowo rozmieszczonymi rozpraszaczami \cite{Frish86,Kuntz01}.

Jednym z zagadnień poruszanych w niniejszej rozprawie będzie krytyczna analiza tezy K\"{u}ntza i Lavall\`{e}e'go oraz próba podania alternatywnego wyjaśnienia obserwowanych przez nich efektów.
\section{Krętość}\label{sec:wstepkretosc}
Dynamika przepływu zależy od wielu czynników, począwszy od siły zewnętrznej, poprzez mikrostrukturę porów ośrodka porowatego, prędkość przepływu, oddziaływania płynu z ośrodkiem, aż po fizyczny mechanizm transportu. Jedną z bardziej interesujących wielkości fizycznych opisujących przepływ jest krętość -- bezwymiarowa wielkość charakteryzująca efektywne wydłużenie drogi, wzdłuż której zachodzi transport.
\begin{figure}[!ht]
  \centering
\includegraphics[scale=1.15]{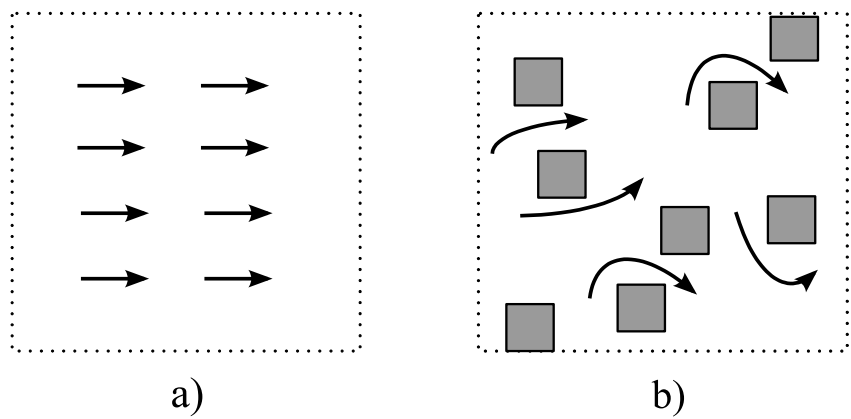}
\caption{Graficzne porównanie transportu przez ośrodek porowaty o różnych wartościach krętości: a) $T=1$ oraz b) $T>1$. \label{pic:kretosc}}
\end{figure}
W klasycznym przepływie cieczy (transportu masy) przez ośrodek porowaty \emph{krętość} $T$ definiuje się jako stosunek:
\begin{equation}
  \label{eq:Tle}
    T = \frac{\langle \lambda \rangle }{L},
\end{equation}
gdzie $\langle \lambda \rangle $ jest średnim efektywnym wydłużeniem drogi przebywanej przez ciecz (często oznaczaną również jako $L_e$), a $L$ jest liniowym rozmiarem układu w kierunku gradientu ciśnienia lub siły zewnętrznej, co implikuje $T\geq 1$ (rysunek \ref{pic:kretosc}). Krętość jest wielkością o tyle interesującą, że zawiera w sobie dwojaką informację: zarówno o \nlb strukturze ośrodka, jak i o charakterze transportu.

Już 46 lat temu w swojej pracy opublikowanej w \emph{Nature} Lorenz zauważył, że krętość w kontekście ośrodków porowatych interpretowana jest na różne, niekiedy wzajemnie sprzeczne sposoby, co dotyczy szczególnie jej relacji do porowatości \cite{Lorenz61}. Obecnie oprócz krętości hydrodynamicznej w literaturze znaleźć można odniesienia do krętości dyfuzyjnej \cite{Nakashima04}, elektrycznej \cite{Lorenz61,Johnson82}, a nawet krętości w transporcie fali dźwiękowej \cite{Johnson82}. Co więcej,  wyznaczane w różny sposób zależności krętości od porowatości $T(\phi)$, bardzo ważne z punktu widzenia zastosowań, różnią się w zależności od przyjętego modelu transportu oraz przyjętej metodyki badań.

W zakresie niskich liczb Reynoldsa, gdy brak jest efektów inercyjnych \cite{Andrade99}, przepływ płynu Newtonowskiego przez ośrodek porowaty podlega prawu Darcy'ego:
\begin{equation}\label{eq:darcy}
    \bm{q} = - \frac{k}{\mu} \bm\nabla P,
\end{equation}
gdzie $\bm{q}$ jest strumieniem płynu, $k$ przepuszczalnością ośrodka, $\mu$ lepkością dynamiczną płynu, a $\nabla P$ gradientem ciśnienia, pod wpływem którego następuje przepływ. Podstawowym problemem zarówno teoretycznym, jak i doświadczalnym, jest zależność przepuszczalności ośrodka $k$ od jego porowatości $\phi$ \cite{Adamson50,Koponen98}. Najbardziej znaną zależnością tego typu jest wyprowadzona na gruncie teorii kapilarnej relacja Kozeny'ego \cite{Bear72}:
\begin{equation}
  \label{eq:Kozeny}
     k = c_0\frac{\phi^3}{S^2},
\end{equation}
gdzie $c_0$ jest stałą Kozeny'ego uważaną za cechę geometryczną układu, a $S$ jest polem powierzchni mikroporów ośrodka porowatego w jednostce objętości (ang. \emph{specific surface area}).

Historycznie, pojęcie krętości w hydrodynamice zostało wprowadzone przez Carmana w 1937 roku \cite{Carman37}, jako poprawka empiryczna do kapilarnej teorii Kozeny'ego. Dzięki niej relacja między przepuszczalnością $k$ a porowatością $\phi$ przyjęła następującą postać:
\begin{equation}
  \label{eq:Kozeny2}
     k = c_0\frac{\phi^3}{T^2S^2},
\end{equation}
gdzie występująca w tym równaniu krętość $T$, początkowo wykorzystywana bardziej jako dodatkowy parametr teorii, okazała się być mierzalną wielkością fizyczną, interpretowaną jako wydłużenie drogi w transporcie przez ośrodek porowaty. Krętość nie jest jednak jedynie cechą geometryczną ośrodka, a jej wartość zależy również od mechanizmu fizycznego transportu \cite{Clennell97}. Co więcej, jak wspomniałem wyżej, rozróżnia się krętość dyfuzyjną, elektryczną oraz hydrodynamiczną \cite{Lorenz61,Zhang95,Nakashima00}. \emph{Krętość dyfuzyjną} $T_\mathrm{d}$ ośrodka porowatego definiuje się wzorem:
\begin{equation}\label{eq:difftort}
T_\mathrm{d}=\frac{\mathrm{D}}{\mathrm{D}_\phi},
\end{equation}
gdzie $\mathrm{D}_\phi$ to współczynnik dyfuzji mierzony w ośrodku porowatym, a $D$ to współczynnik dyfuzji samego płynu bez obecności przeszkód i kanałów. Ze względu na wpływ mikrostruktury ośrodka na dynamikę procesów dyfuzyjnych, $\mathrm{D}\geq\mathrm{D}_\phi$, czyli $T_\mathrm{d}\geq 1$  \cite{Nakashima02}. Analogicznie wprowadza się \emph{krętość elektryczną} $T_\mathrm{e}$ jako:
\begin{equation}\label{eq:electort}
T_\mathrm{e} = \phi \frac{\sigma_\phi}{\sigma},
\end{equation}
gdzie $\phi$ to porowatość ośrodka, a $\sigma_\phi$ oraz $\sigma$ są opornościami właściwymi ośrodka porowatego zanurzonego w elektrolicie oraz opornością samego elektrolitu. Ze względu na spowolnienie procesów transportu przez obecność przeszkód w ośrodku $\sigma_\phi\geq\sigma$, czyli $T_\mathrm{e}\geq 1$.

Jednym z głównych kierunków badań nad krętością są próby skorelowania jej z \nlb łatwiej mierzalną doświadczalnie porowatością $\phi$ ośrodka w możliwie jak najogólniejszej formie. W tym celu, na przestrzeni lat, przeprowadzonych zostało wiele analiz teoretycznych \cite{Johnson81,Johnson82}, badań doświadczalnych \cite{Comiti89,Barrande07}, a w ostatnich latach coraz częściej również obliczeń numerycznych \cite{Koponen96,Koponen97}, których zadaniem było wyznaczenie relacji $T(\phi)$ dla różnych układów fizycznych.

Najbardziej znane i charakterystyczne relacje wiążące $T$ z $\phi$ to\label{str:jeden}:
\begin{subequations}
\begin{eqnarray}
  T(\phi) &=& \phi^{-p},                         \label{fit:archie}   \\
  T(\phi) &=& 1-p\ln \phi,                    \label{fit:comiti}   \\
  T(\phi) &=& 1 + p(1-\phi),                  \label{fit:linear}   \\
  T(\phi) &=& \left[1 + p(1-\phi) \right]^2,  \label{fit:boudreau} \\
  T(\phi) &=& 1+p\frac{(1-\phi)}{(\phi-\phi_c)^m}, \label{fit:koponen}      
\end{eqnarray}
\end{subequations}
gdzie $p$ [oraz $m$ w równaniu (\ref{fit:koponen})] są wolnymi parametrami używanymi w procedurze dopasowania teorii do eksperymentu. Pierwsza z tych relacji została zaproponowana na podstawie badań nad przewodnictwem elektrycznym przez Archiego \lb w 1942 roku \cite{Archie42} i jest często stosowana również w innych kontekstach, np. do opisu przepływu cieczy \cite{Wyllie50,Dias06}. Drugie z równań zostało wyprowadzone na gruncie teoretycznych rozważań nad transportem dyfuzyjnym w układach swobodnie pokrywających się sfer ($p=1/2$) \cite{Weissberg63,Ho81} lub cylindrów ($p=1$ lub $p=3/2$) \cite{Tsai86}. Podobna relacja została również wyznaczona empirycznie (z parametrami $p\approx 0.86$ oraz $p\approx 1.66$) poprzez najlepsze dopasowanie dla krętości hydraulicznej w eksperymencie przepływu przez układy wymieszanych przeszkód z różnym stosunkiem szerokości do grubości \cite{Comiti89} oraz w pomiarach krętości elektrycznej w zawiesinie szklanych kul \cite{Barrande07}. Równanie (\ref{fit:linear}) jest zależnością empiryczną dla przepuszczalności dna naturalnych zbiorników wodnych \cite{Iversen93}. Zależność liniowa (\ref{fit:linear}) była również wyprowadzona na gruncie badań numerycznych nad przepływem z użyciem modelu gazu sieciowego ($p=0.8$) \cite{Koponen96} oraz teoretycznie dla modelu rozpraszania fali dźwiękowej w ośrodku wypełnionym cieczą ($p=1$) \cite{Johnson81,Johnson82}. Równanie (\ref{fit:boudreau}) uzyskano w modelu krętości dyfuzyjnej w osadach dna morskiego, dla którego oszacowana została wielkość $p\approx 1.1$ \cite{Boudreau06}. Z kolei zaproponowana przez Koponena w $1997$ roku relacja (\ref{fit:koponen}), zawiera dodatkowy wolny parametr $m$ i została wyznaczona na gruncie obliczeń w modelu zbliżonym do tego, jakiego użyjemy w analizie przeprowadzonej w \nlb niniejszej pracy ($p=0.65$ i $m=0.19$) \cite{Koponen97}.
\begin{figure}[!ht]
  \centering
\includegraphics[scale=0.55]{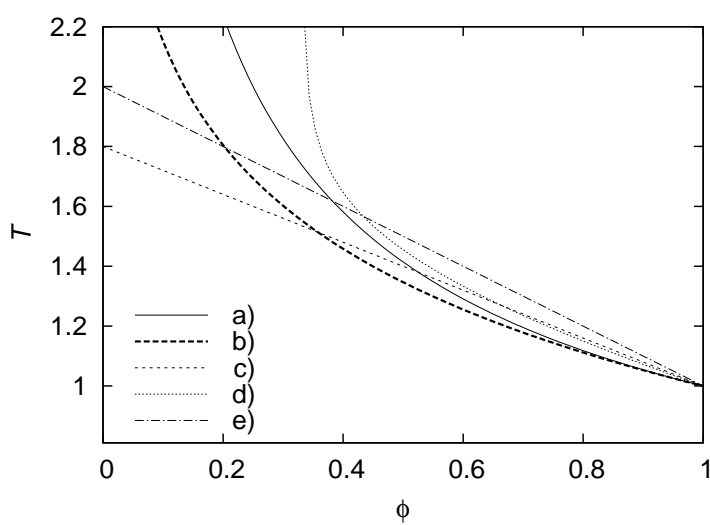}
\caption{Porównanie zależności $T(\phi)$ zebranych z kilku różnych źródeł dla różnych technik pomiarowych / modeli teoretycznych: a) \cite{Dias06} b) \cite{Weissberg63} c) \cite{Koponen96} d) \cite{Koponen97} e) \cite{Johnson81}. \label{pic:tphiporownanie}}
\end{figure}

Powyższe wzory nie są spójne między sobą (rysunek \ref{pic:tphiporownanie}). Różnice wynikają \lb z takich czynników, jak budowa ośrodka porowatego (różny próg perkolacji), metoda badawcza użyta do wyznaczenia krętości, rozmiar fizyczny próbek przyjętych do analizy (efekty skończonego rozmiaru), czy też przyjęty do badań mechanizm transportu.

\section{Cele i struktura rozprawy}\label{sec:budowa}

Celem niniejszej rozprawy jest analiza opisanych powyżej szczegółowych problemów związanych z transportem płynów w ośrodku porowatym.

Praca składa się z $6$ rozdziałów. W rozdziałach \ref{sec:gazsieciowy}. i \ref{sec:chapter3}. opisane zostaną narzędzia numeryczne użyte w trakcie badań -- automat komórkowy gazu sieciowego oraz model gazu sieciowego Boltzmanna. Główną część rozprawy stanowią rozdziały \ref{sec:chapterdyfuzja}. oraz \ref{sec:chapterkretosc}., gdzie przedstawione zostaną oryginalne wyniki. W rozdziale \ref{sec:chapterdyfuzja}. przedstawię analizę problemu transportu w modelu K\"unza i Laval\'ee'go oraz alternatywne wyjaśnienie uzyskanych wyników, nie wymagające odwoływania się do dyfuzji anomalnej. W rozdziale \ref{sec:chapterkretosc}. przeprowadzę szczegółową analizę numeryczną krętości przepływu \lb i jej korelacji z porowatością ośrodka oraz z innymi wielkościami makroskopowymi, które go charakteryzują. Rozdział \ref{sec:chapterpodsumowanie}. zawiera podsumowanie uzyskanych wyników.


\chapter{Model gazu sieciowego FHP}
\label{sec:gazsieciowy}

W tym rozdziale przedstawię teoretyczne podstawy oraz podstawowe zasady działania wprowadzonego w 1986 roku przez U. Frish'a, B. Hasslacher'a i Y. Pomeaou automatu komórkowego FHP \cite{Frish86}.

\section{Wstęp}

Jednym z pierwszych modeli transportu płynów, który nie wymaga jawnego rozwiązywania równań różniczkowych hydrodynamiki, jest opracowany przez Hardy'ego, de Pazzisa i Pomeau dwuwymiarowy model gazu sieciowego HPP \cite{Hardy73,Hardy76,Wolf00}. Jest to automat komórkowy o deterministycznych regułach kolizji opisujący dynamikę dyskretnych cząsteczek rozłożonych w węzłach regularnej sieci kwadratowej. Ze względu na symetrię sieci i tylko dwa wyróżnione kierunki prędkości cząsteczek na sieci, hydrodynamika modelu HPP jest jednak niefizyczna, gdyż otrzymywane rozwiązania nie są izotropowe. Co więcej, w skali makroskopowej nie są one niezmiennicze względem transformacji Galileusza. W roku 1986 Frish, Hasslacher i Pomeau pokazali, że podobny automat komórkowy (nazwany FHP od inicjałów autorów), ale zdefiniowany na sieci trójkątnej, reprezentuje w skali makroskopowej izotropowe równania przepływu Naviera--Stokesa w zakresie niskich liczb Macha \cite{Frish86}. Od tego czasu model FHP był przedmiotem zainteresowania różnych dziedzin nauki i techniki \cite{Biggs98}. Jedną z bardziej interesujących cech modelu jest względna prostota implementacji skomplikowanych (w sensie geometrycznym) warunków brzegowych, a co za tym idzie -- możliwość użycia tej metody do badań transportu w ośrodkach o strukturze porowatej.

\section{Definicja modelu}

\begin{figure}[!h]
 \centering
 \includegraphics[scale=0.7]{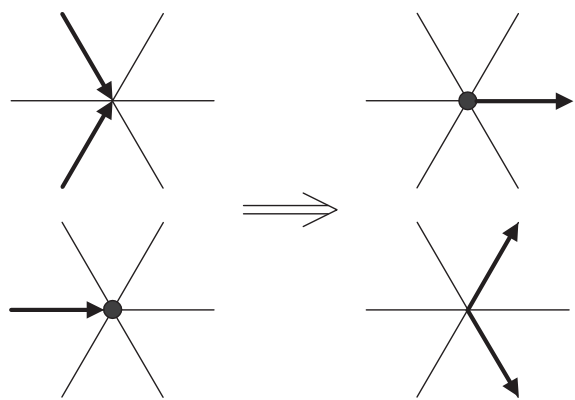}
   \caption{Reguły kolizji dla modelu FHP5. Symbol (\newmoon) oznacza cząstkę spoczywającą, strzałki oznaczają cząsteczki poruszające się w danym kierunku. Kompletny zestaw kolizji zawiera dodatkowo konfiguracje obrócone o wielokrotności kąta $\pi/3$. \label{pic:regulyfhp5}}
\end{figure}

Istnieje kilka wariantów modelu FHP, oznaczanych skrótami FHP1, FHP2 etc. Każdy z nich zdefiniowany jest na sieci trójkątnej, gdzie każdy z $N_x\times N_y$ węzłów posiada sześciu sąsiadów numerowanych od $1$ do $6$. W każdym węźle sieci może się znajdować $0\ldots 6$ cząsteczek o różnych kierunkach pędu. W zależności od wariantu model może dopuszczać dodatkową cząsteczkę spoczywającą o pędzie równym zeru, wtedy maksymalna liczba cząsteczek przypadająca na węzeł wynosi $7$. W żadnym węźle nie mogą znajdować się dwie cząsteczki poruszające się w tym samym kierunku. W \nlb przypadkach kilku cząsteczek zmierzających do tego samego węzła problem kolizji między nimi rozwiązywany jest przez ich proste przekonfigurowanie. W zależności od wariantu modelu ilość i rodzaj konfiguracji, dla których definiowana jest kolizja, różni się między sobą. Na przykład, najbardziej popularny wariant modelu, FHP3, zawiera $70$ kolizji dwu- i trzy- cząsteczkowych \cite{Droz98,Hod03}. W zależności od rodzaju kolizji inna jest lepkość modelowego płynu. W niniejszej rozprawie do symulacji zjawisk propagacji frontu koncentracji użyta zostanie wersja FHP5 o uproszczonych regułach kolizji przedstawionych na rysunku \ref{pic:regulyfhp5}.
\section{Warunki brzegowe i rozpraszacze}

Uwzględnienie warunków brzegowych w modelu gazu sieciowego należy do zadań względnie prostych. Węzeł sieci dostępny dla cząstek może zostać oznaczony jako węzeł rozpraszający. Implementacja warunku brzegowego bez poślizgu (\emph{ang. no slip}) sprowadza się do zmiany zwrotów pędów wszystkich cząsteczek znajdujących się w \nlb węzłach rozpraszających (więcej informacji o implementacji warunków odbicia od rozpraszaczy znajduje się w podrozdziale \ref{sec:modelkuntza}).

\section{Algorytm}

Algorytm działania modelu FHP składa się z dwóch kroków: 1) translacji i 2) kolizji. Cząsteczki rozmieszczone na sieci posiadają określony pęd. Zgodnie z jego kierunkiem w kroku translacji poruszają się one po prostych od węzła, w którym się aktualnie znajdują, do węzła sąsiedniego. Po przesunięciu cząsteczek, w kroku kolizji następuje odbicie cząsteczek od nieruchomych brzegów i rozpraszaczy oraz rozwiązanie kolizji międzycząsteczkowych. To ostatnie wykonuje się najczęściej poprzez umieszczenie konfiguracji ,,po'' zderzeniu w specjalnej tablicy.

\section{Wielkości makroskopowe}

Informacja, jaką otrzymujemy bezpośrednio z automatu komórkowego FHP, to chwilowe konfiguracje cząsteczek (ich położenie w węzłach sieci) oraz ich pędy. W celu wyznaczenia pól makroskopowych gęstości i prędkości w całym obszarze,  należy przeprowadzić procedurę uśredniania. Zdefiniujmy $(N_{\alpha})_i$ jako ilość cząsteczek znajdujących się w węźle $i$ i poruszających się w kierunku $\alpha$ \cite{Balasubramanian87}. Wtedy gęstość cząsteczek $\tilde{\varrho}$ i pęd makroskopowy $\tilde{\varrho}\vect{u}$ zapisać możemy w postaci sum po węzłach i kierunkach:
\begin{equation}
\tilde{\varrho} = \frac{2}{\sqrt{3}}\frac{1}{M}\sum_{i,\alpha} (N_{\alpha})_i,
\end{equation}
\begin{equation}
\tilde{\varrho}\vect{u} = \frac{2}{\sqrt{3}}\frac{1}{M}\sum_{i,\alpha} (N_{\alpha}\vect{c}_{\alpha})_i,
\end{equation}
gdzie $\vect{c}_i$ jest wektorem elementarnym sieci, $M$ jest liczbą węzłów, a czynnik $\frac{2}{\sqrt{3}}$ reprezentuje ilość węzłów na jednostkę powierzchni dla sieci trójkątnej o stałej sieci $c=1$. W dalszej części pracy, a szczególnie w rozdziale \ref{sec:chapterdyfuzja}, zamiast gęstości na jednostkę powierzchni często posługiwać się będziemy pojęciem koncentracji cząstek, którą definiujemy jako $\varrho = \tilde{\varrho}\sqrt{3}/2$.



\chapter{Gaz sieciowy Boltzmanna}
\label{sec:chapter3}


W tym rozdziale przedstawię model gazu sieciowego Boltzmanna oraz omówię jego zastosowanie do symulacji zjawisk transportu w ośrodkach porowatych. Przedstawię podstawy teoretyczne równania transportu Boltzmanna z przybliżeniem jednorelaksacyjnym członu kolizji. Omówię jeden ze sposobów rozwiązania tego równania -- model gazu sieciowego Boltzmanna. W celu weryfikacji opracowanego kodu numerycznego zostanie on użyty do rozwiązania zagadnienia przepływu przez prostokątny kanał, a wyniki zostaną porównane z rozwiązaniem analitycznym.

\section{Wstęp}

Automat komórkowy FHP, pomimo wielu interesujących właściwości w kontekście modelowania przepływów przez ośrodki porowate, posiada również wady, które znacznie ograniczają obszar jego zastosowań praktycznych. Do głównych jego niedoskonałości zaliczyć można szum statystyczny pojawiający się w uśrednianych wielkościach, będący naturalną konsekwencją jego mikroskopowego charakteru \cite{McNamara88}. Sposobem na obejście tego problemu jest zwykle zwiększenie rozmiarów obszarów, po których uśredniane są wielkości makroskopowe. Problem ten ma szczególne znaczenie w układach takich, jak ośrodki porowate, gdzie mamy do czynienia z dwiema skalami przestrzennymi: 1)  skalą mikroporów, w której zachodzi transport i rozwiązywane są równania transportu oraz 2) skalą makroskopową, w której zaniedbuje się detale i wprowadza prawa i właściwości uśrednione po objętości ośrodka. Już samo uwzględnienie obu z nich wymaga użycia dużej siatki obliczeniowej, stąd potrzeba dodatkowego jej zwiększenia ze względu na problem zaszumienia wyników w zasadzie eliminuje metodę FHP z praktycznego użycia do badania transportu hydrodynamicznego w układach porowatych. W roku 1988 G. McNamara i G. Zanetti wprowadzili pojęcie gazu sieciowego Boltzmanna (LBM) jako metody pozwalającej wyeliminować szum statystyczny z wyników otrzymywanych metodami gazów sieciowych \cite{McNamara88,Succi01}. Historycznie metoda ta wywodzi się z klasycznych automatów komórkowych, w których dyskretna zmienna $(N_{\alpha})_{i}$, oznaczająca liczbę cząsteczek znajdujących się w węźle $i$ i poruszających się w kierunku $\alpha$ \cite{Balasubramanian87}, została zastąpiona ciągłą zmienną $(f_{\alpha})_{i}$ przyjmującą dowolne wartości z przedziału $[0;1]$. Wielkość $(f_{\alpha})_{i}$ oznacza w tym kontekście prawdopodobieństwo znalezienia cząsteczki znajdującej się w węźle $i$ i poruszającej się w kierunku $\alpha$. Obecnie podstawę teoretyczną działania modelu LBM stanowi teoria kinetyczna Boltzmanna \cite{Boltzmann95}.

\section{Teoria kinetyczna}
\subsection{Funkcja rozkładu}
Kompletny stan układu mikroskopowego można opisać poprzez podanie zestawu położeń $\v{x}$ oraz pędów $\v{p}$ wszystkich jego cząstek. Teoretycznie śledzenie ewolucji układu można zrealizować poprzez śledzenie trajektorii każdej z nich. Ze względu na ich ogromną ilość, przy rozpatrywaniu dużych układów podejście takie jest jednak niepraktyczne. Możemy jednak przejść z przestrzeni fazowej, w której znajdują się wszystkie cząstki (atomy) reprezentujące badany układ, do przestrzeni konfiguracyjnej położeń i pędów jednej cząstki \cite{Huang78}. Wprowadza się w tym celu pojęcie funkcji rozkładu zdefiniowanej tak, by wyrażenie:
\begin{equation}\label{frozkladu}
f(\v{x},\v{v},t)~d^3\!x~d^3\!v,
\end{equation}
opisywało liczbę cząsteczek w skończonym elemencie przestrzeni konfiguracyjnej położeń i pędów o objętości $d^3\!x\times d^3\!v$.
\begin{figure}[!ht]
\centering
\includegraphics[scale=1.0]{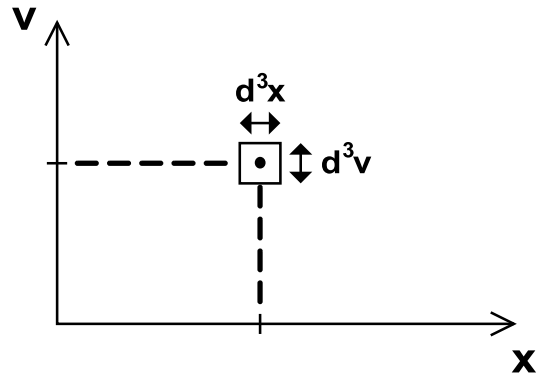}
\caption{Przestrzeń konfiguracyjna położeń i pędów z zaznaczonym jednym punktem wraz z jego otoczeniem o objętości $d^3\!x\times d^3\!v$.\label{pic:konf}}
\end{figure}
Na rysunku \ref{pic:konf} zaznaczony został punkt w \nlb przestrzeni konfiguracyjnej wraz z otoczeniem. Wartość funkcji rozkładu w tym punkcie, $f(\v{x},\v{v},t)$, mówi nam o tym, ile (średnio) cząstek o pędzie $\v{v}$ i położeniu $\v{x}$ znajdziemy w układzie w chwili \emph{t}. Wprowadzenie funkcji rozkładu pozwoliło ograniczyć problemy związane z dyskretnym charakterem modelu gazu sieciowego -- przede wszystkim wyeliminowało potrzebę czasochłonnych uśrednień wielkości makroskopowych. Można wręcz powiedzieć, że proces uśredniania wyników jest tu realizowany już na etapie opisu układu przez funkcję rozkładu.
\subsection{Równanie transportu}
W zagadnieniach, którymi będziemy się zajmować, istotna jest dynamika układu oraz proces jego dochodzenia do stanu równowagi. Dynamikę układu w przestrzeni pędów i położeń opisuje słynne równanie transportu Boltzmanna, które w postaci ciągłej przyjmuje postać \cite{He97}:
\begin{equation}\label{eq:transportcollisions}
\left( \frac{\partial}{\partial t} + \v{v}\cdot\nabla_{\v{r}} + \left(\frac{\v{F}}{m}\right)\cdot\nabla_{\v{v}} \right) f(\v{r},\v{v},t) = \left(\frac{\partial f}{\partial t}\right)_\mathrm{zderz},
\end{equation}
gdzie $\left(\frac{\partial f}{\partial t}\right)_\mathrm{zderz}$ jest członem odpowiadającym za opis kolizji międzycząsteczkowych, $\v{F}$ -- siłą zewnętrzną działającą na cząsteczki układu, $m$ -- masą cząsteczek, a $f(\v{r},\v{v},t)$ -- funkcją rozkładu, której ewolucję badamy \cite{Huang78}. Równanie to było przedmiotem badań na przestrzeni wielu lat, począwszy od prób szukania rozwiązań analitycznych w \nlb prostych przypadkach jednowymiarowych, po badania numeryczne w dwóch \lb i trzech wymiarach przestrzennych. Co ciekawe, historycznie równanie \ref{eq:transportcollisions} nie stanowiło punktu wyjścia do metody gazu sieciowego Boltzmanna. Wyprowadzenie takie zostało przedstawione dopiero w roku 1997 \cite{He97}.
\subsection{Przybliżenie BGK}
Dla wyrażenia odpowiadającego kolizjom w równaniu (\ref{eq:transportcollisions}) stosuje się najczęściej przybliżenie jedno-relaksacyjne opracowane przez Bhatnagara, Grossa i Krooka \lb w 1954 roku (BGK) \cite{Bhatnagar54,Huang78}:
\begin{equation}\label{ruchcollision}
\left(\frac{\partial f}{\partial t}\right)_\mathrm{zderz} \approx -\frac{f-f^\mathrm{eq}}{\tau},
\end{equation}
gdzie $\tau$ to średni czas relaksacji pojedynczego zderzenia. Funkcja rozkładu $f^\mathrm{eq}$ opisuje stan układu w równowadze termodynamicznej i wyznacza się ją zwykle z rozkładu Maxwella--Boltzmanna:
\begin{equation}\label{eq:maxwellrozklad}
	f^\mathrm{eq}(\v{v}) = \frac{\varrho}{(2\pi \kb T)^{d/2}}\exp\left[-\frac{\v{v}^2}{2\kb T}\right],
\end{equation}
gdzie $\kb$ jest stałą Boltzmanna, $T$ -- temperaturą układu, a $\v{v}$ -- prędkością cząsteczek.
Model BGK zdobył uznanie ze względu na prostotę i intuicyjne potraktowanie zderzeń jako procesu lokalnego dochodzenia układu do stanu równowagi ze stałym czasem relaksacji. Istnieje osobna gałąź badań nad wielorelaksacyjnymi modelami zderzeń w gazie sieciowym Boltzmanna -- MRT (\emph{ang. multiple relaxation time}) -- które odznaczają się większą dokładnością w szerszym zakresie numerycznych lepkości płynu. W literaturze znaleźć można wiele porównań modeli BGK i MRT \cite{Dhumieres02,Pan06}. W \nlb niniejszej rozprawie ograniczamy się do stosowania najpopularniejszego przybliżenia BGK, które doskonale radzi sobie z modelowaniem zjawisk dla interesujących nas parametrów \cite{Pan06}.
\section{Model D2Q9}
Równanie (\ref{eq:transportcollisions}) przy braku siły zewnętrznej można zapisać w postaci dyskretnej (przyjmując $\delta t=1$) \cite{Nourgaliev03,He97}:
\begin{equation}\label{transportcollisionsdiscret}
f_i(\v{r}+\v{e}_i,t+1) = f_i(\v{r},t) +\left(\frac{\partial f}{\partial t}\right)_\mathrm{zderz}.
\end{equation}
Wektory $\v{e}_i$ są dyskretnymi wektorami prędkości rozpinającymi sieć o zadanej symetrii. Najczęściej używanym modelem jest wersja dwuwymiarowa o $9$ możliwych wektorach $\v{e}_i$: $(0,0)$, $(1,0)$, $(0,1)$, $(-1,0)$, $(0,-1)$, $(1,1)$, $(-1,1)$, $(-1,-1)$ oraz $(1,-1)$.
Ta wersja modelu funkcjonuje w literaturze pod nazwą D2Q9, gdzie $\mathrm{D}=2$ oznacza wymiar przestrzenny, a $\mathrm{Q}=9$ ilość dozwolonych wektorów prędkości. W literaturze funkcjonuje szereg modeli zdefiniowanych w jednym, dwóch i trzech wymiarach przestrzennych z różną liczbą dozwolonych kierunków wektorów prędkości \cite{He97}.
\subsection{Funkcja rozkładu i wielkości makroskopowe}
Znajomość funkcji rozkładu dla badanego układu umożliwia wyznaczenie hydrodynamicznych wielkości makroskopowych. Wielkości opisujące hydrodynamikę badanego układu wyrazić można przez kolejne momenty funkcji rozkładu.  Korzystając z \nlb jej dyskretnej postaci i zakładając, że model zdefiniowano z $9$ dozwolonymi wektorami prędkości, wielkości makroskopowe takie jak gęstość $\varrho$ oraz prędkość lokalną $\v{u}$ \nlb możemy wyznaczyć poprzez sumy \cite{Sukop06}:
\begin{equation}
\varrho = \sum_{i=1}^{9} f_i,
\end{equation}
\begin{equation}
\varrho \v{u} = \sum_{i=1}^{9} f_i\v{e}_i,
\end{equation}
gdzie $f_i$ jest dyskretną funkcją rozkładu, $\v{e}_i$ są wektorami sieci, a zakresy sumowania w powyższych wzorach zależą od przyjętej symetrii sieci oraz liczby wymiarów w \nlb konkretnej wersji modelu.

Ciśnienie w modelu LBM wyznacza się z równania stanu:
\begin{equation}
p=c_s^2\varrho,
\end{equation}
gdzie stała $c_s=1/\sqrt{3}$, jest prędkością dźwięku dla omawianego modelu \cite{Latt07}.

\subsection{Model LBM w przybliżeniu BGK}
Równanie (\ref{ruchcollision}) wraz z (\ref{transportcollisionsdiscret}) wyrażają ostateczną postać dyskretną równania na ewolucję funkcji rozkładu w modelu gazu sieciowego Boltzmanna w przybliżeniu BGK:
\begin{equation}\label{lbmbgk}
f_i(\v{r}+\v{e}_i,t+1) = f_i(\v{r},t) -\frac{f_i-f_i^\mathrm{eq}}{\tau},
\end{equation}
Postać równowagową funkcji rozkładu otrzymać możemy w zakresie niskich prędkości, rozwijając rozkład Maxwella-Boltzmanna w szereg z dokładnością do wyrazów drugiego rzędu \cite{Latt07}:
\begin{equation}\label{feq_maxwell}
f^\mathrm{eq}_i = \varrho \omega_i \left[ 1 + 3\v{e}_i\cdot\v{u} + \frac{9}{2}(\v{e}_i\cdot\v{u})^2 -\frac{3}{2}u^2 \right],
\end{equation}
gdzie $\omega_0=4/9$, $\omega_{1,3,5,7}=1/9$, a $\omega_{2,4,6,8}=1/36$ \cite{Nourgaliev03,Chen98}.
\subsection{Błąd dyskretyzacji}
Podstawową kwestią jest wielkość błędu, jaki popełniamy, rozwiązując równania transportu metodą LBM, względem dokładnego rozwiązania równań Naviera--Stokesa. Okazuje się \cite{Chen98,Succi01}, że rozwiązanie modelu opisanego równaniami (\ref{lbmbgk}) w przybliżeniu jednorelaksacyjnym ma dokładność drugiego rzędu w czasie i przestrzeni jeśli tylko błąd dyskretyzacji włączy się w lepkość kinematyczną cieczy. W związku z tym, w modelu D2Q9 lepkość kinematyczną definiuje się wzorem:
\begin{equation}\label{lepkosc_kinematyczna}
\nu = \frac{2\tau-1}{6},
\end{equation}
gdzie $\tau$ jest charakterystycznym czasem relaksacji w wyrazie kolizji.
\subsection{Warunki brzegowe}
Poprawna implementacja skomplikowanych geometrycznie warunków brzegowych to jeden z najtrudniejszych problemów w obliczeniach numerycznych. Metoda LBM zdobyła uznanie z powodu zgodnego z intuicją i bardzo prostego w implementacji warunku na ewolucję funkcji rozkładu przy sztywnej ściance w przepływie bez poślizgu (\emph{ang. no slip boundary condition}) \cite{Chen98}. Okazuje się jednak, że stosowanie tego prostego warunku na odbicie funkcji rozkładu przy ściankach powoduje lokalną stratę dokładności rozwiązania i redukcję dokładności do pierwszego rzędu. Dzięki zastosowaniu stałego czasu relaksacji $\tau=1$ użytego w niniejszej pracy i zastosowaniu ulepszonej procedury odbicia (\emph{ang. half way bounce back}) uzyskaliśmy dokładność drugiego rzędu w czasie i przestrzeni \cite{Succi01}.

\subsection{Algorytm}
Typowy algorytm metody gazu sieciowego Boltzmanna sprowadza się do wykonania dwóch kroków:
\begin{itemize}
\item[1)] kolizji, czyli uwzględnienia członu kolizji zgodnie z równaniem:
\begin{equation}
\tilde{f}_i(\v{r},t) = f_i(\v{r},t) - \frac{f_i(\v{r},t)-f_i^\mathrm{eq}(\v{u},\varrho)}{\tau},
\end{equation}
gdzie $\tilde{f}_i(\v{r},t)$ to pewne zmienne pomocnicze,
\item[2)] propagacji, czyli transportu funkcji rozkładu zgodnie z równaniem Boltzmanna (\ref{lbmbgk}):
\begin{equation}
f_i(\v{r}+\v{e}_i,t+1) = \tilde{f}_i(\v{r},t).
\end{equation}
\end{itemize}
Algorytm symulacji sprowadza się więc do rozdzielenia zagadnienia transportu na dwa oddzielne procesy -- transportu i lokalnych kolizji cząstek, czyli tak, jak w przypadku omawianego wcześniej modelu FHP.
\section{Weryfikacja kodu numerycznego}
\begin{figure}[!h]
 \centering
 \includegraphics[scale=0.65]{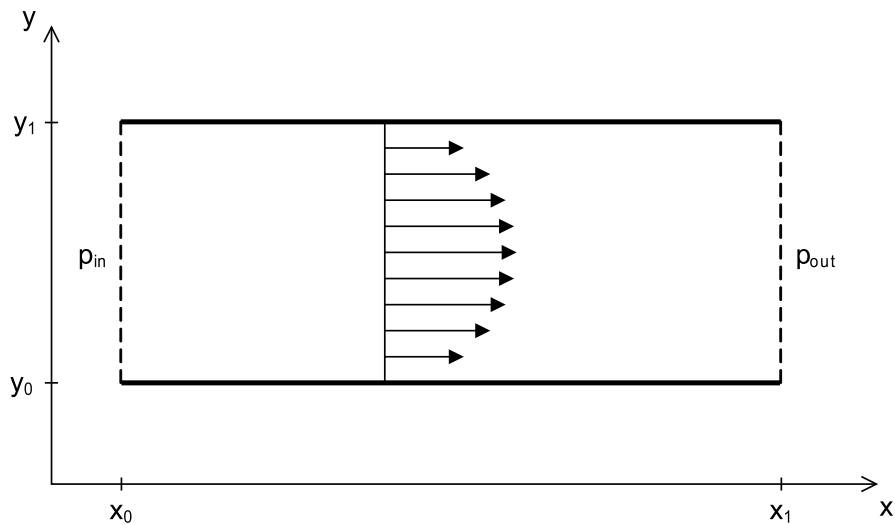}
   \caption{Schemat problemu przepływu przez zamknięty kanał (\emph{ang. Poiseuille flow}). \label{pic:poiseuilleflow}}
\end{figure}

W celu weryfikacji implementacji metody LBM stworzonej na potrzeby niniejszej rozprawy, rozwiązany został problem przepływu przez prosty kanał w obecności siły wymuszającej (lub równoważnej różnicy ciśnień na wlocie i wylocie z kanału). Ze względu na istniejące rozwiązanie analityczne problem ten, znany w literaturze jako przepływ Pouseuille'a (\emph{ang. Poiseuille flow}), jest często używany do weryfikacji poprawności implementacji numerycznej. Na rysunku \ref{pic:poiseuilleflow} przedstawiony został schemat omawianego problemu.
Obszar prostokątny ograniczony punktami $x_0$, $x_1$, $y_0$ oraz $y_1$ jest wypełniony płynem. Ścianki poziome w położeniach $y=y_0$ oraz $y=y_1$ są sztywne (brak poślizgu, t.j. prędkość $v_x=0$ na ściankach). Ze względu na różnicę ciśnień pomiędzy przekrojami kanału dla $x=x_0$ oraz $x=x_1$ w obszarze kanału następuje przepływ cieczy. Obliczenia analityczne i rozwiązanie uproszczonego równania przepływu Naviera--Stokesa prowadzi do wniosku, że profil prędkości w takim kanale ma charakter paraboliczny i przy założeniu, że $y_0=0$ i $y_1=h$ spełnia równanie:
\begin{equation}
\v{v}(x,y)=(v_x,v_y)=\left(0,-\frac{1}{2\nu\varrho}\frac{dp}{dx}y(y - h)\right),
\end{equation}
gdzie $\frac{dp}{dx}$ jest gradientem ciśnienia między wlotem, a wylotem z kanału, $\nu$ -- lepkością kinematyczną, $\varrho$ -- gęstością płynu, a $h$ -- szerokością kanału \cite{Landau94}.

\subsubsection{Warunki brzegowe i początkowe}
Zgodnie z oznaczeniami z rysunku \ref{pic:poiseuilleflow} zakładamy warunki początkowe prędkości oraz ciśnienia w postaci:
\begin{equation}
\begin{array}{c}%
  v_x(x,y,t=0)=v_y(x,y,t=0)=0,\\
  p(x,y,t=0)=p_0,\\
\end{array}
\end{equation}
w całym obszarze kanału, gdzie $p_0 = (p_\mathrm{in}-p_\mathrm{out})/2$ ($\emph{ang. in}$ -- wlot, $\emph{ang. out}$ -- wylot \lb z kanału).

Jako warunki brzegowe ciśnienia na wlocie ($x=x_0$) i wylocie z kanału ($x=x_1$) \lb w dowolnej chwili przyjęte zostały stałe ciśnienia $p_\mathrm{in}$ i $p_\mathrm{out}$, gdzie $p_\mathrm{in}>p_\mathrm{out}$. Komplet warunków brzegowych, włączając te nakładane na pole prędkości, wyrazić można następująco:
\begin{equation}
\begin{array}{c}%
  v_x(x,y_0,t)=v_x(x,y_1,t)=v_y(x,y_0,t)=v_y(x,y_1,t)=0,\\
  p(x_0,y,t)=p_\mathrm{in},\quad t\geq 0, \\
  p(x_1,y,t)=p_\mathrm{out},\quad t\geq 0.\\
\end{array}
\end{equation}
\subsubsection{Porównanie z rozwiązaniem analitycznym}
\begin{figure}
\centering
\includegraphics[scale=0.73]{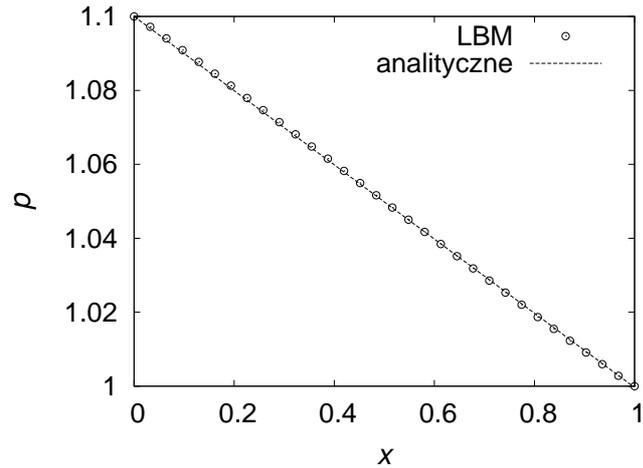}
\caption{Porównanie profilu ciśnienia dla problemu przepływu przez prosty kanał z \protect\nlb rozwiązaniem analitycznym (\ref{porownanieanal}) dla $a=0.1$.\label{pic:pois1}}
\end{figure}

\begin{figure}
\centering
\includegraphics[scale=0.73]{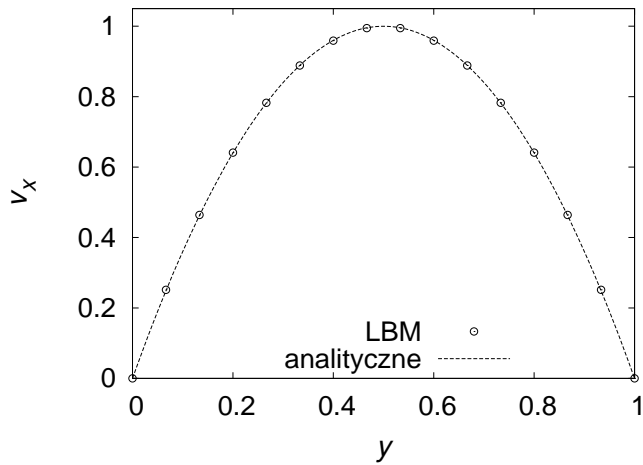}
\caption{Porównanie profilu prędkości dla problemu przepływu przez prosty kanał z \protect\nlb rozwiązaniem analitycznym (\ref{porownanieanal}) dla $b=1.77$.\label{pic:pois2}}
\end{figure}

W zagadnieniu testowym, pomiędzy wlotem, a wylotem z rury, przyjęty został gradient ciśnienia $\Delta p = p_\mathrm{in}-p_\mathrm{out}=0.1$  \emph{mu ts}$^{-2}$ (\emph{ang. mass unit per time step squared}) \cite{Sukop05}.
Problem został rozwiązany na sieci prostokątnej o wymiarach $32\times 16$ dla lepkości kinematycznej $\nu=1.0$ (co daje $\tau=0.4$). Przyjmując $y_0=0$, $y_1=1$, $x_0=0$ i $x_1=1$, rozwiązanie analityczne pól prędkości i ciśnienia dla omawianego problemu można zapisać w wygodnej postaci:
\begin{equation}\label{porownanieanal}
\begin{array}{c}%
  p(x,y,t)=p_\mathrm{in} - a\cdot x,\\
  v_x(x,y,t)=b\cdot y(1-y), \\
  v_y(x,y,t)=0,\\
\end{array}
\end{equation}
gdzie zmienne $a$ oraz $b$ użyte są do dopasowania wzorów do danych pomiarowych \cite{Guo00}. Na rysunkach \ref{pic:pois1} oraz \ref{pic:pois2} przedstawione zostały wyniki odpowiednio dla ciśnienia oraz prędkości w kierunku $x$ odczytanej wzdłuż przekroju przez kanał. 
Paraboliczny profil prędkości na przekroju kanału oraz liniowa zależność ciśnienia od odległości do wlotu rury uzyskane przy pomocy modelu LBM odpowiadają doskonale przedstawionym rozwiązaniom analitycznym.


\addtocontents{toc}{\protect\newpage}
\chapter{Transport w modelu K\"{u}ntza-Lavall\'{e}e'go}
\label{sec:chapterdyfuzja}
W tym rozdziale przedstawię krytyczną analizę występowania dyfuzji anomalnej \lb w modelu K\"{u}ntza-Lavall\'{e}e'go (KL). Uwzględniona i zbadana zostanie rola i znaczenie dryfu hydrodynamicznego w powstawaniu obserwowanych zjawisk ,,anomalnych''. \\
%
Rozdział ten opiera się na wynikach opublikowanych w pracy:\\ \\
M. Matyka, Z. Koza,\\
\textit{Spreading of a density front in the Küntz-Lavallée model of porous media},\\
J. Phys. D: Appl. Phys. 40, 4078-4083~(2007).
\section{Opis zagadnienia}
Jedną z podstawowych właściwości klasycznej dyfuzji pojedynczej cząsteczki jest skalowanie się jej przesunięcia średniokwadratowego (\emph{ang. root-mean-square displacement}) $\sqrt{\langle r^2\rangle}$ z czasem: $\sqrt{\langle r^2\rangle}\propto t^{\alpha}$, gdzie $\alpha=1/2$. Istnieją jednak doniesienia o \nlb wielu naturalnych zjawiskach, w których skalowanie z wykładnikiem $1/2$ nie zachodzi, a w ostatnich latach wysunięto nawet hipotezę, iż zjawisko dyfuzji anomalnej występuje w układach, w których współczynnik dyfuzji zależny od koncentracji dyfundującej substancji \cite{Kuntz03}. Hipoteza ta została uzasadniona zarówno przy użyciu symulacji modelem gazu sieciowego, jak i na gruncie doświadczalnym \cite{Kuntz04,Kuntz05}. Zasadniczym celem dalszej części tego rozdziału jest weryfikacja powyższej hipotezy, powtórzenie otrzymanych wcześniej wyników oraz próba ilościowego wytłumaczenia otrzymanych odchyleń od wykładników klasycznych bez odwoływania się do dyfuzji anomalnej.
\subsection{Model K\"{u}ntza-Lavall\'{e}e'go}\label{sec:modelkuntza}
\begin{figure}[!h]
\centering
\includegraphics[scale=0.8]{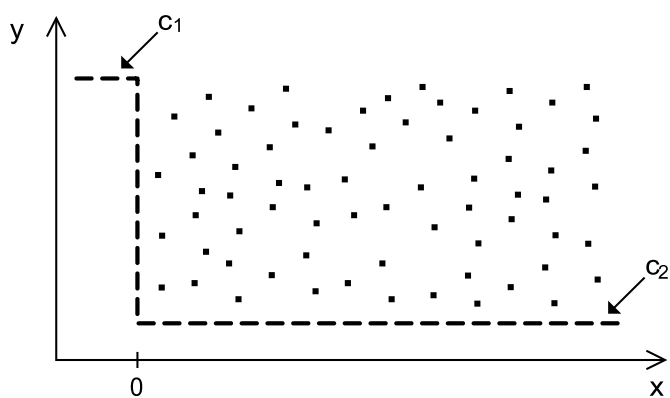}
\caption{Model ośrodka porowatego zbudowanego z rozpraszaczy ($\blacksquare$).  Linia przerywana to schematyczny profil koncentracji początkowej z uskokiem przy $x=0$.  \label{pic:modelzrozpraszaczami}}
\end{figure}

Model K\"{u}ntza-Lavall\'{e}e'go (KL) symulacji transportu płynu przez ośrodek porowaty oparty został na standardowym modelu gazu sieciowego FHP. Płyn reprezentowany jest przez jednakowe, dyskretne cząsteczki rozłożone w węzłach sieci o symetrii heksagonalnej. W każdym węźle sieci znajdować się może maksymalnie do 7 cząsteczek, każda o innym kierunku prędkości (włączając w to cząstki spoczywające o pędzie $p=0$).
Do reprezentacji mikrostruktury ośrodka porowatego wykorzystane zostały komórki rozpraszające rozłożone losowo w obszarze, w którym zachodzi transport substancji (rysunek \ref{pic:modelzrozpraszaczami}), co odróżnia model KL od standardowego gazu sieciowego.
Pojedynczy krok symulacji składa się z dwóch części: 1) translacji wzdłuż wektorów sieci oraz 2) kroku kolizji. Pierwszy jest standardowym balistycznym ruchem cząsteczek wzdłuż wektorów sieci. W drugim kroku, oprócz kolizji cząstek między sobą (opisanych wcześniej w rozdziale \ref{sec:gazsieciowy}),  uwzględnione są odbicia od rozłożonych w obszarze transportu rozpraszaczy. Odbicie to zachodzi całkowicie sprężyście (cząsteczki zachowują swoją energię, zmieniając jedynie kierunek pędu). Przykład rozwiązania kolizji cząsteczki z rozpraszaczem został przedstawiony na rysunku \ref{pic:modelzrozpraszaczami}.
\begin{figure}[!h]
\centering
\includegraphics[scale=0.7]{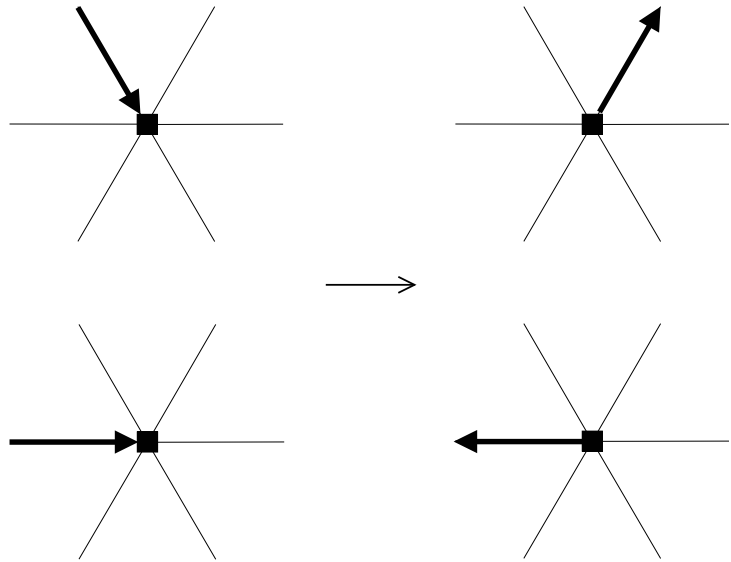}
\caption{Przykład kolizji cząstek z rozpraszaczem ($\blacksquare$) umieszczonym w węźle sieci. Zgodnie z zasadą zachowania energii, cząstka odbija się od przeszkody, zmieniając kierunek pędu i zachowując jego wartość. \label{pic:kolizjerozpraszacze}}
\end{figure}
\subsection{Warunki brzegowe}
W celu zbadania dynamiki rozchodzenia się frontów gęstości, na koncentrację płynu nałożony jest warunek schodkowy reprezentowany równaniem:
\begin{equation}\label{eq:initstep}
\begin{split}
    c(x,t\geq0) & = c_1H(-x), \;x<0, \\
	c(x,t=0) & = c_2H(x), \;x>0,\\
\end{split}
\end{equation}
gdzie $H(x)$ jest funkcją skokową Heaviside'a, a punkt $x=0$ jest granicą oddzielającą obszary o koncentracji początkowej $c_1$ i $c_2$. Za warunek początkowy rozpatrywanego problemu przyjmiemy iloraz koncentracji $c_1/c_2>1$. Dzięki istnieniu gradientu koncentracji na granicy $x=0$, spodziewany jest transport cząsteczek gazu w kierunku obszaru o niższej koncentracji. Na górnej oraz dolnej ściance układu zastosowane zostały warunki periodyczne. Ścianki pionowe ustalone zostały jako sztywne przeszkody od których cząstki odbijają się sprężyście. Zgodnie z równaniem (\ref{eq:initstep}), w obszarze $x<0$ utrzymywana jest koncentracja $c(x)=c_1$ dla wszystkich $t>0$, dzięki czemu zapewniony jest stały dopływ cząsteczek gazu do układu. Dzięki użyciu siatek obliczeniowych o dużej długości układ zachowuje się jak układ nieskończony (brak widocznego wpływu prawego brzegu na otrzymywane rezultaty).

\section{Wyniki}
\subsection{Propagacja frontu koncentracji}

\begin{figure}[!hb]
\centering
\includegraphics[scale=0.74]{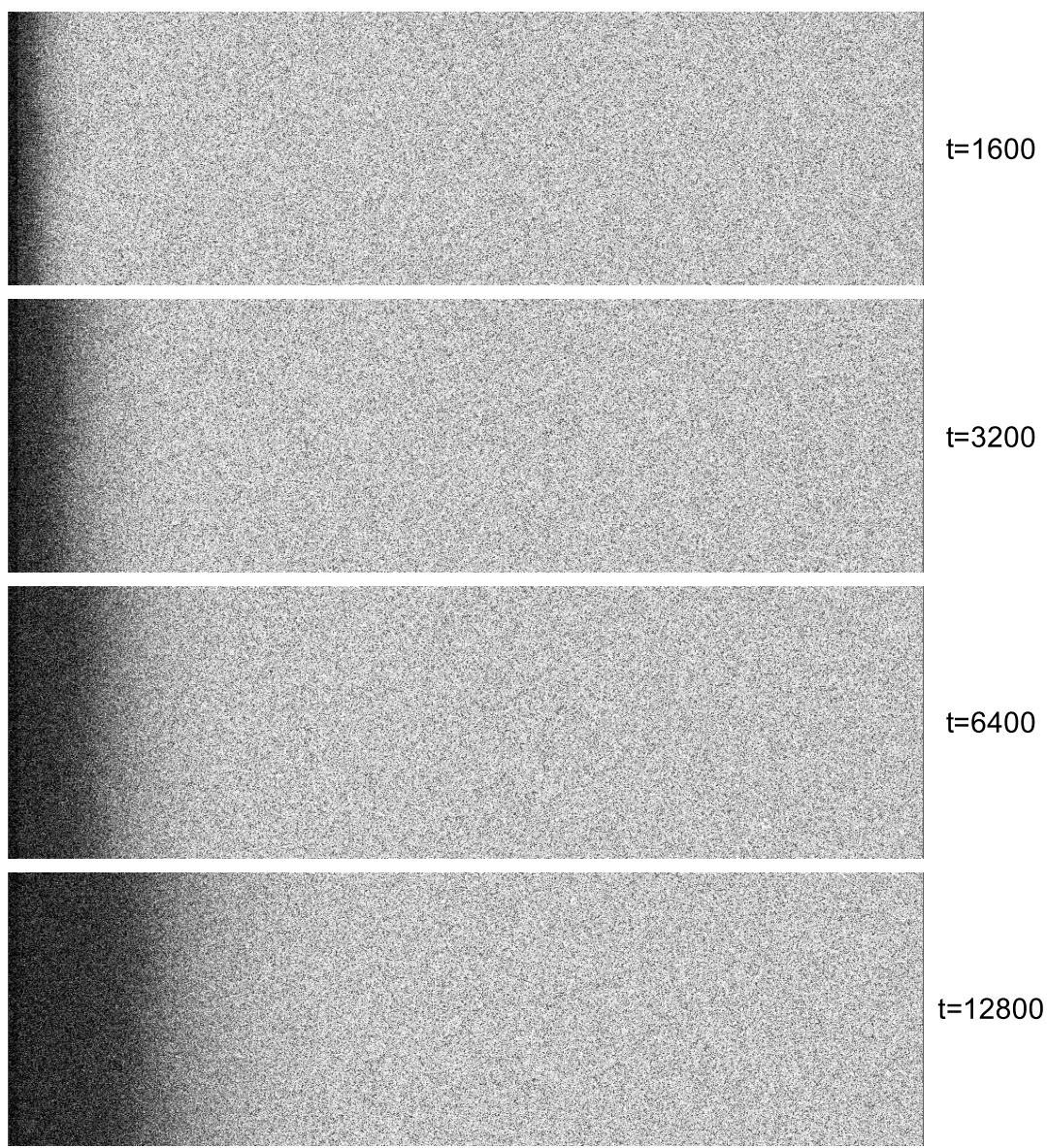}
\caption{Propagacja frontu gęstości dla czasów $t=1600\times2^k$, $k=0,1,2,3$. Ciemniejsze obszary odpowiadają większej koncentracji cząstek. \label{pic:wynikzrzuty}}
\end{figure}

Przeprowadziliśmy symulację rozchodzenia się frontu koncentracji na sieci o rozmiarach $\mathrm{LX}\times\mathrm{LY}=10000\times 1000$ z koncentracją rozpraszaczy równą $c_s=0.08$. Warunki brzegowe zostały ustalone zgodnie z równaniem (\ref{eq:initstep}), w którym przyjęto $c_1=0.9$ i $c_2=0.2$. Rozkłady cząstek dla $t=1600\times2^k$, $k=0,1,2,3$ zostały przedstawione na rysunku \ref{pic:wynikzrzuty}.

\begin{figure}[!h]
\centering
\includegraphics[scale=0.58]{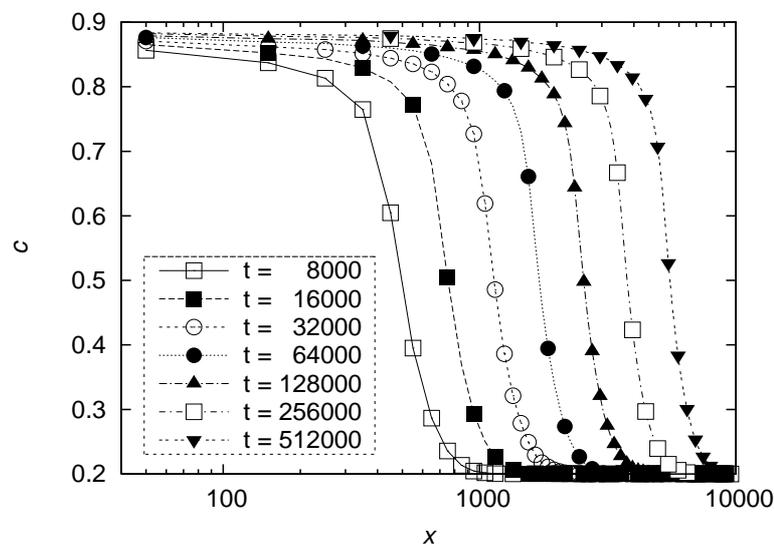}
\caption{Profile koncentracji $c(x,t)$ dla czasów $t=8000\times2^k$, $k=0,1,\ldots,6$. Warunki brzegowe koncentracji $c_1=0.9$ i $c_2=0.2$. \label{pic:frontygestosci}}
\end{figure}

Profile gęstości dla $t=8000\times2^k$, $k=0,1,\ldots,6$ zostały pokazane na rysunku \ref{pic:frontygestosci} w pół-logarytmicznym układzie współrzędnych.
\begin{figure}[!h]
\centering
\includegraphics[scale=0.58]{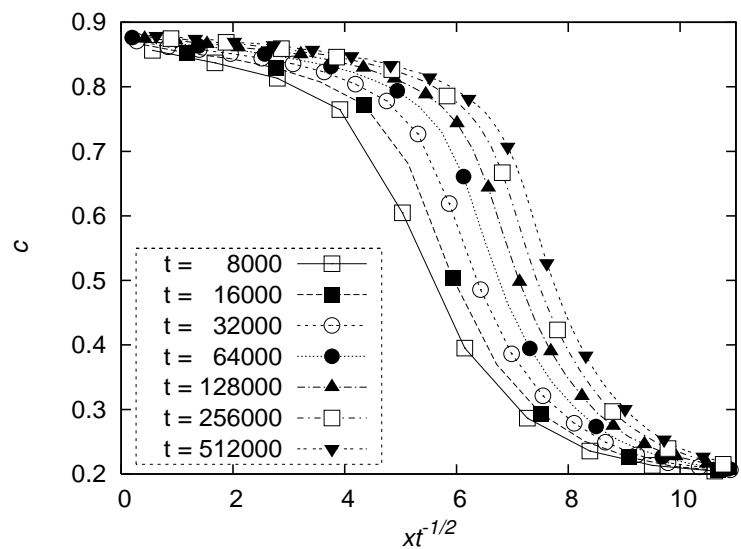}
\caption{Profile koncentracji $c(x,t)$ z rysunku \ref{pic:frontygestosci} w funkcji $xt^{-1/2}$. \label{pic:frontygestoscialpha05}}
\end{figure}
Kształty oraz odległości pomiędzy krzywymi koncentracji dla poszczególnych pomiarów nie różnią się bardzo między sobą, co pozwala wysunąć hipotezę, że profil $c(x,t)$ można opisać jako funkcję jednej zmiennej $x/t^\alpha$. Można to łatwo zweryfikować, rysując profile koncentracji dla różnych czasów w funkcji $x/t^\alpha$. Jeśli hipoteza o skalowaniu jest prawdziwa -- profile powinny ułożyć się jeden na drugim w pojedynczą krzywą. W naszym przypadku próba skalowania profili z wykładnikiem $\alpha=1/2$ nie jest udana, co widać wyraźnie na rysunku \ref{pic:frontygestoscialpha05} (profile nie układają się na jednej krzywej).
Analiza taka została przeprowadzona wcześniej w pracy \cite{Kuntz03}, w której pokazano, że najlepsze dopasowanie uzyskuje się dla $\alpha\approx 0.55$, co zinterpretowano jako super-dyfuzję o $\alpha>0.5$. Założenie o super-dyfuzji powoduje jednak dość paradoksalny wynik przyspieszenia transportu poprzez dodanie rozpraszaczy. K\"{u}ntz i Lavall\'{e}e w \cite{Kuntz03} tłumaczyli to na dwa sposoby. Pierwszy to super-dyfuzyjny charakter transportu w układach, w których współczynnik dyfuzji zależy silnie od koncentracji substancji (z takim właśnie układem mamy do czynienia w przypadku modelu FHP). Drugie wytłumaczenie mówi, że anomalna dyfuzja jest tylko zjawiskiem przejściowym i dla dostatecznie dużych czasów $t\rightarrow\infty$ wykładnik $\alpha$ będzie bardzo wolno zbiegał do klasycznej wartości $1/2$.

\begin{figure}[!h]
\centering
\includegraphics[scale=0.83]{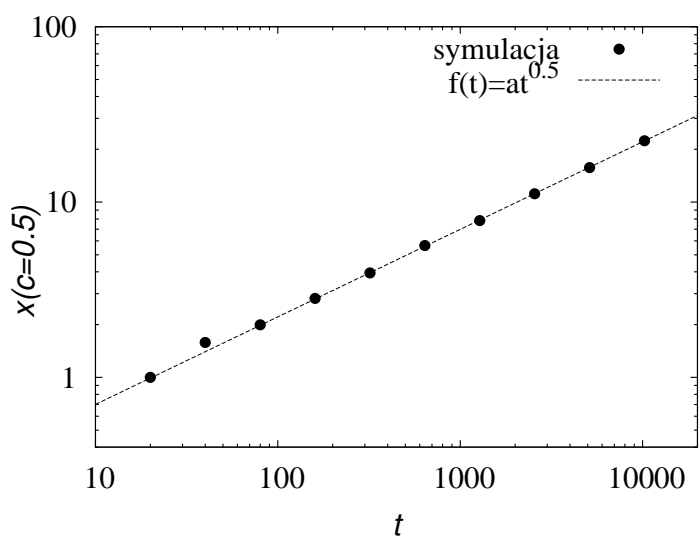}
\caption{Zależność położenia wartości $c=0.5$ frontu koncentracji w funkcji czasu. Rozwiązanie numeryczne równania (\ref{eq:fickdc}) ze współczynnikiem $D(c)$ odczytanym z\protect\nlb~pracy \cite{Kuntz03}. \label{pic:rozwiazanieficka}}
\end{figure}

Jednak zgodnie z analizą Botzmanna-Matano \cite{Gomer90,Crank56}, każde rozwiązanie równania dyfuzji postaci:
\begin{equation}
 \label{eq:fickdc}
  \frac{\partial c(x,t)}{\partial t}
     =
  \frac{\partial}{\partial x}
     \left(
       D(c)\frac{\partial c(x,t)}{\partial x}
     \right),
\end{equation}
z warunkami brzegowymi (\ref{eq:initstep}) skaluje się z wykładnikiem $\alpha=1/2$ dla każdego $t$, niezależnie od postaci $D(c)$ \cite{Crank56}. W celu weryfikacji tego twierdzenia rozwiązałem numerycznie równanie (\ref{eq:fickdc}) z uwzględnieniem dokładnej postaci funkcji $D(c)$ dla modelu FHP5 bezpośrednio z \cite{Kuntz03}. Wynik został przedstawiony na rysunku \ref{pic:rozwiazanieficka}.
Widać wyraźnie, że otrzymane rozwiązanie równania dyfuzji skaluje się klasycznie w całym obszarze badanych czasów.

\begin{figure}[!h]
\centering
\includegraphics[scale=0.6]{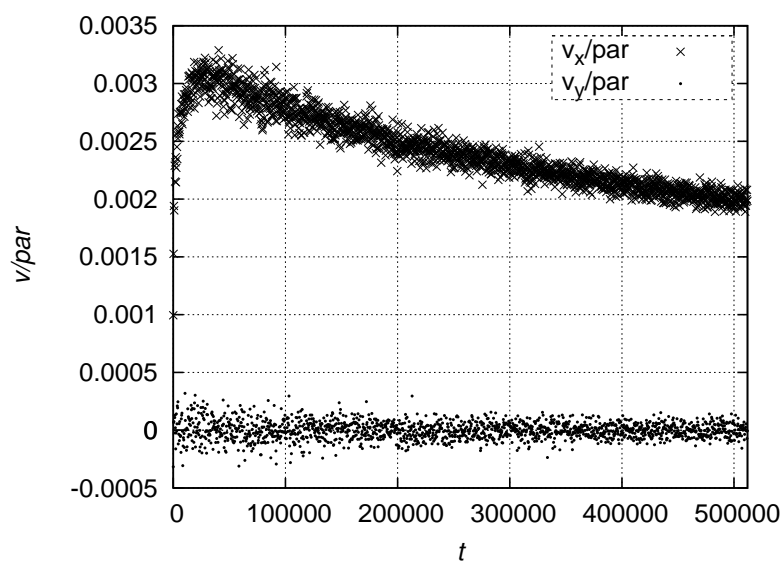}
\caption{Prędkość kierunku $x$ oraz $y$ przypadająca na jedną cząsteczkę płynu od czasu. Wartości podane zostały w jednostce \emph{lu tu}$^{-1}$ (stała sieci / krok symulacji). \label{pic:prędkość}}
\end{figure}

W związku z powyższymi uwagami, brak skalowania frontu koncentracji z wykładnikiem $1/2$ w modelu KL oznacza, że równanie klasycznej dyfuzji (\ref{eq:fickdc}) nie może skutecznie opisać tego zjawiska. Oprócz dyfuzji, w obserwowanej propagacji frontu koncentracji, musi brać udział również inny mechanizm transportu powodujący przyspieszoną migrację cząsteczek z obszaru o wyższej koncentracji wgłąb ośrodka porowatego. Aby zrozumieć dynamikę transportu, wyznaczona została średnia prędkość (uśredniona po całym obszarze symulacji) przypadająca na cząsteczkę płynu odpowiednio dla kierunków $x$ oraz $y$. Na rysunku \ref{pic:prędkość} przedstawiona została zależność tych wielkości od czasu w badanym układzie.
Widać wyraźnie, że zgodnie z \nlb oczekiwaniami, prędkość cieczy w kierunku $y$ oscyluje wokół $0$. Inaczej jest w przypadku prędkości w kierunku $x$, która rośnie bardzo szybko do wartości około $3\cdot10^{-3}$ \emph{lu tu}$^{-1}$ (stała sieci / krok symulacji), by następnie bardzo powoli maleć, utrzymując wartość dodatnią. W poprzednich badaniach innych autorów \cite{Kuntz03} wpływ niezerowego pędu cząsteczek został pominięty (ze względu na jego pozornie niewielkie wartości). W dalszej części tego rozdziału będę starał się pokazać, że efekt ten jest istotny. Spróbuję odpowiedzieć na pytanie, czy poszukiwanym mechanizmem odpowiadającym za anomalne zachowanie może być hydrodynamiczny przepływ cieczy spowodowany wysokim gradientem ciśnienia występującym pomiędzy obszarami o różnych koncentracjach.
\subsection{Średnia prędkość frontu}
\begin{figure}[!h]
\centering
\includegraphics[scale=0.58]{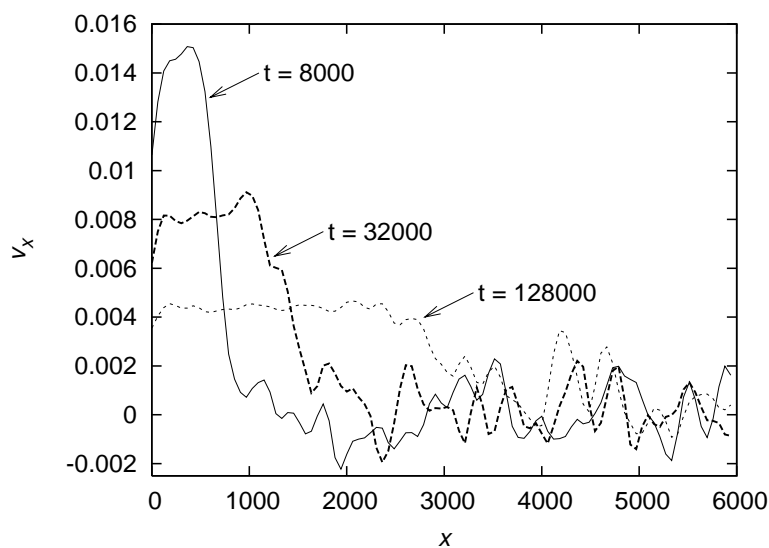}
\caption{Składowa $x$ wektora prędkości -- $v_x(x,y,t)$ -- uśrednionego po $y$ dla trzech różnych chwil czasu $t=8000\times4^k$, $k=0,1,2$. \label{pic:predkoscvx}}
\end{figure}

\begin{figure}[!h]
\centering
\includegraphics[scale=0.58]{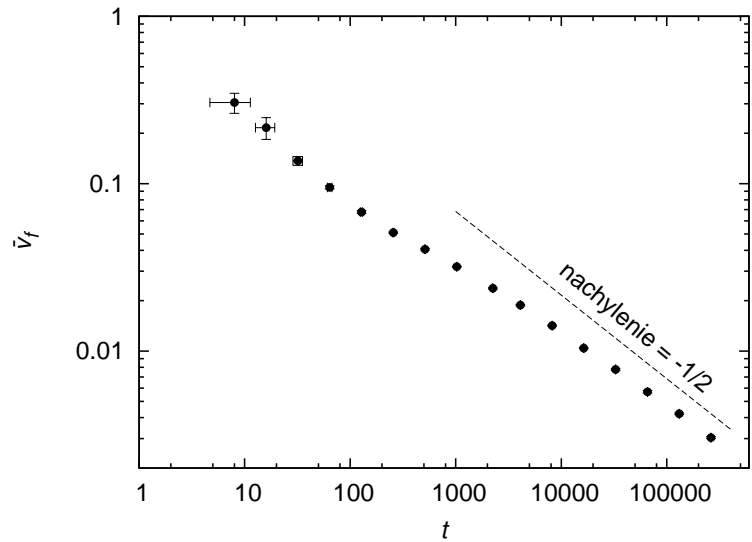}
\caption{Prędkość $\vf$ frontu koncentracji w funkcji kroku czasowego. Dla orientacji zaznaczona została linia o nachyleniu z wykładnikiem $-1/2$. \label{pic:predkoscvf}}
\end{figure}

W celu sprawdzenia poprawności hipotezy o przepływie hydrodynamicznym w układzie, przeprowadzona została analiza ilościowa przestrzennego rozkładu składowej $x$ wektora %
prędkości, $v_x(x,t)$, uśrednionego po wszystkich komórkach w danej kolumnie. Na rysunku \ref{pic:predkoscvx} przedstawione zostały profile prędkości dla trzech różnych chwil czasu.
Z rysunku widać wyraźnie, że dla każdego z czasów obszar $x>0$ może zostać podzielony na dwa charakterystyczne podobszary. W jednym z nich pęd przepływającej substancji jest mniej więcej stały i jest to obszar penetracji substancji w \nlb głąb ośrodka porowatego. To w tym obszarze układ odpowiada na istniejący gradient ciśnienia zwiększeniem pędu. Druga część układu nie została spenetrowana przez wpływającą substancję i średnie wartości $v_x(x,t)$ oscylują wokół zera, co odpowiada warunkowi początkowemu. Z wykresu widać, że z upływem czasu wartość maksymalna pędu w obszarze frontu maleje, zwiększa się za to szerokość tego obszaru.
W \nlb celu ilościowej analizy dynamiki propagacji frontu, zdefiniuję prędkość frontu koncentracji $v_f$. W ogólności wielkość ta jest funkcją $x$ i $t$, ale nas interesować będzie jej przybliżona wartość w czasie $t$. Można ją oszacować jako średnią ważoną z wagą $c(x,t)-c_2$ wzdłuż kierunku propagacji frontu:
\begin{equation}
 \label{eq:def:v-mean}
  \vf(t) =
     \frac{\displaystyle \int_0^\infty v_x(x,t)[c(x,t) - c_2]\,\mathrm{d}x}%
          {\displaystyle \int_0^\infty [c(x,t) - c_2]\,\mathrm{d}x}.
\end{equation}
Dzięki wykorzystaniu uśredniania ważonego z wagami pochodzącymi bezpośrednio z profili koncentracji znacznie zmniejszony został wpływ szumu pochodzącego z modelu gazu sieciowego w obszarach niskich koncentracji, gdzie $c(x,t)\approx c_2$ (rysunek \ref{pic:predkoscvx}, $x>3000$). Korzystając z definicji wyrażonej wzorem (\ref{eq:def:v-mean}) wyznaczona została zależność $\vf(t)$ (rysunek \ref{pic:predkoscvf}). Z rysunku widać wyraźnie i ilościowo, że prędkość średnia frontu w chwilach początkowych jest bardzo wysoka i bliska wartości maksymalnej dla modelu sieciowego $\vf=1$. Dla większych czasów $\vf$ maleje jak funkcja wykładnicza $t^{-\beta}$ z wykładnikiem $\beta \lesssim 1/2$ bardzo słabo zależącym od czasu.

Znajomość średniej prędkości frontu koncentracji pozwoliła na wyznaczenie średniej drogi $s(t)$ przebytej przez front w czasie $t$, jako bezpośredniej konsekwencji istnienia niezerowej prędkości (przepływu):
\begin{equation}
 \label{eq:def-s}
   s(t) = \int_0^t \vf(\tau)\,\mathrm{d}\tau,
\end{equation}
która dla dużych czasów rośnie jak $t^{1-\beta(t)}$, czyli szybciej niż byłoby to w przypadku klasycznej dyfuzji z wykładnikiem $1/2$. Bezpośrednio z wyników numerycznych odczytać można pozycję $x_\mathrm{f}$ frontu dla której $c(x_\mathrm{f})=0.5$, która wynosi $x_\mathrm{f}\approx 3750$ \emph{lu} (jednostek sieci) po czasie $t=256 000$ \emph{tu} (jednostek czasu) (rysunek \ref{pic:frontygestosci}). Jednocześnie całka (\ref{eq:def-s}) dla tego samego $t$ daje $s(t) \approx 1440$ \emph{lu}. Oznacza to, że około $38\%$ z przesunięcia $x_\mathrm{f}$ było spowodowane przepływem cieczy, a reszta dyfuzją. Dlatego przepływ hydrodynamiczny nie może być pominięty w analizie dynamiki układu.

\subsection{Ruchomy układ odniesienia i skalowanie}

\begin{figure}[!h]
\centering
\includegraphics[scale=0.58]{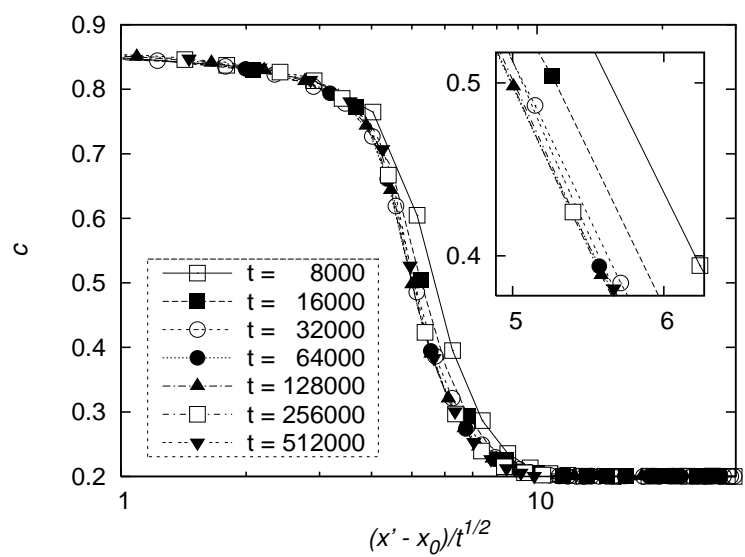}
\caption{Profile koncentracji $c$ w funkcji $(x' - x_0)/\sqrt{t}$ dla $x_0=-190$. Wstawka pokazuje centralną część krzywych w osiach nielogarytmicznych i powiększeniu. \label{pic:frontygestoscifinal}}
\end{figure}

W celu oddzielenia mechanizmu transportu hydrodynamicznego od dyfuzyjnego przeprowadzone zostało klasyczne przejście do ruchomego układu odniesienia $(x',t)$, gdzie:
\begin{equation}
 \label{eq:def-xprime}
  x' \equiv x - s(t).
\end{equation}
Spodziewany rezultat takiej transformacji to odseparowanie lub zminimalizowanie efektów pochodzących od przepływu z niezerową prędkością i obserwacja klasycznej dyfuzji z wykładnikiem $1/2$. Jeśli powyższa hipoteza o klasycznej dyfuzji byłaby prawdziwa, profile koncentracji w ruchomym układzie odniesienia powinny spełniać relację skalowania:
\begin{equation}
  \label{eq:scaling-moving}
  c(x',t) = f\left((x' - x_0)/\sqrt{t}\right),
\end{equation}
gdzie $f$ jest funkcją podobieństwa, a $x_0$ jest stałą. Korzystając z wartości pozycji frontu dla $c(x',t)=0.5$ oszacowana została stała $x_0\approx-190$, a następnie profile koncentracji $c$ w funkcji $(x' - x_0)/\sqrt{t}$ wyznaczono dla $t=8000\times2^k$, $k=0,1,\ldots,6$ (rysunek \ref{pic:frontygestoscifinal}).
Na rysunku widać wyraźnie, że przejście do ruchomego układu odniesienia pozwoliło wyodrębnić klasyczną dyfuzję, a krzywe profili koncentracji asymptotycznie układają się na jednej krzywej. Dla czasów większych niż $t=64000$ przeskalowane profile są praktycznie nierozróżnialne.
\subsection{Nieciągłość profili przy $x = 0$}
\begin{figure}[!h]
\centering
\includegraphics[scale=0.58]{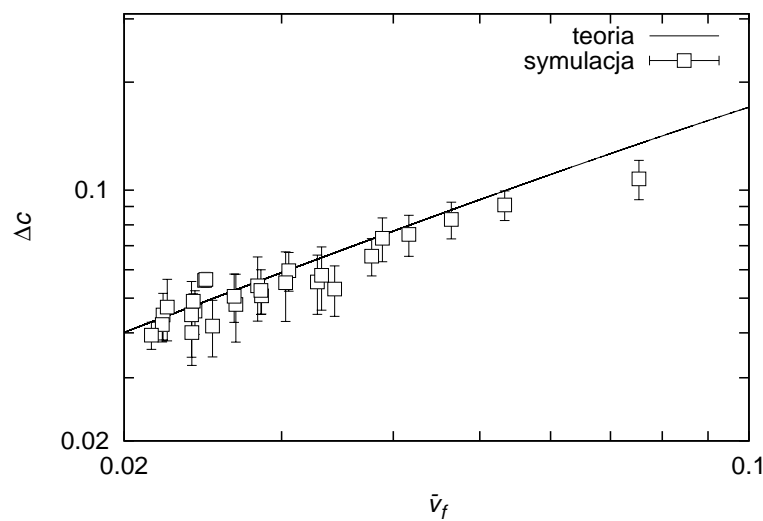}
\caption{Wielkość spadku koncentracji $\Delta c$ na brzegu $x = 0$ w funkcji prędkości frontu $\vf$, symulacja ($\square$) i teoria (--). \label{pic:c}}
\end{figure}

Oddzielenie zjawiska przepływu od dyfuzji w modelu KL pozwala również wytłumaczyć kolejną z jego charakterystycznych cech -- nieciągłość profilu koncentracji w \nlb okolicach $x=0$. Mimo że warunki brzegowe wyrażone równaniem (\ref{eq:initstep}) utrzymują stałą koncentrację $c_1$ w obszarze $x < 0$, obserwuje się nieciągłość koncentracji i dość znacznie obniżoną wartość (dużo niższą niż $c_1$) zaraz za punktem brzegowym (rysunek \ref{pic:frontygestosci} dla $c_1=0.9$). Nieciągłość koncentracji $\Delta c$ jest związana z nieciągłością prędkości średniej $v_x$ przepływu na brzegu $v_x=0$ dla $x < 0$ i $v_x\approx\vf$ dla $x>0$. Oznaczmy koncentrację w punkcie $x = 1$ \emph{lu} (stałych sieci) jako $c^{+}=c_1-\Delta c$. Tempo transportu cząstek przez płaszczyznę $x=0$ jest w przybliżeniu równe $3c_1 - 3c^{+}=3\Delta c$ (czynnik 3 jest liczbą możliwych kierunków z których może zachodzić transport w kierunku poziomym ze strony lewej na prawą). Wartość ta musi być zrównoważona przez hydrodynamiczny strumień cząstek, który może być przybliżony przez $7c^{+}\vf$ (czynnik 7 jest liczbą możliwych kierunków wektora prędkości). Stąd, $3c_1-3c^{+}\approx7c^{+}\vf$, co pozwala oszacować wielkość nieciągłości $c$:
\begin{equation}
  \label{eq:Dc-Dv}
    \Delta c \approx  \frac{7c_1\vf}{3 + 7\vf}.
\end{equation}
Porównanie wyprowadzonej relacji z wynikami z symulacji zostało przedstawione na rysunku \ref{pic:c}.

\section{Omówienie wyników}
Wiele właściwości modelu użytego w tym rozdziale do badania propagacji frontów koncentracji zasługuje na bliższą analizę. Po pierwsze, do reprezentowania ośrodka porowatego i redukcji efektów hydrodynamicznych w modelu użyte zostały rozpraszacze punktowe \cite{Kuntz03}. Działanie tych rozpraszaczy jest jednak osobliwe: składowa $x$ prędkości zmienia się jedynie przy kolizji z rozpraszaczem cząstek poruszających się dokładnie wzdłuż kierunków równoległych do osi $x$. W omawianym modelu KL rozpraszacze zostały rozłożone losowo w obszarze ośrodka porowatego. Okazuje się więc, że taka konfiguracja nie ma dużego wpływu na dyfuzyjność w kierunku osi $x$. Wiadomo również, że średnia droga swobodna cząsteczek, a co za tym idzie dyfuzyjność w oryginalnym modelu FHP (bez rozpraszaczy), dąży do $\infty$ z koncentracją dążącą do $0$ lub $1$ \cite{Kuntz03}. Biorąc pod uwagę dwa wyżej wspomniane fakty, łatwo jest wytłumaczyć zależność współczynnika dyfuzji od koncentracji cząsteczek, przedstawioną wcześniej jako argument na rzecz rzekomej anomalnej dyfuzji transportowanej substancji \cite{Kuntz03}.
Model KL jest oparty w głównej mierze o hydrodynamiczny model gazu sieciowego FHP w którym lokalne zasady kolizji i translacji spełniają zasady zachowania pędu i masy i prowadzą do równań przepływu w skali makroskopowej. Jedną z \nlb cech tego typu modeli jest przepływ od obszarów o wyższym do obszarów o \nlb niższym ciśnieniu, przy czym w modelu FHP ciśnienie jest proporcjonalne do koncentracji \cite{Wolf00}.

Można oszacować skalę efektów hydrodynamicznych. Zakładając niską prędkość frontu koncentracji, strumień cząsteczek $\v{q}$ przepływających przez ośrodek porowaty jest proporcjonalny do gradientu ciśnienia, co wyraża prawo Darcy'ego\cite{Bear72}:
\begin{equation}
  \v{q} = -\frac{\kappa}{\mu} \nabla P,
\end{equation}
gdzie $\kappa$ jest stałą przepuszczalności (cecha ośrodka porowatego), $\mu$ jest lepkością dynamiczną cieczy, a $\nabla P$ jest gradientem ciśnienia między wlotem a wylotem cieczy. Skoro $q\propto v_x$ \cite{Bear72}, $\mu\propto D^{-1}$ (przybliżenie Stokesa-Einsteina opływu cieczy wokół kuli) oraz $P\propto c$ \cite{Droz98}, więc:
\begin{equation}
\label{eq:vx_propto_c}
  v_x \propto D \nabla c.
\end{equation}
Z powyższego wzoru wynika, że dla dużych wartości dyfuzyjności $D$ (które, jak wspominałem wyżej, występują w modelu KL) nawet niewielki gradient koncentracji $\nabla c$ cząstek może indukować znaczną prędkość unoszenia.

Kolejną ciekawą cechą w badanym modelu jest duża wartość stałej $x_0$ w relacji (\ref{eq:scaling-moving}). Co więcej, czas po którym relacja ta jest spełniona jest również dość duży. Można podejrzewać, że oba te efekty są wynikiem ograniczenia z góry prędkości $v_x \leq 1$ i relacji wyrażonej równaniem (\ref{eq:vx_propto_c}). Z warunków brzegowych wiadomo bowiem, że w obszarze frontu koncentracji w chwilach początkowych panują wysokie wartości $\nabla c$ oraz $D$, natomiast $v_x$ jest ograniczone z góry przez $1$ \emph{lu/tu}. Dlatego prawo Darcy'ego może być spełnione dopiero po pewnym czasie relaksacji układu potrzebnym do redukcji gradientu $\nabla c$ do wartości rzędu $1/D$, co odpowiada warunkowi na maksymalną wartość $v_x\approx 1$ \emph{lu/tu}. To spostrzeżenie tłumaczy, dlaczego w poprzednich doniesieniach anomalną propagację frontów koncentracji obserwowano w przypadku, gdy $c_1$ było bliskie $1$, a rozpraszacze miały formę przedstawioną w tym rozdziale \cite{Kuntz03,Kuntz04,Kuntz05}. Warto zauważyć, że dość duża wartość $x_0$ oraz duże wartości czasu relaksacji układu są zgodne z aktualnymi badaniami nad zwilżaniem materiałów budowlanych, gdzie krótkozasięgowe odchylenia od skalowania z czasem jak $t^{1/2}$ zostały zaobserwowane już wcześniej \cite{Lockington03}. Co więcej, z rysunku \ref{pic:predkoscvf} wynika, że wartości prędkości $\vf$ dla małych czasów są dużo mniejsze niż wartości oczekiwane z ekstrapolacji prawa Darcy'ego. Konsekwencją tego jest ujemna wartość $x_0$ i ,,anomalny'' transport profilu koncentracji w początkowej fazie. Można się spodziewać, że w układach w których ekstrapolacja wartości prędkości frontu daje mniejsze wartości od tych mierzonych, wartość $x_0$ byłaby dodatnia, a transport profili zachodził z wykładnikiem $\alpha < 1/2$.
Podsumowując, z rysunku \ref{pic:predkoscvf} wynika, że dla dostatecznie dużych czasów symulacji ($t>10^3$) prędkość średnia frontu $\vf\propto1/\sqrt{t}$, co prowadzi do oszacowania charakterystycznej długości hydrodynamicznej $\int_{\tau=0}^t 1/\sqrt{\tau}\,\mathrm{d}\tau \propto \sqrt{t}$. Prowadzi to do wniosku, że skale długości zjawisk dyfuzyjnych i hydrodynamicznych w omawianym modelu są takie same i nie jest możliwe rozróżnienie pomiędzy nimi na gruncie badań nad asymptotycznym zachowaniem frontów koncentracji.

\section{Podsumowanie}

W tym rozdziale udało się pokazać, że w modelu gazu sieciowego FHP użytego do analizy propagacji frontu gęstości oprócz  klasycznej dyfuzji Ficka występuje niezerowy pęd materii, a położenia profili gęstości skalują się zgodnie z prawem potęgowym jak $t^\frac{1}{2}$ (gdzie $t$ oznacza czas) w ruchomym układzie odniesienia. Podejście to okazało się pomysłem prostszym i bardziej efektywnym niż opublikowane wcześniej w literaturze doniesienia o dyfuzji anomalnej \cite{Kuntz03,Kuntz04,Kuntz05}, pozwalając wytłumaczyć m.in. spadek koncentracji przy brzegu $x=0$.
Analiza przeprowadzona w tym rozdziale pozwala spojrzeć całościowo na przebieg transportu w modelu KL. Nałożony warunek brzegowy -- początkowy gradient koncentracji wzdłuż brzegu $x=0$ -- jest równoważny gradientowi ciśnienia. W konsekwencji prowadzi to do powstania sił wymuszających przepływ. Przepływ ten jest blokowany przez rozłożone losowo rozpraszacze, stąd można się spodziewać, że w takim układzie efekty hydrodynamiczne będą asymptotycznie (w czasie) zaniedbywalne. Okazało się jednak, że sposób rozpraszania, wyróżnienie kierunków równoległych do osi $x$ oraz nieefektywność rozpraszaczy dla koncentracji bliskich $1$ powodują, iż w modelu występują efekty hydrodynamiczne, które wpływają na dynamikę frontów koncentracji. Efekty te mają istotny wpływ na czas relaksacji układu do stanu, w którym spełnione jest prawo Darcy'ego, a co za tym idzie -- na przesunięcie $x_0$ (równanie \ref{eq:scaling-moving}). Propagacja frontu koncentracji jest więc realizowana przy pomocy dyfuzji oraz przepływu cieczy, przy czym dla dostatecznie dużych czasów oba procesy dają przesunięcie proporcjonalne do czasu jak $\sqrt{t}$. W konsekwencji dla dostatecznie dużych czasów profile koncentracji mogą zostać opisane jako funkcje pojedynczej zmiennej $x/\sqrt{t}$, co może prowadzić do fałszywego wniosku o wyłącznie dyfuzyjnym charakterze omawianego procesu.


\chapter{Krętość przepływu}
\label{sec:chapterkretosc}

W rozdziale tym przedstawię badania nad krętością $T$ przepływu cieczy przez mikroskopowy model ośrodka porowatego. Wprowadzę definicję krętości jako bezwymiarowego parametru fizycznego opisującego stopień wydłużenia torów wybieranych przez płyn podczas przepływu. Wyznaczę numeryczną zależność $T$ od porowatości $\phi$ z uwzględnieniem efektów skończonego rozmiaru sieci. Wyniki porównam do zależności empirycznych.
Rozdział ten opiera się na wynikach opublikowanych w \lb pracy:\\ \\
    M. Matyka, A. Khalili, Z. Koza,\\
    \textit{Tortuosity-porosity relation in porous media flow},\\
    Phys. Rev. E 78, 026306 (2008).\\ \\
Dodatkowo, w podrozdziałach \ref{sec:progperkol}, \ref{rozdz:nowe1} oraz \ref{rozdz:nowe2}, przedstawione zostaną nowe wyniki dotyczące progu perkolacji modelu oraz korelacji krętości z powierzchnią charakterystyczną i porowatością efektywną układu.

\section{Wprowadzenie}

Jak pokazaliśmy w poprzednim rozdziale, w przypadku procesów transportu przez ośrodek porowaty, bardzo istotne jest, aby oprócz zjawisk dyfuzyjnych brać pod uwagę hydrodynamiczny przepływ cieczy. Transport ten zachodzi w mikroskali przez system kanalików, których struktura i skomplikowana sieć połączeń może wpływać w sposób znaczący na jego właściwości. Zwykle do opisu ośrodków porowatych używa się dwóch parametrów: porowatości $\phi$ i przepuszczalności $\kappa$ \cite{Bear72}. Parametry te nie mówią jednak nic konkretnego o detalach mikrostruktury układu, ani o tym, jak zachodzi transport. W tym kontekście atrakcyjnym pomysłem wydaje się zdefiniowanie dodatkowego, bezwymiarowego parametru, który opisywałby średnie wydłużenie drogi cząsteczek transportowanych przez ośrodek porowaty, nazwanego w literaturze krętością (\textit{ang. tortuosity}). W tej części zajmę się badaniami numerycznymi nad krętością oraz zależnością krętości od porowatości w wybranym modelu ośrodka porowatego.

\section{Model}

\begin{figure}[!h]
  \centering
\begin{tabular}{cc}
\includegraphics[width=0.35\columnwidth]{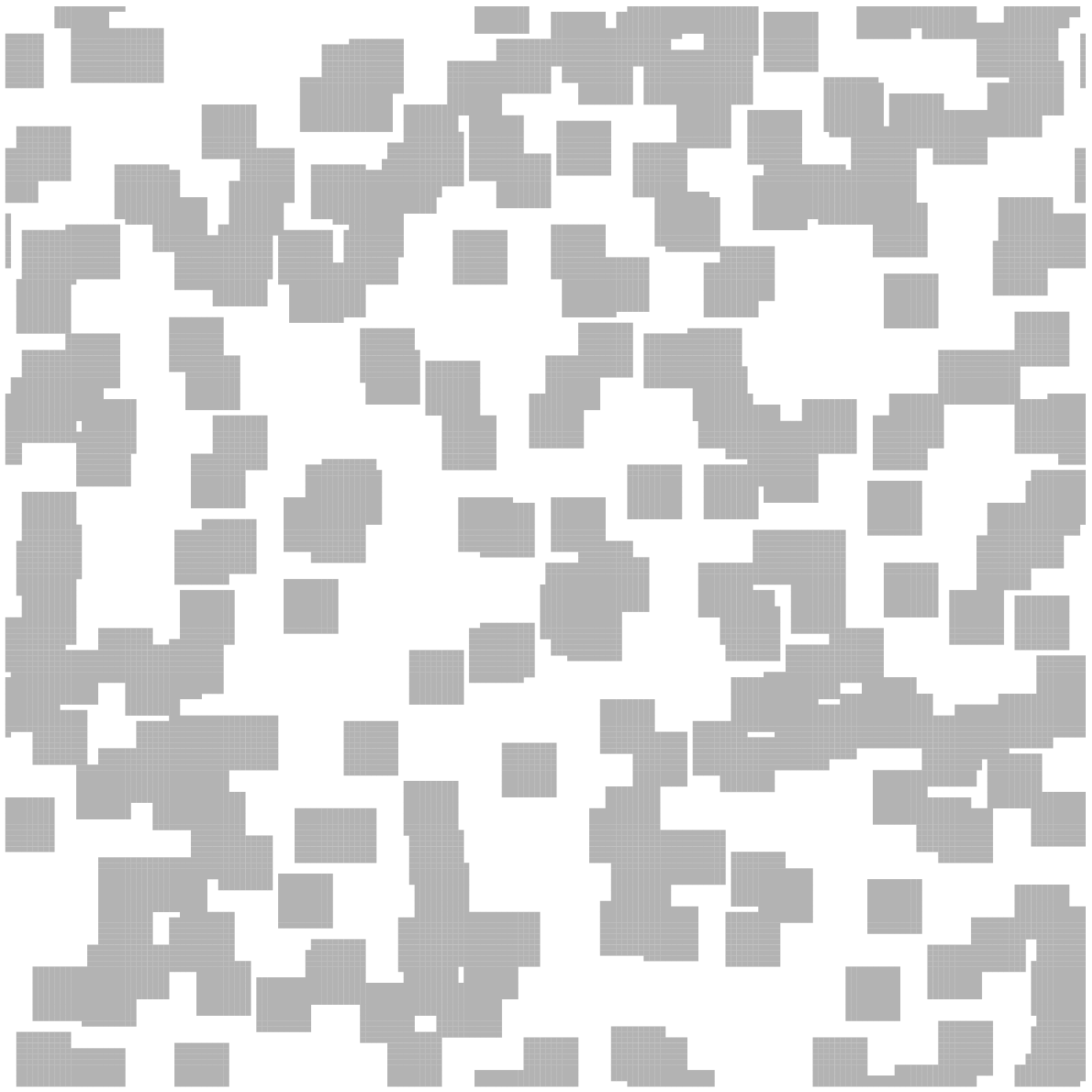}
  \hspace{0.1cm}
    &
  \hspace{0.1cm}
\includegraphics[width=0.35\columnwidth]{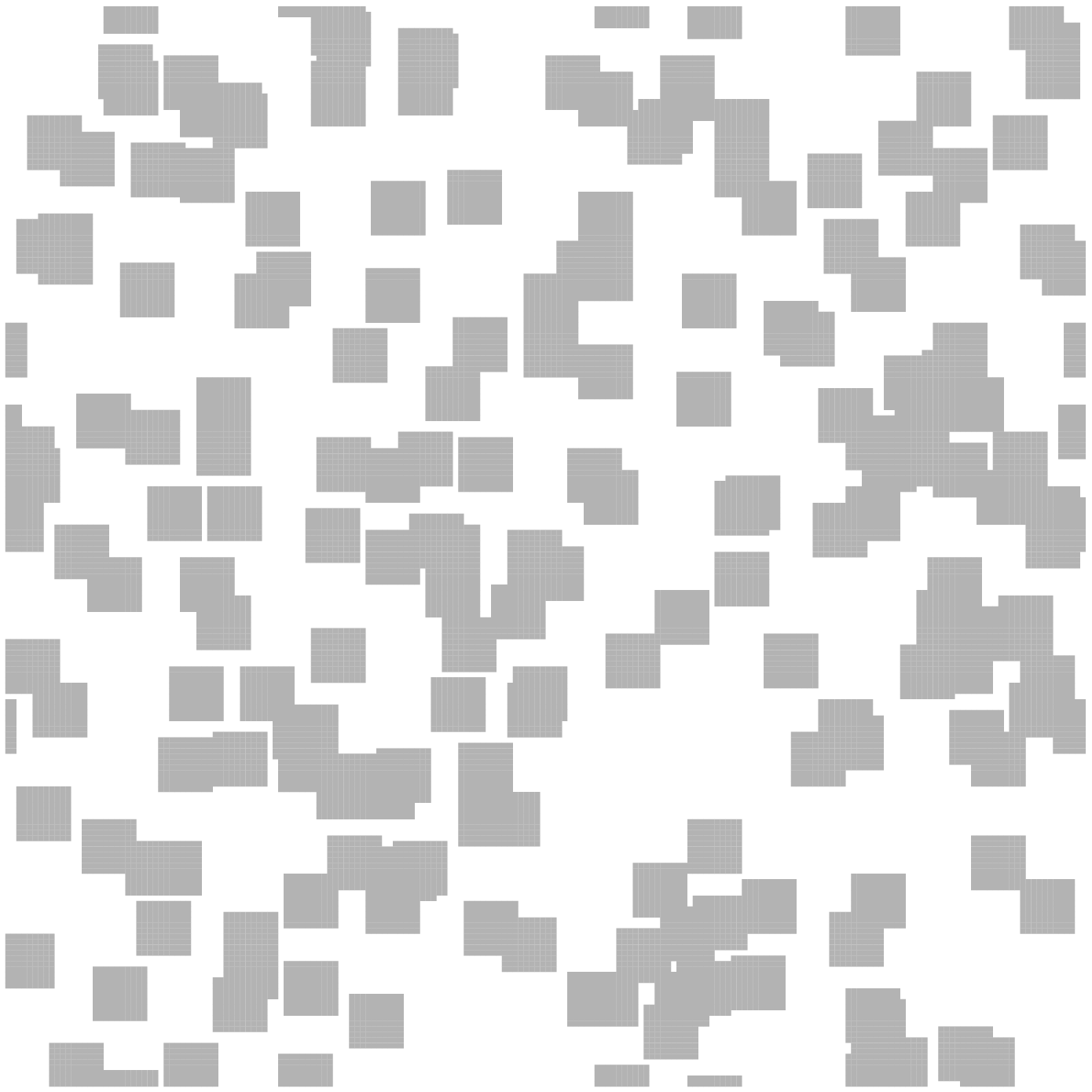} \\
 & \\
a) $\phi=0.5$ \hspace{0.1cm}&\hspace{0.1cm} b) $\phi=0.6$ \\
 & \\
\includegraphics[width=0.35\columnwidth]{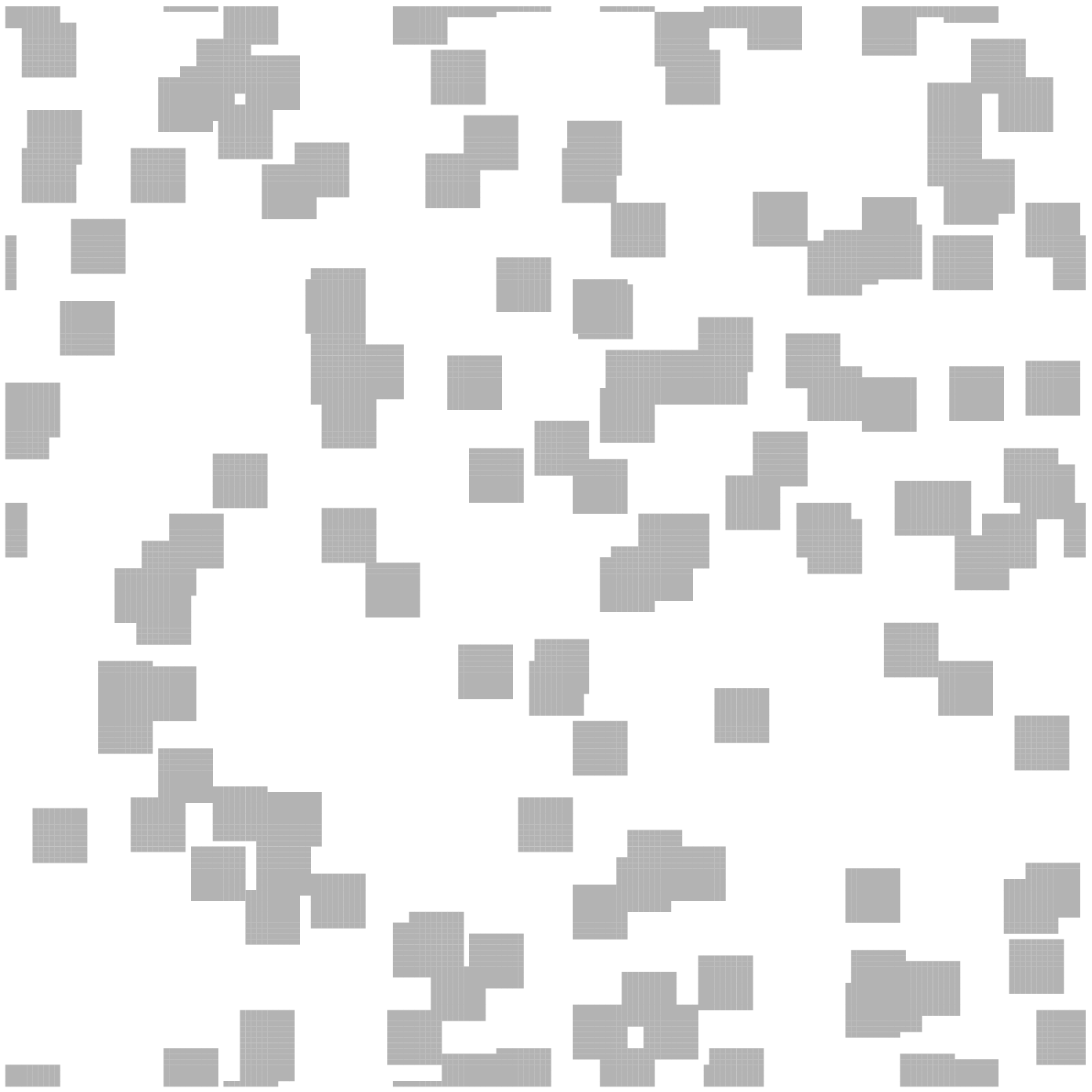}
  \hspace{0.1cm}
    &
  \hspace{0.1cm}
\includegraphics[width=0.35\columnwidth]{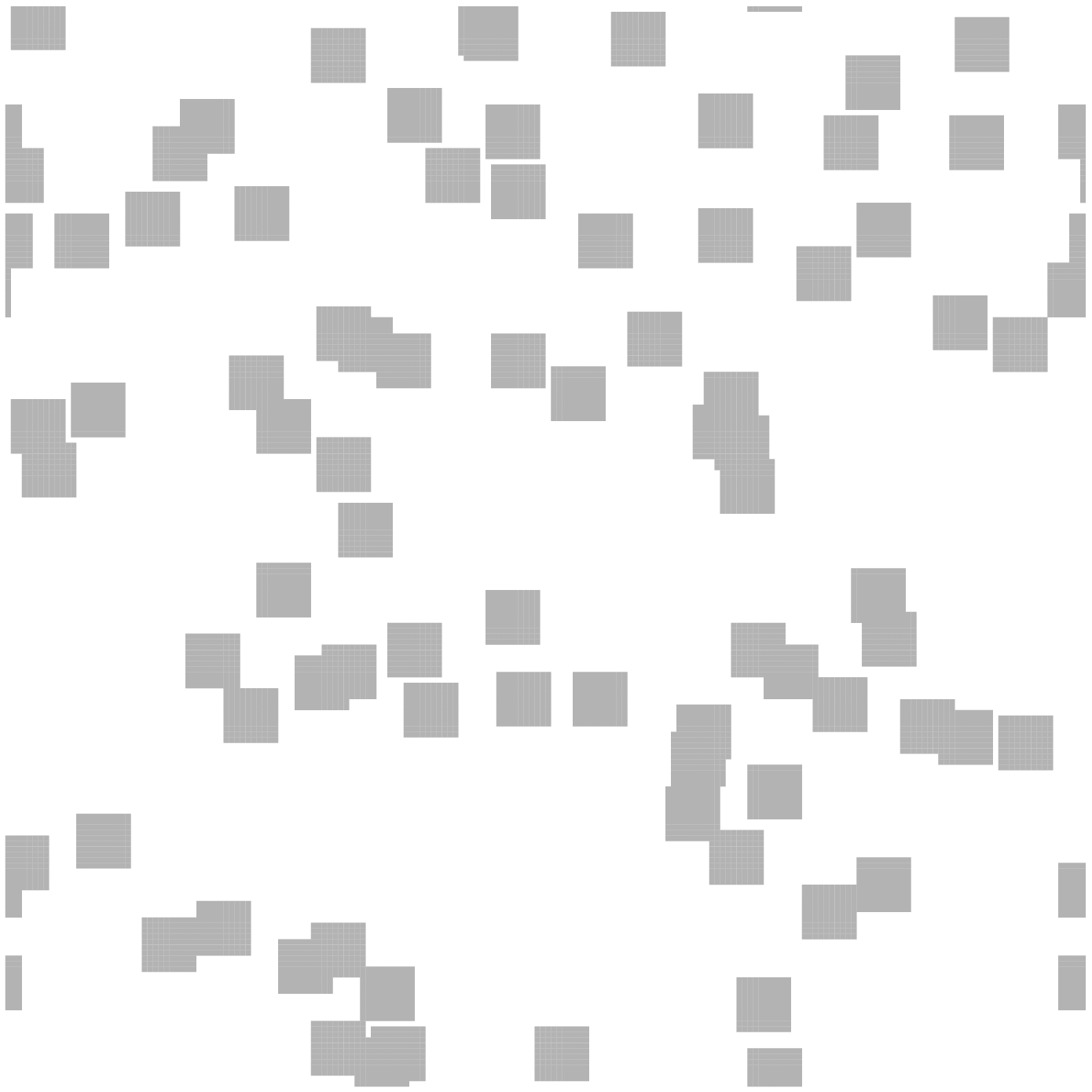} \\
 & \\
c) $\phi=0.7$ \hspace{0.1cm}&\hspace{0.1cm} d) $\phi=0.8$ \\
 & \\
\end{tabular}
\caption{Przykłady konfiguracji na sieci o rozmiarach $200\times 200$ (\emph{lu}\@.~$\times$ \emph{lu}) (jednostek sieci) utworzonych przez losowe rozmieszczenie pokrywających się kwadratów o \protect\nlb rozmiarach $10  \times 10 $ (\emph{lu}~$\times$ \emph{lu}) dla różnych porowatości $\phi$.\label{pic:geometry}}
\end{figure}

Model ośrodka porowatego, dla którego liczyć będę krętość, zdefiniowany jest na sieci kwadratowej o rozmiarze $L\times L$ \emph{lu} (stałych sieci). Na sieci rozłożone zostały losowo identyczne przeszkody w kształcie kwadratów o rozmiarach $a \times a$ \emph{lu}, gdzie $1\leq a\leq L$. Przeszkody te mogą dowolnie pokrywać się nawzajem i są całkowicie nieprzepuszczalne dla cieczy. Pozostały obszar jest dla cieczy dostępny i tu zachodzi jej transport. Na rysunku \ref{pic:geometry} przedstawione zostały układy dla kilku wybranych porowatości, przy czym kolorem szarym zaznaczono przeszkody reprezentujące część nieprzepuszczalną ośrodka porowatego.
W celu minimalizacji wpływu efektów brzegowych na obliczenia, założyliśmy periodyczne warunki brzegowe w obu kierunkach \cite{Koponen96,Koponen97}. W celu wymuszenia przepływu cieczy przez mikrostrukturę, na układ nałożona została stała siła zewnętrzna (grawitacja) równoległa do jednej z osi układu.

\subsection{Próg perkolacji}\label{sec:progperkol}

\begin{figure}[!h]
  \centering
\includegraphics[scale=0.65]{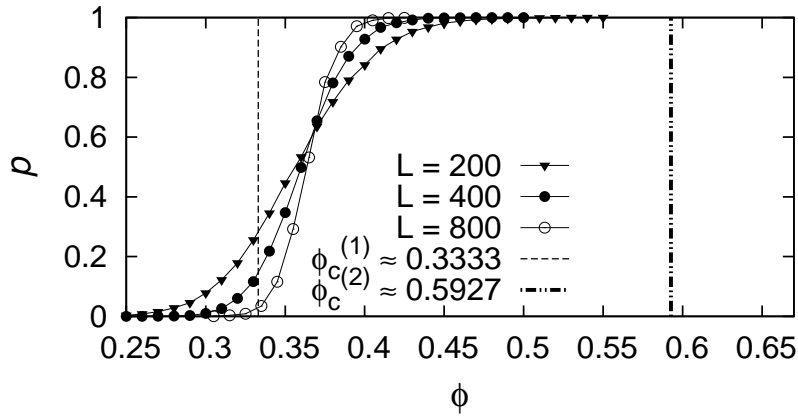}
\caption{Jakościowa analiza progu perkolacji dla $a=10$ i trzech różnych rozmiarów sieci. Na osi $y$ zaznaczone zostało prawdopodobieństwo, że układ perkoluje. Na osi $x$ odłożono porowatość układu. Widać wyraźnie, że skalowanie wskazuje na próg perkolacji w okolicach $\phi\approx0.367$ (punkt przecięcia krzywych). Wartość $\phi_c^{(1)} \approx 0.33(3)$ obliczona została dla perkolacji ciągłej w \cite{BakerStanley04}, a wartość $\phi_c^{(2)}\approx0.5927$ wyznaczono dla standardowej perkolacji węzłów \cite{Sahimi93}.\label{pic:prog}}
\end{figure}

\begin{figure}[!h]
  \centering
\includegraphics[scale=0.83]{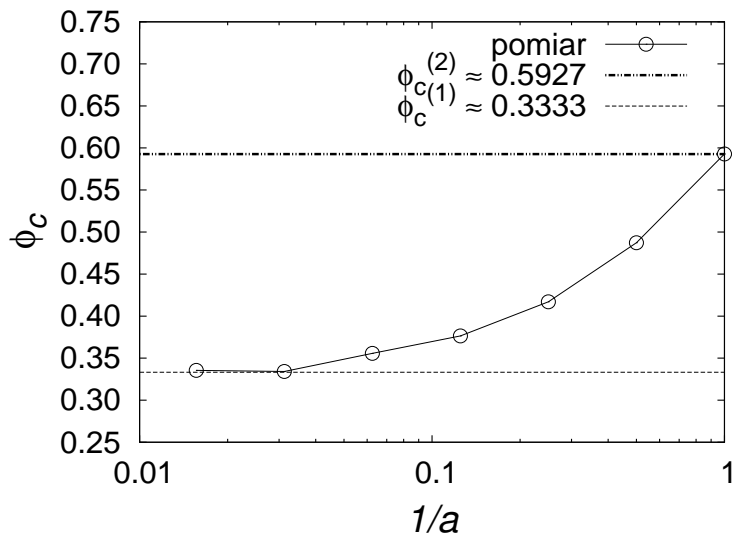}
\caption{Zależność progu perkolacji $\phi_c$ od odwrotności długości boku przeszkody $1/a$. Symbole (o) reprezentują numeryczną wartość $\phi_c$ w zależności od rozmiaru przeszkody $a=2^k$ \emph{lu} (stałych sieci), gdzie $k=0,1,2\ldots 6$. Liniami przerywanymi zaznaczone zostały progi perkolacji ciągłej $\phi_c^{(1)}$ oraz standardowej perkolacji sieciowej $\phi_c^{(2)}$. \label{pic:perkolacja}}
\end{figure}

Ze względu na obliczenia krętości w funkcji porowatości układu, istotnym parametrem jest upakowanie krytyczne, czyli próg perkolacji $\phi_c$. Okazuje się, że wielkość przeszkód użytych do budowy modelu ma znaczący wpływ na położenie progu perkolacji. W naszych badaniach, podobnie jak w \cite{Koponen96,Koponen97}, we wszystkich symulacjach przyjęto $a=10$. Przeprowadzone zostały obliczenia numeryczne progu perkolacji przy pomocy algorytmu numerowania klastrów Hoshena-Kopelmana \cite{Hoshen76} oraz metody Kirkpatricka wyznaczania punktu krytycznego \cite{Oleksy99} dla $a=10$ i otrzymaliśmy próg perkolacji $\phi_c\approx 0.367$.  Wartość ta leży pomiędzy wartością $\phi_c^{(1)} \approx 0.33(3)$ \cite{BakerStanley04} (obliczoną dla perkolacji ciągłej, $a\rightarrow\infty$), a wartością $\phi_c^{(2)}\approx0.5927$ \cite{Sahimi93} (standardowa perkolacja węzłów, $a=1$). Dodatkowo, na rysunku \ref{pic:prog} przedstawiona została prosta analiza położenia progu perkolacji z użyciem metody skalowania rozmiaru.
Na rysunku wyraźnie widać, że $\phi_c^{(1)} \le \phi_c \le \phi_c^{(2)}$. Warto zauważyć, że obliczona przez nas wartość $\phi_c$ jest różna od wartości podawanych w \cite{Koponen96,Koponen97}, gdzie autorzy błędnie powiązali $\phi_c$ dla $a=10$ z $\phi_c^{(1)}$ dla $a\rightarrow\infty$ i używali $\phi_c = \phi_c^{(1)} \approx 0.33(3)$. Dodatkowo na rysunku \ref{pic:perkolacja} przedstawiam zależność progu perkolacji $\phi_c$ od odwrotności długości boku pojedynczej przeszkody $a$ z użyciem osi półlogarytmicznych. Wartości numeryczne wyznaczone zostały algorytmem Hoshena-Kopelmana dla przeszkód wielkości $2^k$, gdzie $k=0,1,2\ldots 6$.
Widać wyraźnie, jak sukcesywne zwiększanie rozmiaru przeszkody prowadzi do przesunięcia progu perkolacji od $\phi_c^{(2)}$ dla klasycznej perkolacji węzłów (ang. \emph{site percolation}) do wartości $\phi_c^{(1)}$ dla perkolacji ciągłej.

\section{Algorytmy numeryczne}

W celu wyznaczenia krętości w badanym modelu musimy użyć szeregu metod numerycznych. Począwszy od generowania mikrostruktury, przez rozwiązanie równań przepływu, wyznaczenie linii prądu, właściwe obliczenia krętości, aż po weryfikację czasu relaksacji w układzie, ekstrapolację części rezultatów do stanów stacjonarnych oraz analizę błędów numerycznych.

\subsection{Konstrukcja ośrodka porowatego} \label{sec:konstrukcja}

Konstrukcja ośrodka porowatego o określonej porowatości jest realizowana przy pomocy prostego algorytmu losowego osadzania swobodnie pokrywających się przeszkód na badany obszar \cite{Koponen96,Koponen97}. Rozpoczynając z pustego obszaru ($\phi=1$), nieprzepuszczalne kwadraty są na niego nakładane tak długo, jak długo $\phi>\phi_0$, gdzie $\phi_0$ jest oczekiwaną porowatością.

%

\subsection{Przepływ cieczy}\label{sec:refinement}

Na potrzeby rozwiązania problemu przepływu cieczy przez układ porowaty zaadoptowany został model gazu sieciowego Boltzmanna (LBM) \cite{Succi01} w przybliżeniu BGK operatora kolizji \cite{Bhatnagar54} (patrz rozdział \ref{sec:chapter3}). Metoda ta potwierdziła swoją użyteczność w symulacjach przepływu cieczy w różnych warunkach \cite{Koponen98,Pan01}.

\begin{figure}[!h]
 \centering
 \includegraphics[width=0.9\columnwidth]{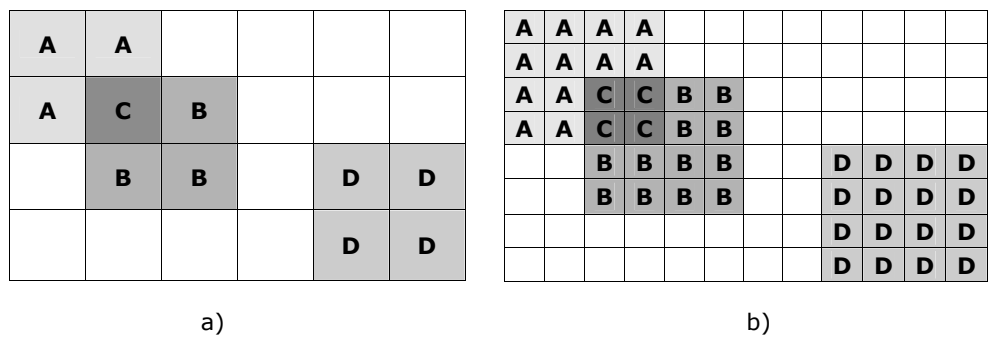}
   \caption{Procedura dodatkowego podziału sieci ośrodka porowatego. Stopień podziału: a) $k_\mathrm{ref}=2$, b) $k_\mathrm{ref}=4$. Litery A, B i C użyte zostały do odróżnienia odpowiadających sobie elementów. \label{pic:refinement}}
\end{figure}

Jednym z jej ograniczeń jest minimalna skala przestrzenna $4$ \emph{lu}, na której jest możliwe uzyskanie rozwiązania równań przepływu Naviera--Stokesa z zadowalającą dokładnością. Ograniczenie to ma dość duże znaczenie w przypadku porowatości $\phi\rightarrow\phi_c$, gdyż w tym przypadku znacznie rośnie liczba wąskich kanałów o szerokości $d<4$ \emph{lu}. Z tego powodu na układy skonstruowane wg opisanego powyżej algorytmu nałożona jest standardowa procedura zwiększenia podziału sieci na większą ilość komórek (rysunek \ref{pic:refinement}).
Procedura ta dzieli każdą z $L^2$ komórek sieci obliczeniowej na kwadraty $k_\mathrm{ref}\times k_\mathrm{ref}$, gdzie $k_\mathrm{ref}=1,2,\ldots$ jest stopniem podziału. Otrzymana sieć ma zatem rozmiar $k_\mathrm{ref}L\times k_\mathrm{ref}L$.

\begin{figure}[!h]
  \centering
\begin{tabular}{cc}
 \includegraphics[width=0.45\columnwidth]{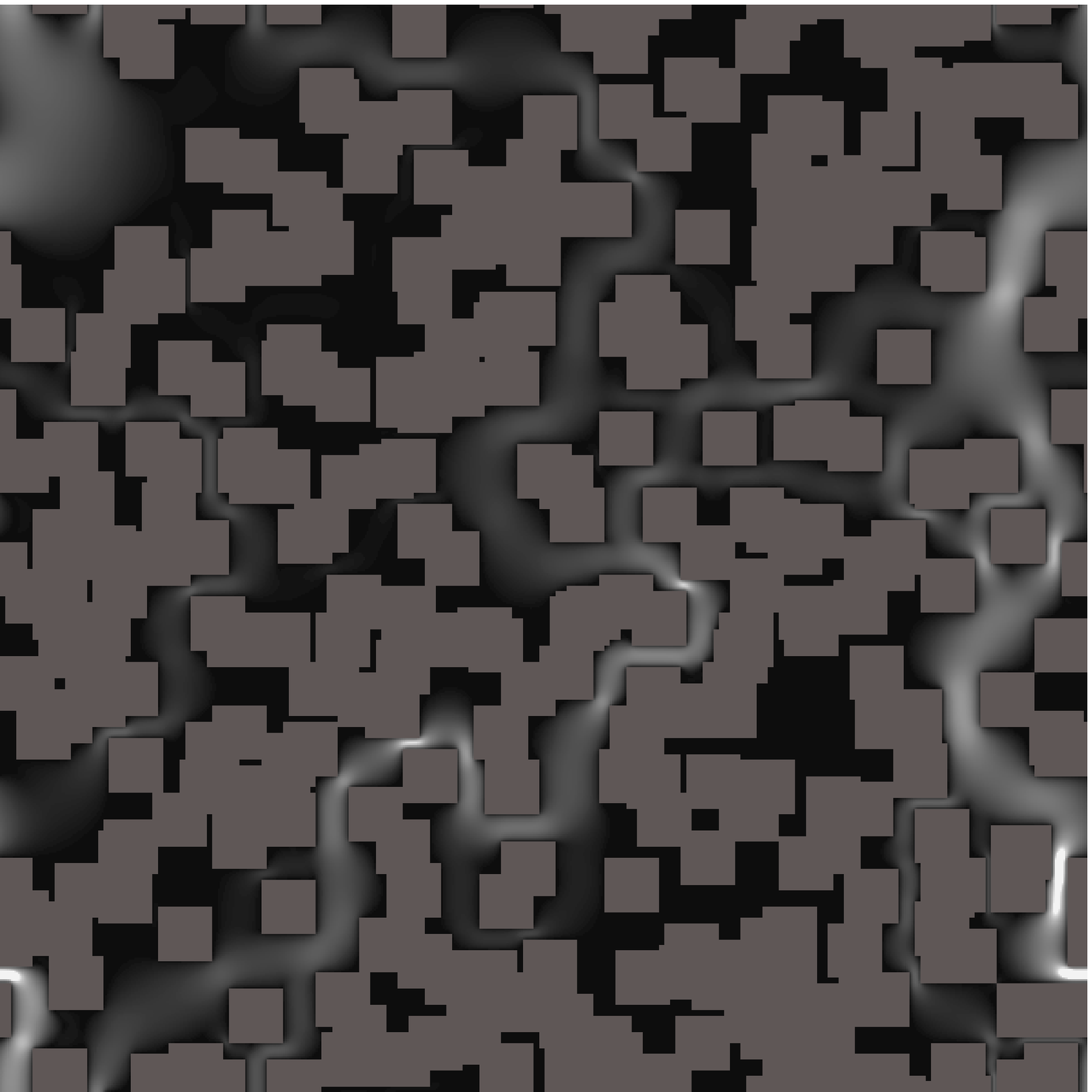}
  \hspace{0.1cm}
    &
  \hspace{0.1cm}
 \includegraphics[width=0.45\columnwidth]{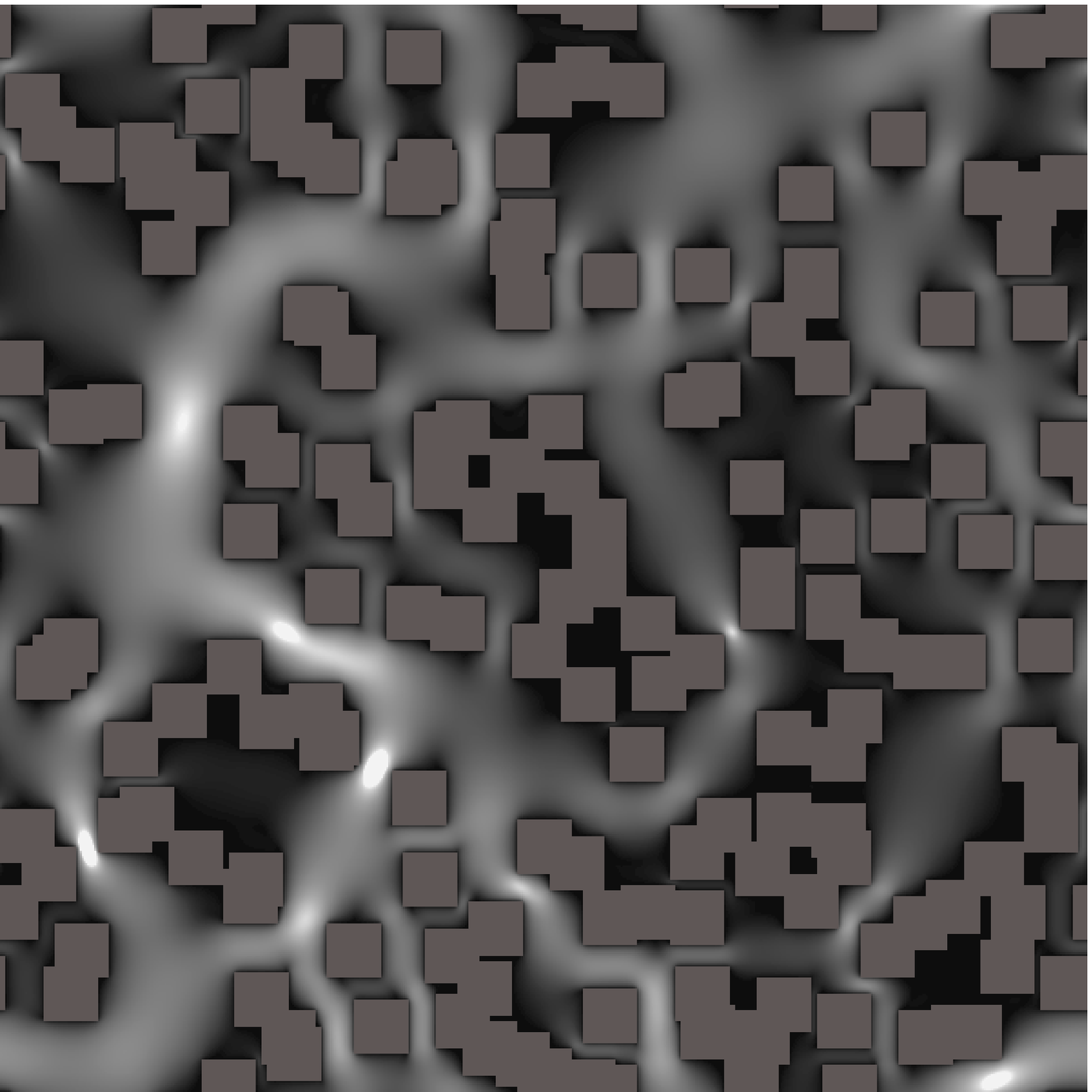}\\
 &  \\
a) $\phi=0.45$ \hspace{0.1cm}&\hspace{0.1cm} b) $\phi=0.65$ \\
 &  \\
 \includegraphics[width=0.45\columnwidth]{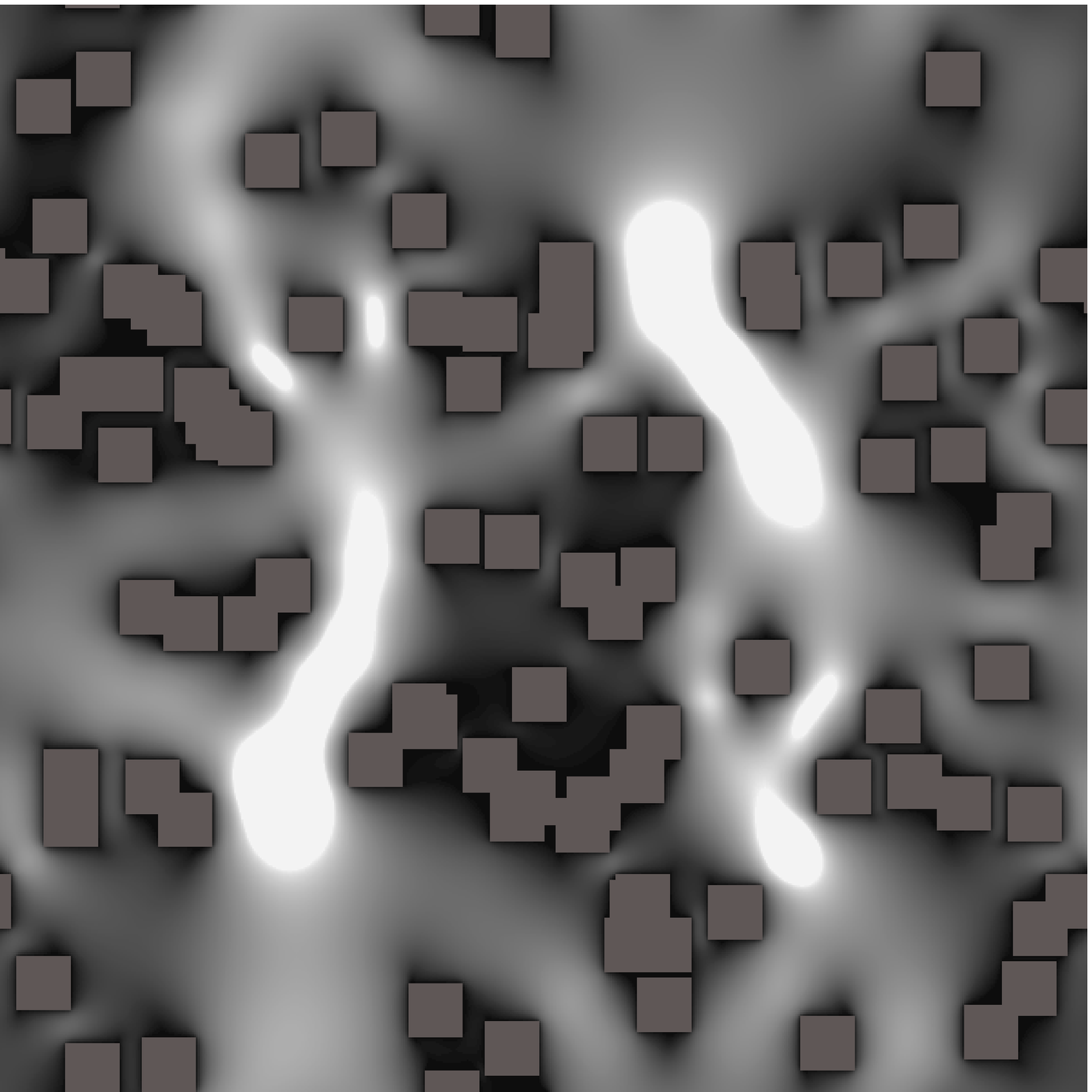}
  \hspace{0.1cm}
    &
  \hspace{0.1cm}
 \includegraphics[width=0.45\columnwidth]{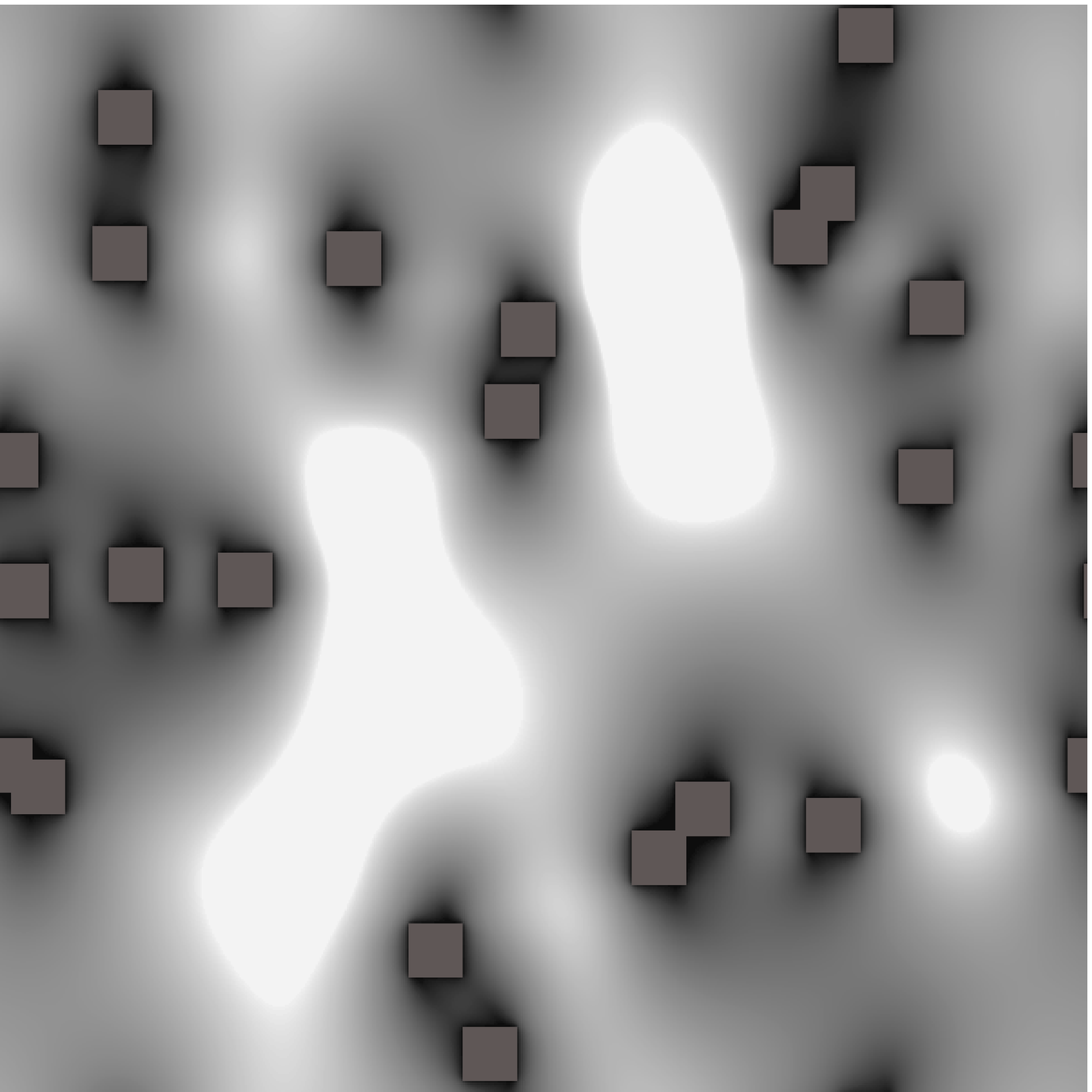}\\
 &  \\
c) $\phi=0.80$ \hspace{0.1cm}&\hspace{0.1cm} d) $\phi=0.95$ \\
 &  \\
\end{tabular}
   \caption{Prędkość cieczy $u=|\v{u}|$ obliczona na sieci obliczeniowej o rozmiarach $600\times600$ odpowiadającej sieci fizycznej $200$
        \emph{lu} $\times$ $200$ \emph{lu} (poziom podziału $\ensuremath{k_\mathrm{ref}} = 3$). Wielkość przeszkód:
                $10$ \emph{lu} $\times10$ \emph{lu} (t.j., $30\times30$ węzłów sieci). Porowatość układu:\newline a) $\phi = 0.45$, b) $\phi = 0.65$, c) $\phi = 0.8$, d) $\phi = 0.95$. Warunki periodyczne zastosowane zostały w obu kierunkach. Szare kwadraty reprezentują obszar niedostępny dla cieczy, pozostały obszar jest dozwolony dla przepływu. Zewnętrzna siła grawitacyjna działa w kierunku pionowym. \label{pic:magn1}}
\end{figure}

Po inicjalizacji układu, pętla obliczeniowa modelu jest kontynuowana $t_\mathrm{max}$ kroków czasowych, gdzie $t_\mathrm{max}$ jest wybrane wg kryterium, które omówimy w dalszej części pracy (patrz podrozdział \ref{sec:time}). Dzięki użyciu odpowiednich warunków brzegowych na komórkach nieprzepuszczalnych (\emph{ang. midgrid}\thinspace) otrzymane rozwiązania mają dokładność drugiego rzędu w czasie i przestrzeni \cite{Succi01}. Przykładowe pola prędkości uzyskane za pomocą omawianej metody dla kilku różnych wartości porowatości przedstawione zostały na rysunku \ref{pic:magn1} .

\subsection{Linie prądu}

Kluczową procedurą całej operacji liczenia krętości przepływu jest dokładne wyznaczenie linii prądu \cite{Landau94}. Aby je wyznaczyć, potrzebna jest znajomość pola prędkości $\v{u}(\v{r})$ w całej objętości ośrodka. W naszym przypadku pole $\v{u}(\v{r})$ zostało uzyskane za pomocą interpolacji dwuliniowej (\emph{ang. bilinear interpolation}) \cite{Press86} z prędkości węzłowych $\v{u}$ otrzymanych metodą LBM. Na tak otrzymanym polu prędkości rozwiązywane jest równanie ruchu bezmasowych cząstek (znaczników):
\begin{equation}
 \label{eq:MaselessMotion}
  \frac{d\v{r}}{dt} = \v{u}(\v{r}),
\end{equation}
których tory ruchu odpowiadają, w przypadku stanu stacjonarnego, liniom prądu przepływającej cieczy \cite{Landau94,Harlow65}. Ze względu na skomplikowaną postać brzegów oraz znaczne różnice w lokalnych wartościach interpolowanego pola prędkości $\v{u}(\v{r})$, do rozwiązania równań typu (\ref{eq:MaselessMotion}) została wykorzystana metoda Runge'go-Kutty czwartego rzędu ze zmiennym krokiem czasowym \cite{Press86}.

\subsection{Krętość}

Zgodnie z równaniem (\ref{eq:Tle}), krętość przepływu zdefiniowana jako stosunek średniej długości linii prądu przechodzących przez dany przekrój w jednostce czasu do szerokości próbki \cite{Bear72}, co prowadzi do:
\begin{equation}
  \label{eq:finaltortuosity}
     \ensuremath{T} = \frac{1}{L}
              \frac{\displaystyle \int_A u_y(x) \lambda(x) dx }{\displaystyle \int_A u_y(x) dx},
\end{equation}
gdzie $A$ jest dowolnym przekrojem prostopadłym do osi $y$, a $L$ oznacza długość układu. W równaniu tym $x\in A$, $\lambda(x)$ jest długością linii prądu przecinającej $A$ w punkcie $x$, a $u_y(x)$ -- składową wektora prędkości $\v{u}$ w punkcie $x$ prostopadłą do przekroju $A$.
W literaturze całkowania w równaniu (\ref{eq:finaltortuosity}) definiującym krętość wykonywane były albo za pomocą metody Monte Carlo (MC) \cite{Koponen96,Koponen97,Alam06}, albo poprzez kwadratury \cite{Knackstedt94}. W metodzie MC długości linii prądu przechodzących przez losowo wybrane punkty na wybranym przekroju są uśredniane z użyciem odpowiednich wag. Z kolei metoda kwadratur jest realizowana przez przybliżenie równania (\ref{eq:finaltortuosity}):
\begin{equation}
  \label{eq:DiscreteTortuosity}
     \ensuremath{T} \approx
       \frac{1}{L}
       \frac{\displaystyle  \sum_j u_y(x_j) \lambda(x_j) \Delta x_j }{\displaystyle
                            \sum_j u_y(x_j) \Delta x_j },
\end{equation}
gdzie przedziały $\Delta x_j = x_{j+1} - x_j$ są dyskretnymi podziałami przekroju. W ogólności oba powyższe podejścia do obliczenia całek w równaniu (\ref{eq:finaltortuosity}) są równoważne, okazuje się jednak, że mogą być one w różny sposób implementowane. Na przykład w \nlb \cite{Koponen96,Alam06} całkowanie Monte Carlo zostało przeprowadzone na liniach prądu przechodzących przez losowo wybrane węzły leżące w obszarze porów ośrodka, w \cite{Koponen97} linie prądu przecinały dokładnie wszystkie komórki w obszarze porów, tymczasem w \cite{Zhang95,Knackstedt94} za punkty startowe wybrane zostały węzły brzegowe na wlocie cieczy do układu. Wszystkie wymienione podejścia łączy jednorodny rozkład punktów startowych wybieranych losowo bądź regularnie w obszarze całego ośrodka. Sposób wybrania komórek startowych dla linii prądu może mieć ogromne znaczenie dla obliczanych wartości krętości, szczególnie ze względu na selektywny charakter przepływu w obszarze niskich porowatości, gdyż w tym przypadku ciecz wybiera kilka głównych kanałów przez które przepływa niemal cała jej objętość (rysunek \ref{pic:magn1}). Z tego powodu sumy w równaniu (\ref{eq:DiscreteTortuosity}) mogą zawierać wiele wyrazów o znikomym wpływie na wartość krętości. Aby ominąć ten problem, na równanie (\ref{eq:DiscreteTortuosity}) nałożyliśmy warunek stałości strumienia pomiędzy dwiema sąsiadującymi liniami prądu, co można wyrazić wzorem:
\begin{equation}
 \label{eq:cf-constraint}
  u_y(x_j)\Delta x_j = \mathrm{const}.
\end{equation}
Wartości punktów $x_j$ na przekroju wyznaczone zostały ze wzoru rekurencyjnego:
\begin{equation}
 \label{eq:implicit-xj}
    \int_{x_{j-1}}^{x_j} u_y(x) dx = \frac{1}{N} \int_{0}^{L} u_y(x) dx,\quad j=1,\ldots,N,
\end{equation}
gdzie $x_0 = 0$. Zastosowanie warunku stałego strumienia pomiędzy sąsiednimi liniami prądu powoduje, że wyrażenie (\ref{eq:DiscreteTortuosity}) upraszcza się do:
\begin{equation}
 \label{eq:finalaverage}
   \ensuremath{T} \approx \frac{1}{L} \frac{1}{N} \sum_{j=1}^N \lambda(x_j),
\end{equation}
gdzie $N$ jest liczbą linii prądu wziętą do uśredniania. We wzorze tym wszystkie wyrazy sumy są tego samego rzędu wielkości.

\begin{figure}[!ht]
  \centering
\begin{tabular}{cc}
 \includegraphics[width=0.45\columnwidth,angle=-90]{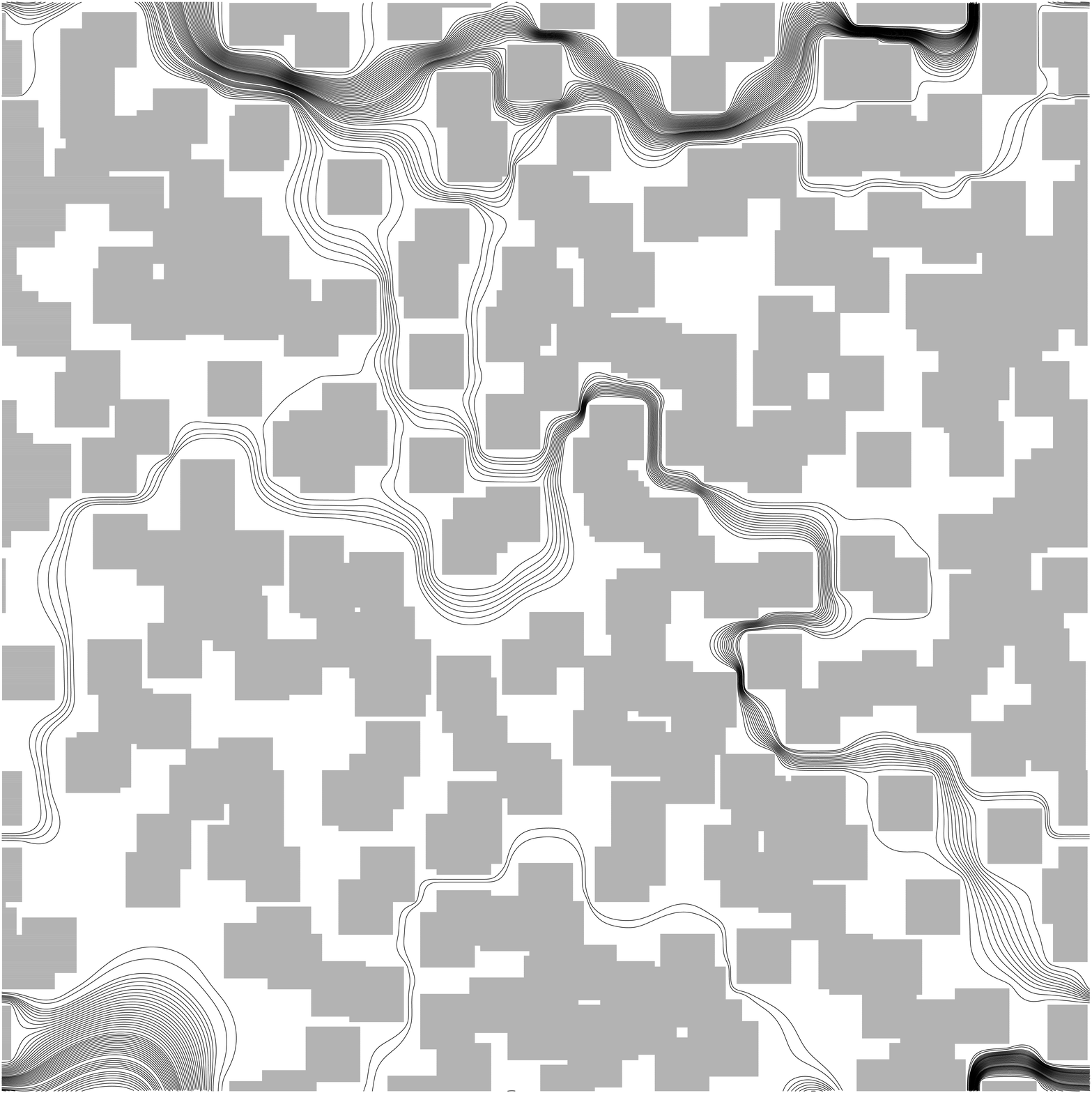}
  \hspace{0.1cm}
    &
  \hspace{0.1cm}
 \includegraphics[width=0.45\columnwidth,angle=-90]{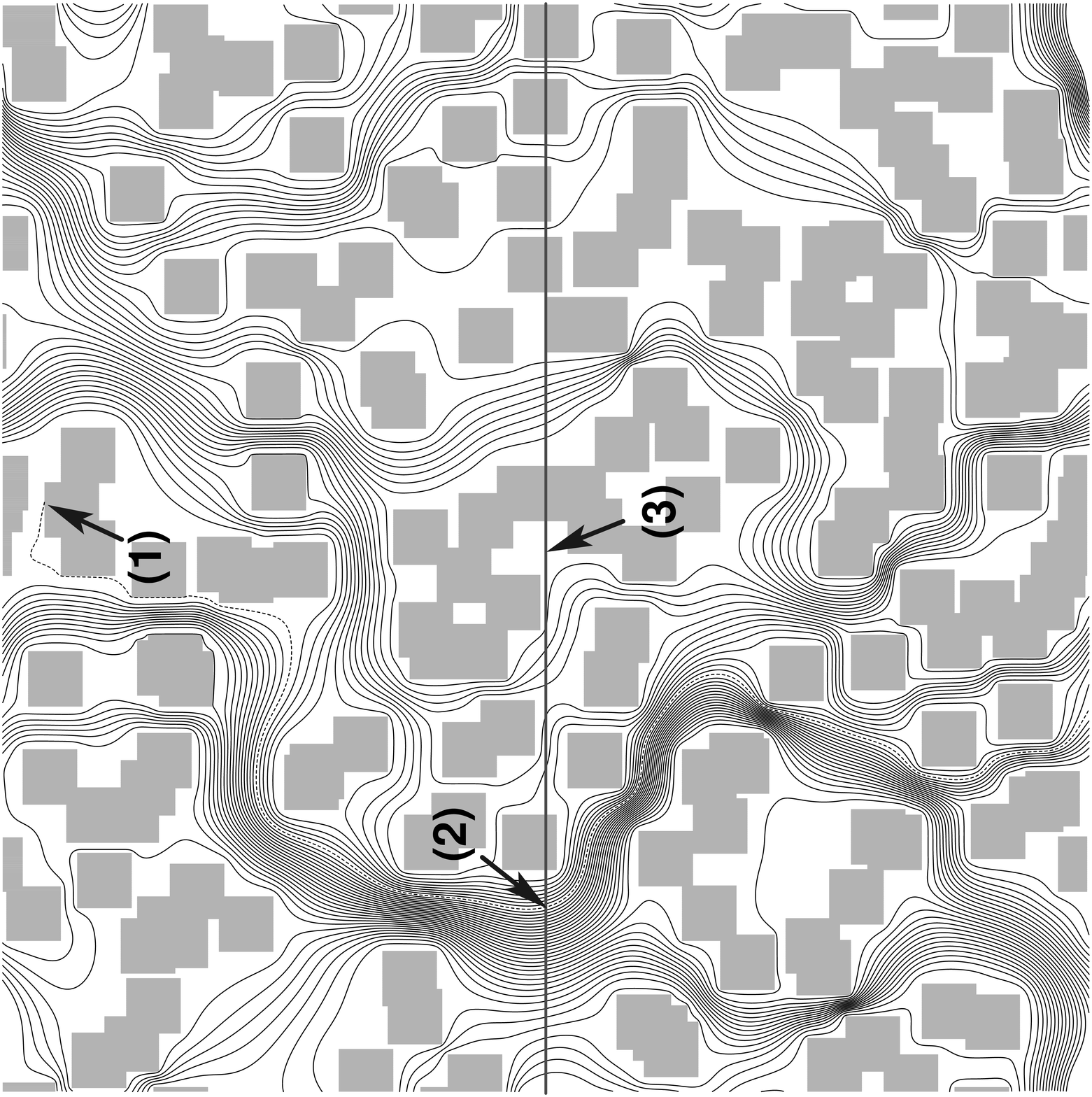}\\
 &  \\
a) $\phi=0.45$ \hspace{0.1cm}&\hspace{0.1cm} b) $\phi=0.65$ \\
 &  \\
 \includegraphics[width=0.45\columnwidth,angle=-90]{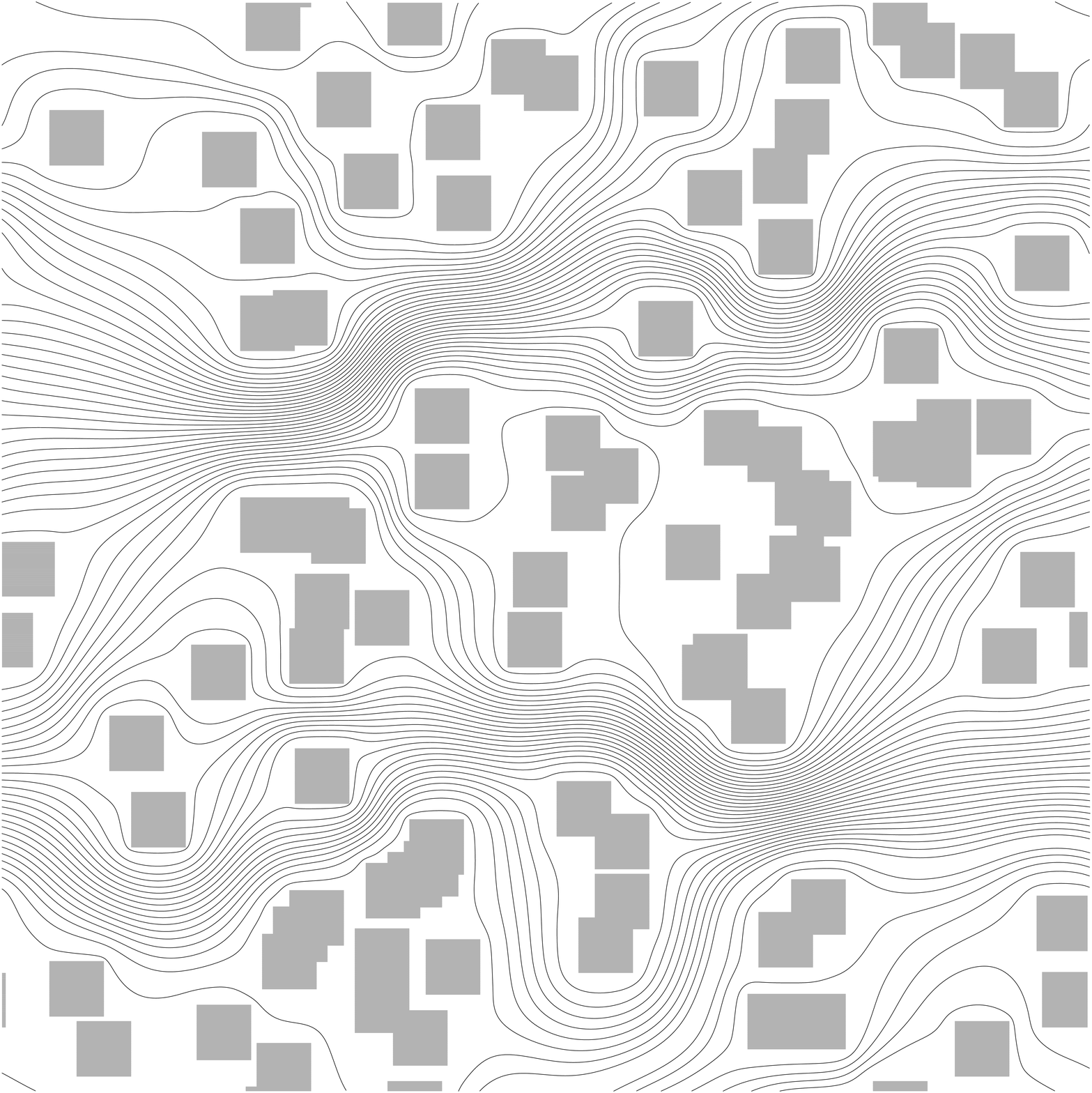}
  \hspace{0.1cm}
    &
  \hspace{0.1cm}
 \includegraphics[width=0.45\columnwidth,angle=-90]{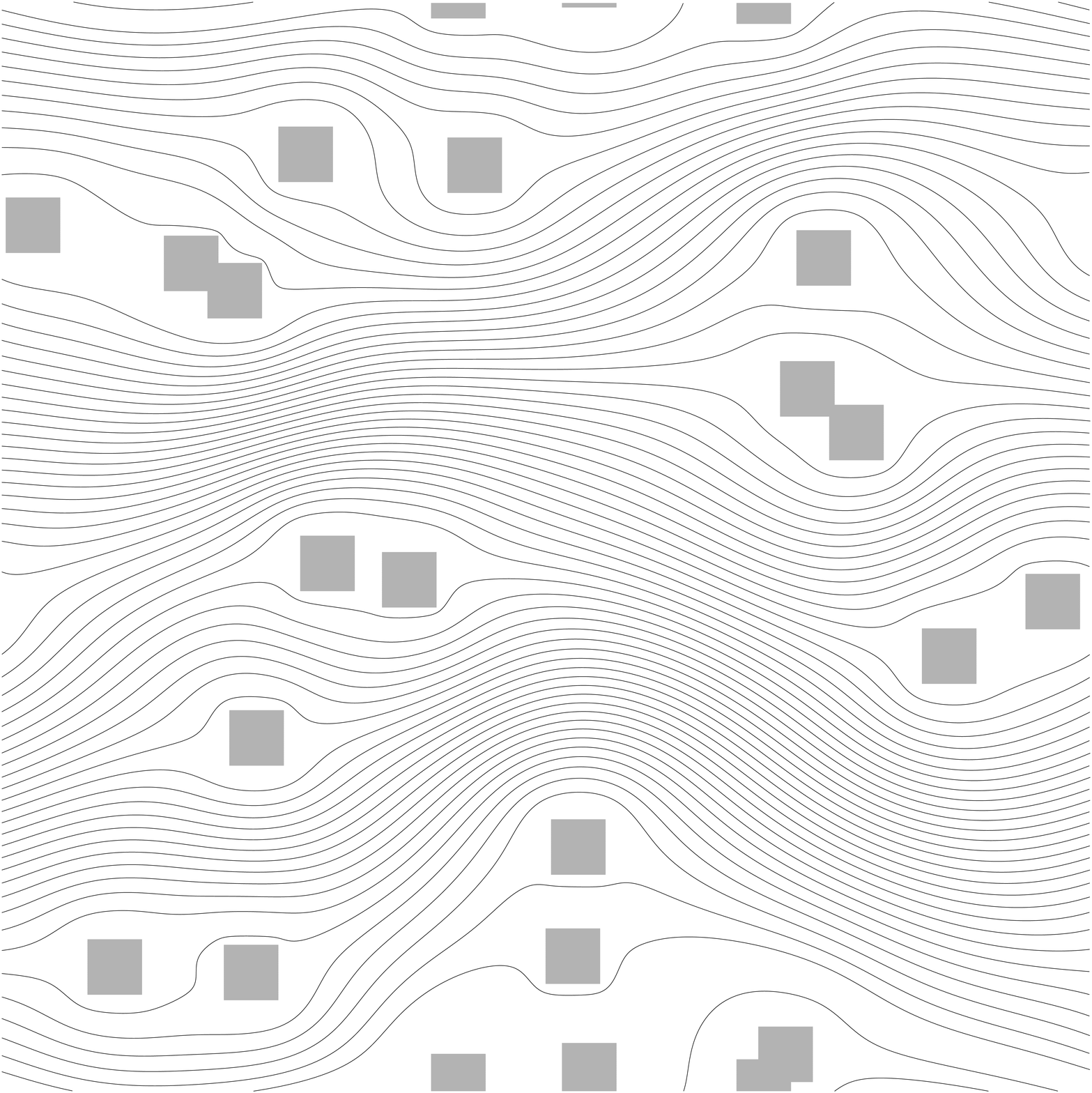}\\
 &  \\
c) $\phi=0.80$ \hspace{0.1cm}&\hspace{0.1cm} d) $\phi=0.95$ \\
 &  \\
\end{tabular}
   \caption{Linie prądu wyznaczone z warunku stałego strumienia (równanie \ref{eq:cf-constraint}) pomiędzy sąsiadującymi liniami prądu dla tych samych układów jak na rysunku \ref{pic:magn1}. Pozioma linia zaznaczona dla układu o $\phi=0.65$ reprezentuje przekrój $y=L/2$, na którym znalezione zostały punkty startowe $x_j$ dla linii prądu. Etykiety wskazują: 1) punkt końcowy niekompletnej linii prądu, 2) punkt startowy niekompletnej linii prądu, 3) ,,martwy'' obszar ośrodka, w którym nie następuje przepływ cieczy. \label{pic:str1}}
\end{figure}

Ostatecznie, w celu obliczenia krętości przepływu przez ośrodek porowaty o znanym kierunku makroskopowym przepływu, wybierany jest przekrój prostopadły do tego kierunku. Następnie, z równania (\ref{eq:implicit-xj}) znajdowany jest rozkład punktów $x_j$ na przekroju, tak aby spełniony był warunek (\ref{eq:cf-constraint}). Linie prądu są wyznaczone przez śledzenie toru ruchu cząstek (równanie \ref{eq:MaselessMotion}) o masie $m=0$ startujących w dwóch przeciwległych kierunkach ze znalezionych punktów $x_j$ na przekroju. Ruch cząstek śledzony jest tak długo, aż dotrą one do brzegów układu $x=0$ lub $x=L$. Z tak wyznaczonych linii prądu obliczana jest średnia wyrażona równaniem (\ref{eq:finalaverage}).
Przykład linii prądu wyznaczonych wg omawianej procedury został przedstawiony na rysunku \ref{pic:str1}.

Na tym rysunku zaznaczona została również linia prądu, która nie znajduje drogi przez cały obszar ośrodka porowatego od brzegu $x=0$ do $x=L$. Może się to zdarzyć w przypadku linii prądu, które przecinają obszary o bardzo niskich prędkościach lokalnych. Przykładowa linia prądu z rysunku \ref{pic:str1} startuje z punktu (2) o bardzo wysokiej wartości prędkości lokalnej i próbuje przejść przez obszar (1), gdzie prędkości są bardzo niskie. Linie, które nie znajdują drogi przez cały ośrodek, nie są brane pod uwagę w sumach w równaniu (\ref{eq:DiscreteTortuosity}). Dzięki zastosowaniu procedury uwzględniającej warunek stałego strumienia pomiędzy sąsiadującymi liniami prądu, ,,martwe'' obszary ośrodka porowatego są praktycznie całkowicie pomijane przy wyborze punktów startowych $x_j$.

\subsection{Ekstrapolacja do stanów stacjonarnych\label{sec:estrapolacja}}
\begin{figure}[!h]
  \centering
\includegraphics[scale=0.83]{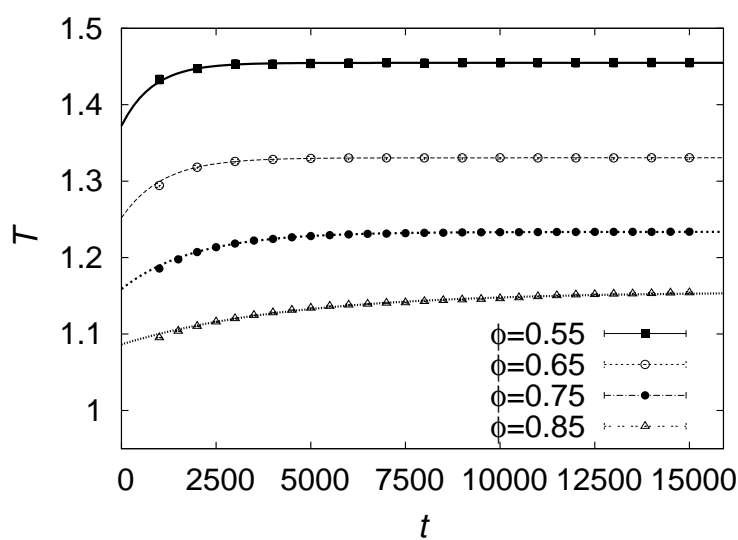}
\caption{Zależność $T$ od kroku czasowego w procesie relaksacji do stanu stacjonarnego dla kilku różnych porowatości układu $\phi$. Każdy z punktów przedstawionych na wykresie stanowi wynik uśredniania po $25$ układach. Liniami zaznaczone zostały dopasowane relacje wykładnicze postaci (\ref{eq:temporal-extrapolation}).\label{pic:asym}}
\end{figure}

Obliczenia metodą LBM prowadzone dla warunków i wg procedury omówionej w \nlb poprzednich podrozdziałach prowadzą do uzyskania stanu stacjonarnego przepływu. Dopiero stan asymptotyczny (stacjonarny) pozwala w efektywny sposób wyznaczyć linie prądu potrzebne do obliczeń krętości. W celu wyznaczenia minimalnej ilości kroków symulacji potrzebnych do otrzymania stanu stacjonarnego, badałem chwilowe wartości $T$ w kolejnych krokach czasowych (rysunek \ref{pic:asym}).
Każdy z punktów pomiarowych przedstawionych na rysunku \ref{pic:asym} jest krętością uśrednioną po $N=25$ różnych konfiguracjach o określonej porowatości $\phi$. Okazało się, że po $t_0$ krokach, zależność $T(t)$ może być przybliżona relacją wykładniczą:
\begin{equation}
 \label{eq:temporal-extrapolation}
    T(t) \approx T_\mathrm{s} - c \exp(-t/t_\mathrm{rel}), \qquad t > t_0,
\end{equation}
gdzie $T_s$, $c$ oraz $t_\mathrm{rel}$ są pewnymi parametrami i różnią się wartościami dla różnych porowatości $\phi$. Z rysunku widać wyraźnie, że czas relaksacji układu rośnie wraz ze zbliżaniem się  porowatości $\phi$ do $1$, co przedyskutuję dokładniej w dalszej części pracy.

\section{Wyniki}

\subsection{Wartości lokalne na przekroju}

\begin{figure}[!ht]
  \centering
\includegraphics[scale=0.59]{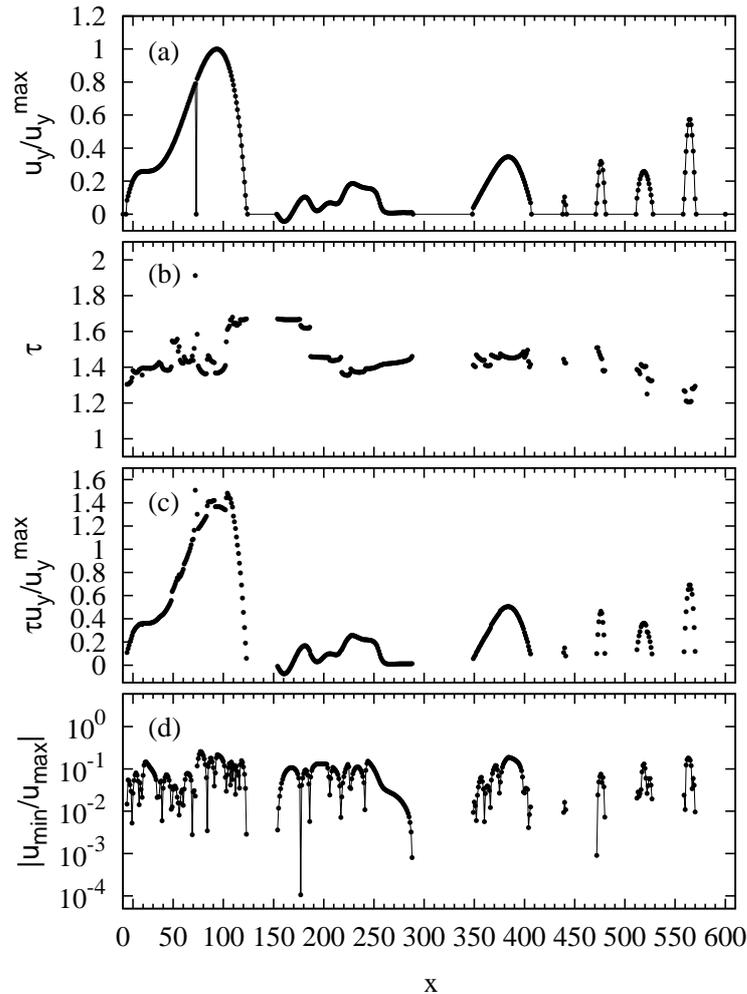}
\caption{Lokalne wartości obliczane na przekroju $A$ używane do obliczeń krętości: (a) składowa prędkości $u_y(x)$ znormalizowana do wartości maksymalnej $u_y^\mathrm{max}\approx 5\times10^{-5}$ \emph{lu} \emph{tu}$^{-1}$; (b) lokalna krętość $\tau(x)=\lambda(x)/L$; (c) iloczyn $\tau(x)u_y{x}/u_y^\mathrm{max}$ i\protect\nlb~(d) stosunek prędkości minimalnej do maksymalnej (obie wartości znalezione wzdłuż całej linii prądu). Wartości z tego rysunku zostały zebrane z przekroju dla konfiguracji o porowatości $\phi=0.65$ z rysunku \ref{pic:str1}. \label{pic:crossdata1}}
\end{figure}

\begin{figure}[!ht]
  \centering
\includegraphics[scale=0.59]{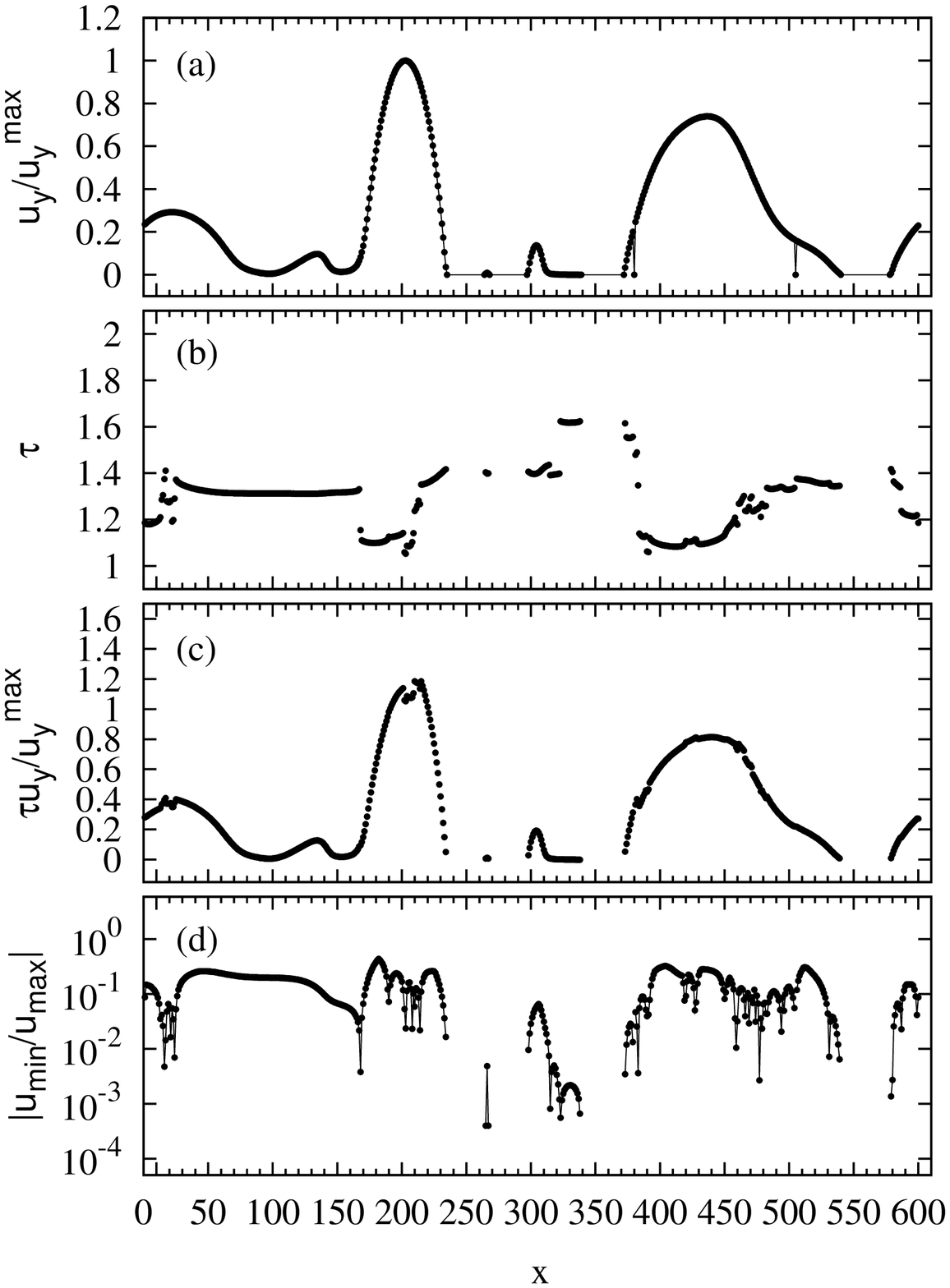}
\caption{Lokalne wartości obliczane na przekroju $y=L/2$ używane do obliczeń krętości dla konfiguracji o porowatości $\phi=0.80$ z rysunku \ref{pic:str1}. Pozostałe parametry jak na rysunku \ref{pic:crossdata1}. \label{pic:crossdata3}}
\end{figure}

\begin{figure}[!ht]
  \centering
\includegraphics[scale=0.59]{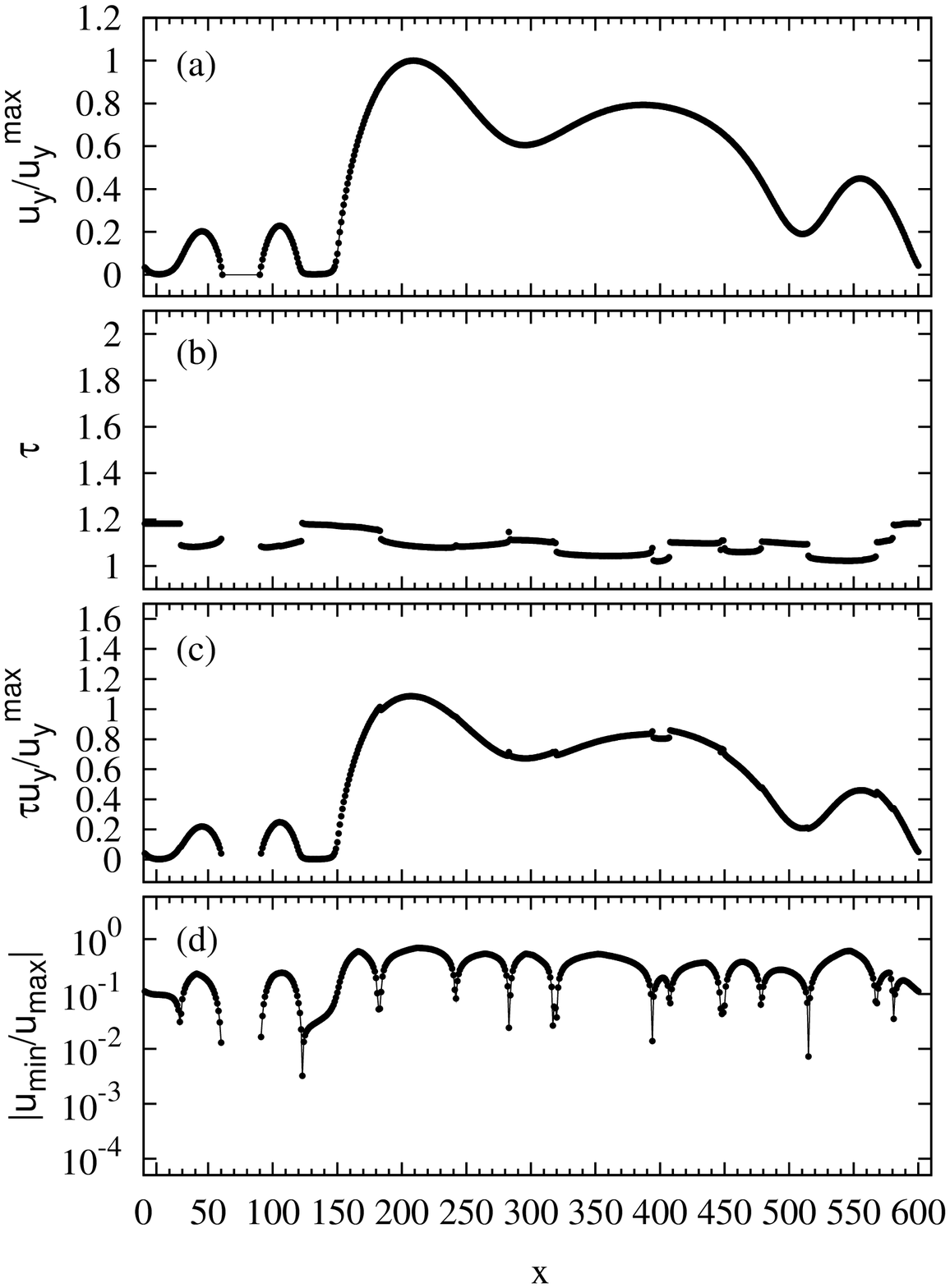}
\caption{Lokalne wartości obliczane na przekroju $y=L/2$ używane do obliczeń krętości dla konfiguracji o porowatości $\phi=0.95$ z rysunku \ref{pic:str1}. Pozostałe parametry jak na rysunku \ref{pic:crossdata1}. \label{pic:crossdata4}}
\end{figure}

\begin{figure}[!h]
  \centering
\includegraphics[scale=0.83]{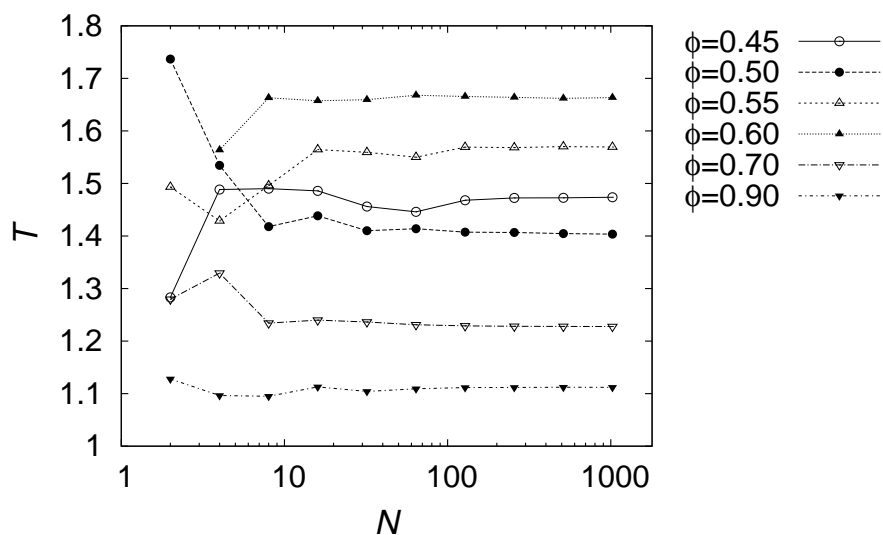}
\caption{Krętość w funkcji liczby linii prądu dla układu $200\times 200$ \emph{lu} i podziału $k_\mathrm{ref}=3$. \label{pic:kretoscwfunkcjilinii}}
\end{figure}

Wartości wielkości potrzebnych do wyznaczenia całek w równaniu (\ref{eq:finaltortuosity}) zostały przedstawione na rysunkach \ref{pic:crossdata1}--\ref{pic:crossdata4}. Na przekrojach poziomych obu układów obliczono:
\begin{enumerate}
\item znormalizowaną prędkość $u_y(x)/u_y^\mathrm{max}(x)$, gdzie $u_y^\mathrm{max}(x)$ jest największą prędkością wyznaczoną wzdłuż wszystkich linii prądu w układzie,
\item krętość pojedynczej linii prądu $\tau(x)=\lambda(x)/L$,
\item iloczyn $\tau(x)u_y(x)/u_y^\mathrm{max}(x)$,
\item stosunek minimalnej do maksymalnej prędkości wzdłuż linii prądu.
\end{enumerate}
Wszystkie wielkości zostały zebrane z pól prędkości dla układów z rysunków \ref{pic:magn1} oraz \ref{pic:str1}.
%
%
Zależności przedstawione na wykresach (\ref{pic:crossdata1}--\ref{pic:crossdata4}, a--c) są silnie zależne od pozycji, na której był wyznaczony przekrój. W naszym przypadku przekroje były wzięte dla $y=L/2$. Jak można było się spodziewać, profile prędkości $u_y(x)/u_y^\mathrm{max}(x)$ są ciągłe i kawałkami różniczkowalne, czego nie można powiedzieć o przebiegach krętości $\tau(x)$. Co oczywiste, iloczyn $\tau(x)u_y(x)/u_y^\mathrm{max}(x)$ również nie jest ciągły i jest to jeden z \nlb powodów trudności, jakie pojawiają się przy całkowaniu numerycznym równania (\ref{eq:finaltortuosity}).
Każda nieciągłość funkcji $\tau(x)$ przedstawionej na omawianych wykresach w części b) odpowiada łączeniu się lub rozwidlaniu dwóch sąsiadujących linii prądu. Odpowiada to sytuacji, kiedy dwie strugi cieczy napotykające przeszkodę opływają ją z dwóch różnych stron. Poprzez liczbę punktów nieciągłości funkcji $\tau(x)$ oszacować można liczbę tego typu ,,wysp'' w układzie. Porównując profile dla różnych porowatości od $\phi=0.45$ do $\phi=0.9$ można zauważyć, że liczba punktów nieciągłości maleje dla $\phi\rightarrow\phi_c$ oraz dla $\phi\rightarrow 1$.
Problem znajdowania miejsc nieciągłości funkcji $\tau(x)$ na przekroju nie jest dobrze określony numerycznie. Wybór miejsc startowych dla linii prądu blisko punktów nieciągłości powoduje, że wartości krętości mogą być obarczone dużymi błędami, gdyż niewielka niedokładność w wyznaczeniu $x_j$ może skutkować dużą niedokładnością (skokiem) wartości $\lambda(x_j)$. Okazuje się również, że stosunek prędkości maksymalnej do minimalnej rośnie znacząco w punktach bliskich punktom nieciągłości.
Jednym z czynników wpływających na dokładność związaną z \nlb nieciągłościami w $\tau(x)$ może być liczba linii prądu wziętych do obliczeń. Aby sprawdzić wpływ tego parametru na otrzymywane wartości, dla kilku układów o rozmiarach $200\times 200$ \emph{lu} (stałych sieci) z podziałem $k_\mathrm{ref}=3$ o różnych $\phi$ wyznaczona została wartość $T$ obliczona z $N=2^k$ linii prądu, gdzie $k=1,2,\ldots,10$. Wyniki przedstawione zostały na rysunku \ref{pic:kretoscwfunkcjilinii}.
Widać wyraźnie, że w okolicach $N\approx 100$ krętość przestaje się zmieniać wraz z zagęszczaniem linii prądu. Można stąd wnioskować, że liczba $N\approx L$ linii prądu jest wystarczająca do obliczeń krętości, a jej zwiększenie nie zmieniłoby znacząco otrzymywanych wartości $T$.

\subsection{Czas relaksacji $t_\mathrm{rel}$}\label{sec:time}

\begin{figure}[!h]
  \centering
\includegraphics[scale=0.57]{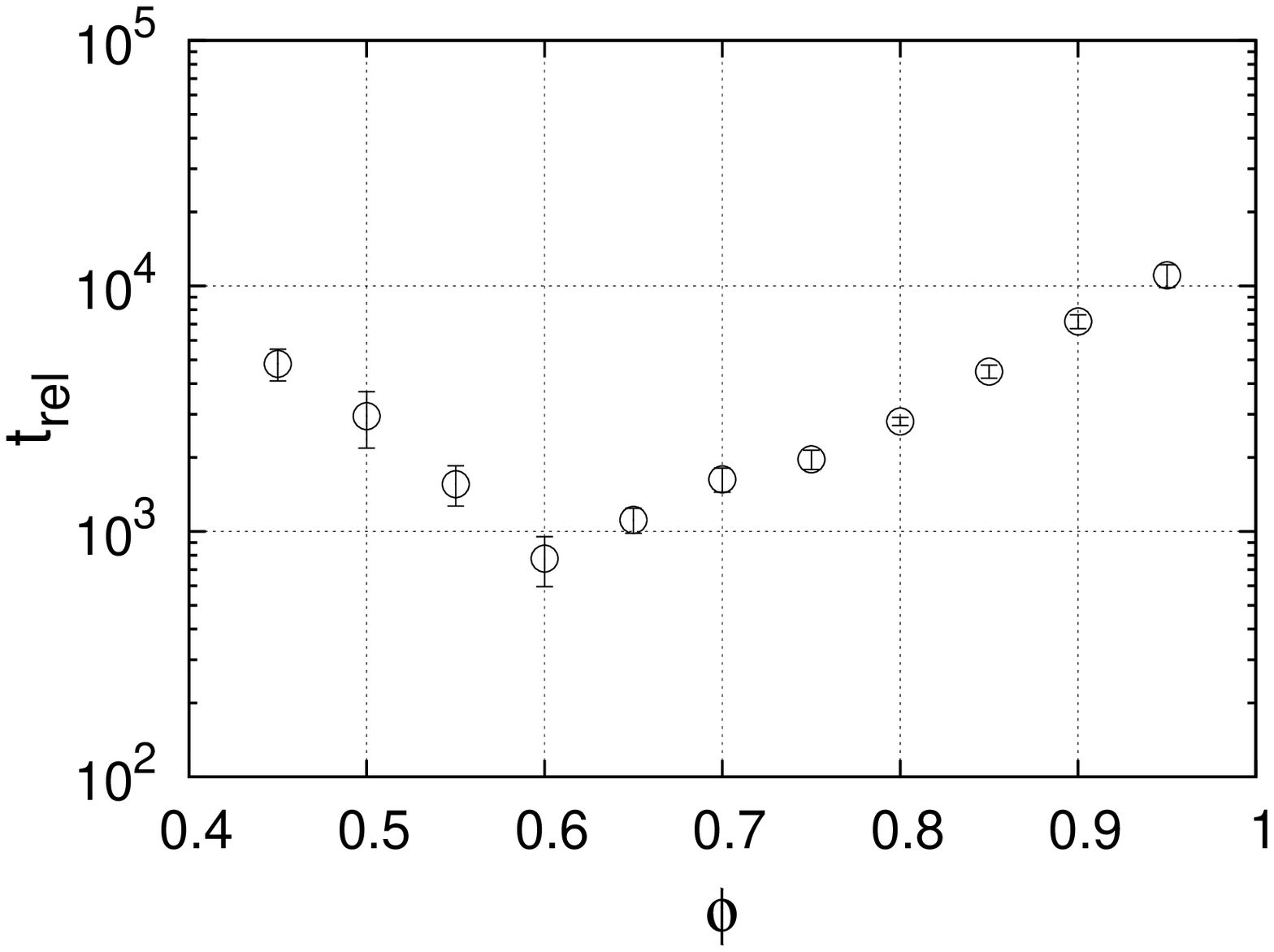}
\caption{Czas relaksacji $t_\mathrm{rel}$ w funkcji porowatości $\phi$ dla $L=200$ \emph{lu}, $k_\mathrm{ref}=3$ i $t_\mathrm{max}=30000$ \emph{tu}. \label{pic:trel}}
\end{figure}

Jak pokazaliśmy w podrozdziale \ref{sec:estrapolacja} (rysunek \ref{pic:asym}) wartość $T$ jest zależna od kroku czasowego $t$ symulacji. Charakterystyczny czas relaksacji tego zjawiska $t_\mathrm{rel}$ jest różny dla różnych porowatości oraz różnych konfiguracji porów w ośrodku. Zależność $t_\mathrm{rel}(\phi)$ przedstawiona została na rysunku \ref{pic:trel}, gdzie każdy punkt reprezentuje średnią po co najmniej $25$ układach.
Jak widać z wykresu, czas relaksacji ma minimum dla $\phi\approx 0.6$ i od tej wartości rośnie w obu kierunkach. Wzrost w kierunku wysokich porowatości ($\phi\rightarrow 1$) jest jasny i zgadza się z innymi doniesieniami o takim zachowaniu modelu LBM w zakresie niskich liczb Macha \cite{Guo04}. Wzrost $t_\mathrm{rel}$ w kierunku wysokich porowatości ($\phi\rightarrow \phi_c \approx 0.367$) jest prawdopodobnie związany ze wzrostem liczby i długości martwych pól w ośrodku porowatym, które są wypełnione cieczą, ale praktycznie nie dają żadnego wkładu do przepływu \cite{Andrade97}. Czysto dyfuzyjny charakter transportu w tych porach powoduje znaczący wzrost $t_\mathrm{rel}$.

\begin{figure}[!h]
  \centering
\includegraphics[scale=0.83]{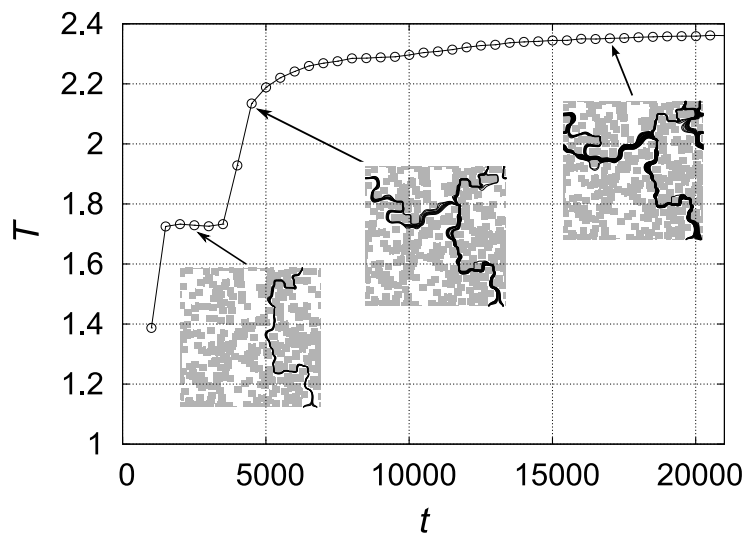}
\caption{Przykład zbieżności wartości $T$ do stanu asymptotycznego. Na rysunkach wstawionych do wykresu widać konfiguracje linii prądu odpowiadające czasom $t=2500, 4500$ i $17000$. Siła zewnętrzna skierowana jest w kierunku pionowym. \label{pic:reorganization}}
\end{figure}

Jak pokazaliśmy w podrozdziale \ref{sec:estrapolacja} w zależności $T(t)$ występuje czynnik $t_0$. Okazuje się, że czynnik ten spełnia warunek $500 < t_0 < 1000$ w całym zakresie $\phi$. Dlatego, korzystając z danych z rysunku \ref{pic:trel}, do obliczeń krętości założyliśmy maksymalną liczbę $t_\mathrm{max}=1.5\times 10^4$ kroków dla $\phi\le 0.8$ i $3\times10^4$ dla $\phi\ge 0.8$. Okazało się jednak, że dla kilku układów z porowatością $\phi=0.45$ $t_0$ było znacznie większe od $10^3$ \emph{tu}. Jeśli przyjrzeć się bliżej strukturze tych układów, zauważyć można, że są one zbudowane w odmienny sposób. Posiadają one dwa lub więcej głównych kanałów, którymi zachodzi transport wzdłuż siły wymuszającej przepływ i są one połączone ze sobą kanałem mniej więcej prostopadłym do kierunku przepływu głównego. W \nlb układzie takim, w procesie relaksacji do stanu stacjonarnego dojść może do ,,przełączenia'' przepływu z równoległych kanałów do kanału łączącego, co widoczne jest na wykresie $T(t)$ jako nagły uskok wartości $T$.  Przykład relaksacji krętości dla przykładowego układu o porowatości $\phi$ przedstawiony został na wykresie \ref{pic:reorganization}.
Ze wstawionych do wykresu wizualizacji linii prądu dla trzech różnych chwil czasu widać wyraźnie, że wybrany do przepływu kanał pionowy, po około $4000$ kroków przestaje być jedynym dozwolonym dla przepływu, a w okolicach $t=15000$ przestaje dominować. Co ciekawe, przez wybranie początkowo najkrótszej ścieżki, układ wydawał się być stabilny ze względu na krętość aż do około $t=4000$, a następnie z powodu znalezienia dodatkowej drogi dla przepływu, wartość $T$ zaczęła gwałtownie rosnąć, by następnie kontynuować wzrost z relaksacją wykładniczą. Jak widać z powyższej analizy, w tego typu układach trudno jest zdefiniować ściśle kryterium zbieżności wartości $T$ do stanu asymptotycznego. Dlatego nie ma pewności, że we wszystkich badanych układach w okolicy $\phi=0.45$ osiągnięty został stan stacjonarny. Wartości $T(\phi=0.45)$ mogą być więc obarczone niewielkim, systematycznym błędem, którego wielkość oszacować można na mniej niż $1\%$ (na podstawie liczby układów, które zdradzają wyżej wymienione cechy). Minimalna wartość czasu relaksacji w okolicach $\phi\approx 0.6$ może być związana z przeprowadzoną w poprzednim podrozdziale analizą ilości punktów nieciągłości w zależności od $\phi$. W okolicach $\phi\approx 0.6$ nieciągłości jest bardzo dużo (najwięcej) co odpowiadać może selekcji kanałów o niskiej krętości, których ilość jest dość duża (każdy punkt nieciągłości to taki właśnie kanał). Można się spodziewać, że relaksacja w kanałach o takiej charakterystyce będzie najszybsza.

\subsection{Znaczenie wielkości układu}
\begin{figure}[!h]
  \centering
\includegraphics[scale=0.58]{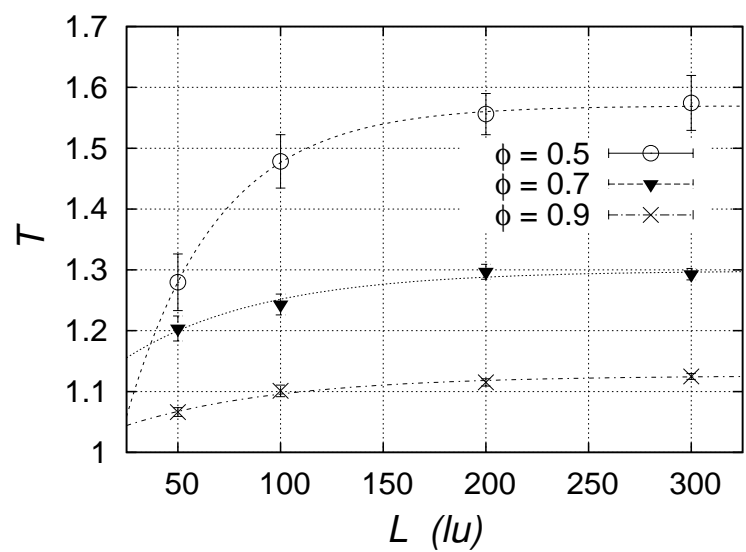}
\caption{Zależność $T(L)$ dla $\phi=0.5$, $0.7$ i $0.9$. Punkty reprezentują wartości średnie, uśrednione po $N\geq 24$ układach. Krzywe są dopasowaniami do wzoru (\ref{eq:fit}), a słupki błędów reprezentują błąd standardowy wartości średniej. \label{pic:TL}}
\end{figure}

Jedną z ważniejszych cech krętości układu, jaką udało mi się zauważyć, a jaka potraktowana została pobieżnie w poprzednich pracach innych autorów, jest jej zależność od wielkości sieci $L$. W celu ilościowej analizy tej zależności, średnia krętość $T$ \nlb obliczona została dla trzech porowatości: $\phi=0.5$, $0.7$ i $0.9$ oraz czterech wielkości układu $L=50$, $100$, $200$ i $300$. Wyniki przedstawione zostały na rysunku \ref{pic:TL}.
Każdy z punktów wykresu reprezentuje średnią wartość $T$ uśrednioną po $N\geq 24$ układach ($N$ dobrane zostało tak, by błędy statystyczne widoczny na wykresie były tego samego rzędu). Krzywe widoczne na wykresie są dopasowanymi zależnościami wykładniczymi wyrażonymi wzorem:
\begin{equation}
  \label{eq:fit}
   \ensuremath{T}(L) = \ensuremath{T}_{\infty} - b\exp(-cL),
\end{equation}
gdzie $T_{\infty}$ jest wartością asymptotyczną krętości (dla układu nieskończonego) i razem z $b$ oraz $c$ tworzy grupę parametrów swobodnych dopasowania. Można zauważyć, że w całym zakresie porowatości $T$ jest rosnącą funkcją rozmiaru układu. Co więcej, wielkość charakterystyczna $L^{*}$ powyżej której $T$ nie zmienia się znacząco wraz ze wzrostem $L$ wynosi $L^{*}\approx 200$. Ponadto, rysunek \ref{pic:TL} nie pozostawia wątpliwości, że zignorowanie efektów związanych z rozmiarem sieci prowadzi do niedoszacowania wartości $T$. Wielkość niedoszacowania jest zależna od porowatości $\phi$ i spodziewać się można, że większe różnice wystąpić mogą dla niższych porowatości.

\begin{figure}[!h]
  \centering
\includegraphics[scale=0.58]{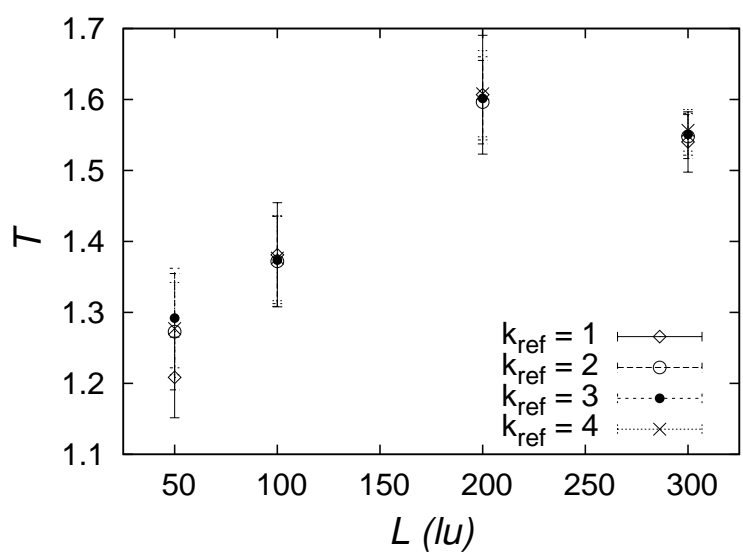}
\caption{Zależność $T(L)$ dla $\phi=0.5$ i czterech stopni podziału $k_\mathrm{ref}=1,2,3$ i $4$. Punkty reprezentują wartości średnie, uśrednione po $n=9$ układach. Słupki błędów reprezentują błąd standardowy wartości średniej. \label{pic:TREF}}
\end{figure}

Kolejnym krokiem było sprawdzenie, jak na otrzymywane wartości wpływa stopień podziału sieci $k_\mathrm{ref}$. Na rysunku \ref{pic:TREF} przedstawione zostały wartości $T$ w funkcji wielkości układu dla $\phi=0.5$ oraz czterech stopni podziału sieci $k_\mathrm{ref}=1,2,3$ i $4$. Okazało się, że mimo znanego z literatury faktu, że model LBM potrzebuje przynajmniej cztery jednostki sieci do poprawnego opisania hydrodynamiki \cite{Succi01} (co w naszych warunkach odpowiada podziałowi $k_\mathrm{ref}=3$), w przypadku obliczeń krętości stopień podziału nie ma tak dużego znaczenia.
Z wykresu (\ref{pic:TREF}) widać wyraźnie, że już $k_\mathrm{ref}=1$ wydaje się być wystarczające do tego typu obliczeń. Ten niespodziewany wynik (część układu na przewężeniach posiada przecież lokalne rozwiązania niezgodne z równaniami Naviera--Stokesa) można wytłumaczyć zauważając, że dla porowatości dużo wyższych od progu perkolacji główny przepływ cieczy przez układ zachodzi szerokimi kanałami o dość dużej średnicy w porównaniu z komórką elementarną sieci. Dlatego wpływ niedokładności w wyznaczeniu pola prędkości w okolicach przewężeń nie ma istotniejszego znaczenia. Spodziewać się jednak można, że wraz ze zbliżaniem się z porowatością do progu perkolacji ($\phi\rightarrow \phi_c$), występowanie szerokich kanałów dominujących przepływ będzie rzadsze, a co za tym idzie -- znaczenie stopnia podziału może być większe.

\subsection{Zależność krętości od porowatości}

\begin{figure}[!h]
\centering
\includegraphics[scale=0.58]{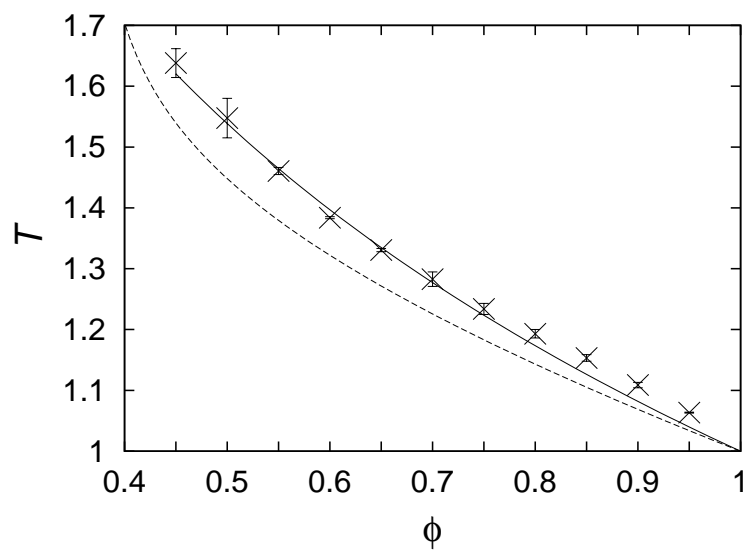}
\caption{Zależność krętości $T$ od porowatości $\phi$. Dane otrzymane zostały wg omówionej procedury z użyciem modelu LBM i równań (\ref{eq:finalaverage}) oraz (\ref{eq:temporal-extrapolation}) (krzyże ze słupkami błędów). Dodatkowo zaznaczone zostały: 1) krzywa teoretyczna zaproponowana dla takiego samego modelu przez Koponena i in. \protect\cite{Koponen97} (linia przerywana) oraz najlepsze dopasowanie do równania (\ref{fit:comiti}) (linia ciągła). \label{pic:TPHI}}
\end{figure}

Dzięki znajomości minimalnych wymagań odnośnie rozmiaru sieci, stopnia podziału oraz długości czasu relaksacji do stanu stacjonarnego, jesteśmy w stanie wyznaczyć zależność krętości $T$ od porowatości $\phi$ w szerokim zakresie $\phi$. Dla porowatości $\phi=0.45,0.5,\ldots,0.95$ wybrane zostały: $L=L^{*}$ oraz stopień podziału $k_\mathrm{ref}=3$. Dla każdej porowatości, krętość $T$ wyznaczona została dla $M$ różnych konfiguracji przeszkód, gdzie $M$ zmieniało się od $25$ dla $\phi=0.95$ do $100$ dla $\phi=0.45$. Średnie wartości $T$ przedstawione zostały na wykresie \ref{pic:TPHI} wraz z relacją wyrażoną równaniem (\ref{fit:koponen}) dla tego samego zagadnienia zaproponowaną przez Koponena i innych w \cite{Koponen97}.
Przyczyny różnic pomiędzy wynikami prezentowanymi tutaj a pracą \cite{Koponen97} można tłumaczyć nieuwzględnieniem przez autorów tamtej pracy efektów wielkości sieci, wyborem $L<L^{*}$ oraz brakiem dokładnej analizy czasów relaksacji krętości.

Dane z rysunku \ref{pic:TPHI} dopasowane zostały do czterech jednoparametrowych wzorów empirycznych z jednym wolnym parametrem zaproponowanych w innych opracowaniach (rozdział \ref{sec:wstepkretosc}, str. \pageref{str:jeden}). W przypadku naszych obliczeń krętości wyznaczonej dla modelu pokrywających się prostokątów przy pomocy procedury omówionej powyżej, najlepsze dopasowanie dał wzór empiryczny (\ref{fit:comiti}) z parametrem $p=0.77\pm 0.03$. Najlepsze dopasowanie zostało narysowane linią ciągłą na rysunku \ref{pic:TPHI}. Okazuje się, że dla $\phi\approx 1$ dane odstępują nieznacznie od wzoru \ref{fit:comiti}. Niezbyt dobre dopasowanie danych w tym zakresie porowatości może być m.in. rezultatem nieuwzględnienia anizotropii badanych układów spowodowanej periodycznymi warunkami brzegowymi.

\subsection{Korelacja krętości z powierzchnią charakterystyczną}\label{rozdz:nowe1}

\begin{figure}[!h]
  \centering
\includegraphics[scale=0.83]{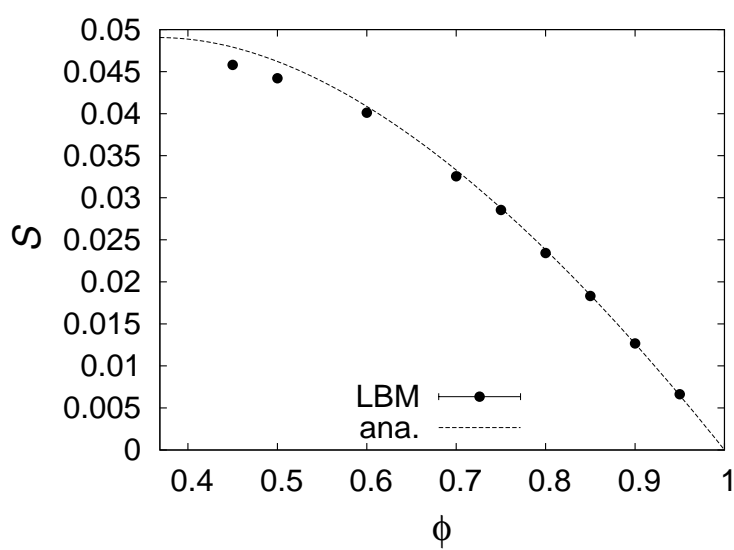}
\caption{Zależność powierzchni charakterystycznej $S$ od porowatości dla kwadratów $10\times 10$ i $k_\mathrm{ref}=3$. Linią przerywaną zaznaczony został wynik analityczny (\ref{eq:specific}) z uwzględnieniem wartości $d=2$ oraz $R=15$. \label{pic:SPHI}}
\end{figure}

Fakt, że najlepszym dopasowaniem do danych numerycznych zależności $T(\phi)$ okazało się prawo logarytmiczne, ma dość ciekawe konsekwencje dla korelacji tej wielkości z innymi parametrami makroskopowymi opisującymi strukturę ośrodka. Wiadomo bowiem, że powierzchnia charakterystyczna $S$ ośrodka złożonego z losowo rozmieszczonych, pokrywających się kwadratowych przeszkód o polu powierzchni $V_0$ i obwodzie $A_0$, charakteryzuje się logarytmiczną zależnością od porowatości (\cite{Koponen97}):
\begin{equation}
  \label{eq:specific}
   S = -\frac{d}{R} \phi\ln\phi,
\end{equation}
gdzie $R$ jest tzw. promieniem hydraulicznym przeszkód, a $d$ jest wymiarem przestrzeni (tu $d=2$). Promień hydrauliczny definiujemy jako $R=d\cdot V_0/A_0$, gdzie $V_0$ jest objętością przeszkody, a $A_0$ jej powierzchnią. Na rysunku \ref{pic:SPHI} przedstawione zostało porównanie powyższej relacji z wartościami numerycznymi.
Punkty na wykresie reprezentują średnie po $N=14$ układach dla każdej z przedstawionych wartości $\phi$. Ogólna zgodność jest bardzo dobra, a niewielkie odstępstwa od wartości analitycznej w okolicy progu perkolacji mogą być związane z efektami rozmiaru sieci oraz użyciem $k_\mathrm{ref}=3$.

\begin{figure}[!h]
  \centering
\includegraphics[scale=0.83]{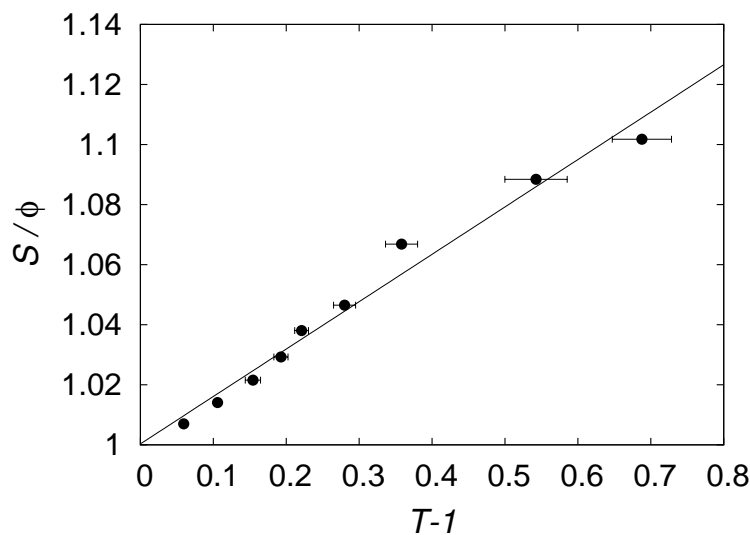}
\caption{Graficzna weryfikacja proporcjonalności wyrażonej wzorem \ref{eq:lindep}. Linia prosta reprezentuje dopasowanie liniowe postaci $y=a\cdot x+b$, przy czym najlepsze dopasowanie uzyskane zostało dla $a=0.16$ i $b=1.0$. \label{pic:ST}}
\end{figure}

Przy pomocy prostych rachunków, można pokazać, że z relacji (\ref{fit:comiti}) oraz (\ref{eq:specific}) wynika, że \cite{Matyka08}:
\begin{equation}
   \label{eq:lindep}
      \ensuremath{T} - 1\propto \frac{S}{\phi}.
\end{equation}
Na rysunku \ref{pic:ST} przedstawiony został wykres zależności $T-1$ w funkcji $\frac{S}{\phi}$, gdzie stała proporcjonalności $p=0.77$ wyznaczona została z procedury dopasowania do wzoru (\ref{fit:comiti}).
Okazuje się, że proporcjonalność wyrażona wzorem (\ref{eq:lindep}) dobrze opisuje zależność między $T$, $S$ i $\phi$ w całym badanym zakresie porowatości i krętości. Oznacza to, że w badanym modelu wzrost długości linii brzegowej (powierzchni charakterystycznej) $S$ prowadzi do wzrostu krętości $T$, a efekt ten wzmacnia się wraz ze zmniejszaniem porowatości ($\phi$ w mianowniku). Na tym etapie nie można jednak stwierdzić, czy relacja (\ref{eq:lindep}) jest w jakimkolwiek stopniu ogólna. Warto jednak zauważyć, że nawet jeśli relacja pomiędzy krętością a porowatością w jakimś układzie nie jest logarytmiczna, to zależność (\ref{eq:lindep}) może być nadal spełniona. Jest to jeden z kierunków dalszych badań nad korelacjami parametrów makroskopowych w ośrodkach porowatych.

\subsection{Korelacja krętości z porowatością efektywną}\label{rozdz:nowe2}

\begin{figure}[!h]
  \centering
\includegraphics[scale=0.83]{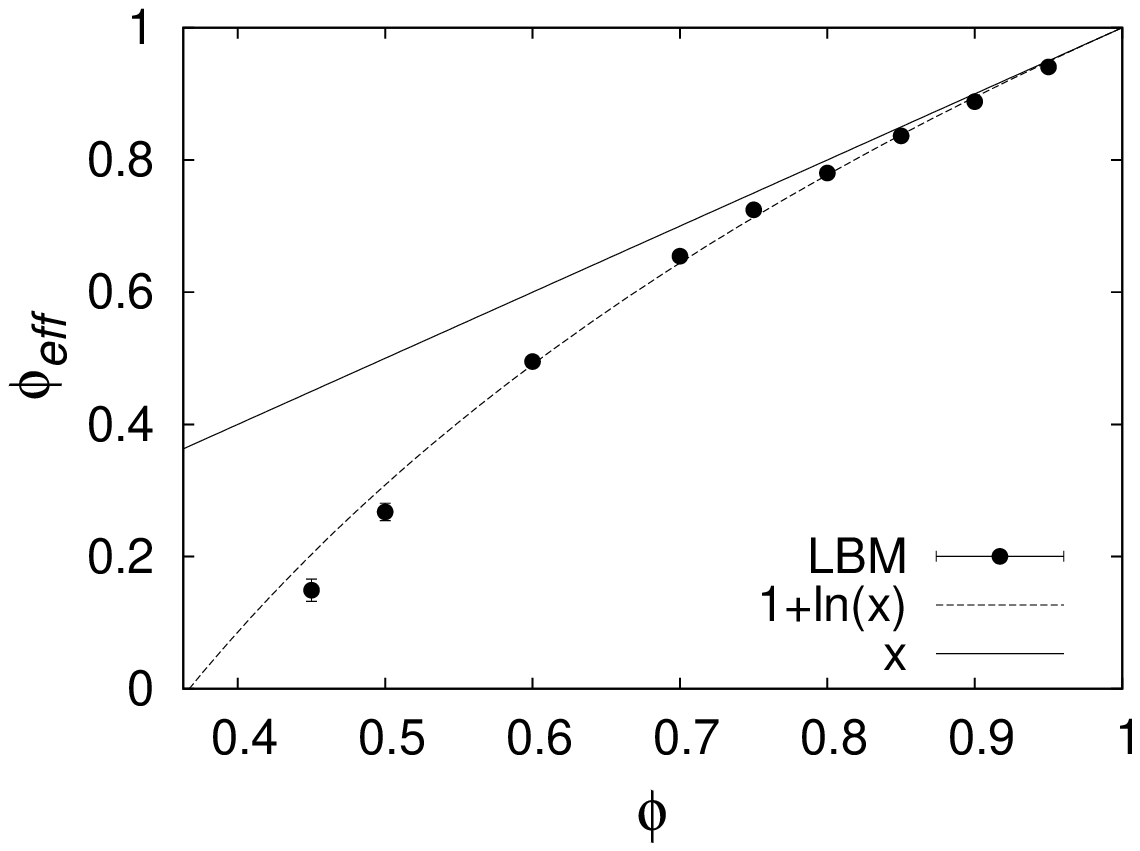}
\caption{Zależność porowatości efektywnej $\phi_\mathrm{eff}$ od porowatości całkowitej $\phi$. Czarne punkty to rezultat procedury numerycznej. Dodatkowo narysowane zostały funkcje $y=x$ oraz $\phi_\mathrm{eff}=1+\ln(\phi)$ -- postulowana zależność $\phi_\mathrm{eff}(\phi)$. \label{pic:PHIPHIEFF}}
\end{figure}

\begin{figure}[!h]
  \centering
\includegraphics[scale=0.83]{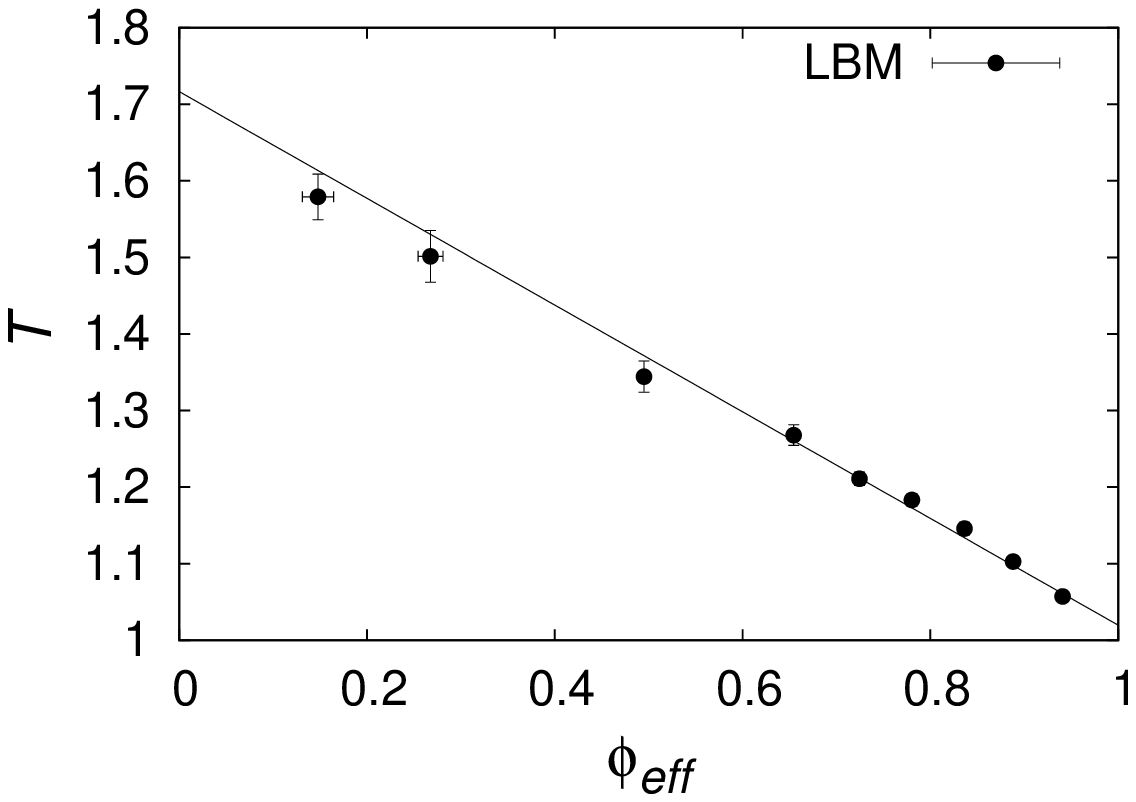}
\caption{Liniowa zależność krętości $T$ od porowatości efektywnej $\phi_\mathrm{eff}$. \label{pic:TPHIEFF}}
\end{figure}

Analizę korelacji porowatości z krętością można też przeprowadzić dla porowatości efektywnej $\phi_\mathrm{eff}$, którą tu definiujemy jako stosunek ilości węzłów sieci, przez które zachodzi efektywnie cały transport cieczy, do reszty węzłów \cite{Koponen97}. Dzięki tej definicji wszelkie strefy ,,martwe'', które z punktu widzenia transportu hydrodynamicznego są pomijalne (ale wypełnione cieczą, dlatego wchodzące do wartości porowatości klasycznej), nie są wliczane do wartości $\phi_\mathrm{eff}$. Wartość $\phi_\mathrm{eff}$ jest więc z założenia mniejsza niż klasyczna porowatość zdefiniowana jako stosunek objętości porów dostępnej dla cieczy do objętości zajmowanej przez cały materiał porowaty.
W \nlb celu numerycznego wyznaczenia porowatości efektywnej, zastosowałem procedurę przedstawioną w \nlb \cite{Koponen97} uwzględniając wprowadzoną w niniejszej rozprawie zmodyfikowaną procedurę wyznaczania linii prądu. Za komórki sieci, które aktywnie biorą udział w \nlb transporcie cieczy, brane były te komórki, przez które przechodziły linie prądu wyznaczone metodą całkowania znaczników o zerowej masie unoszonych przez pole prędkości. Wyznaczona w ten sposób zależność porowatości efektywnej dla kilku różnych porowatości przedstawiona została na rysunku \ref{pic:PHIPHIEFF}.
W tym miejscu warto zauważyć, jakie konsekwencje miało uwzględnienie błędnej wartości progu perkolacji $\phi_c$ przez autorów pracy \cite{Koponen97} (patrz podrozdział \ref{sec:progperkol}), którzy też zauważyli, że zależność $\phi_\mathrm{eff}(\phi)$ zachowuje się jak zależność logarytmiczna. Ze względu na przyjęcie błędnego $\phi_c$, nie byli jednak w stanie wyjaśnić, dlaczego proponowana zależność $\phi_\mathrm{eff}=1-\ln(\phi)/\ln(\phi_c)$ daje wartości niefizyczne (tzn. $\phi_\mathrm{eff}>\phi$) dla wysokich porowatości. Jeśli zauważyć, że prawdziwe $\phi_c$ przyjmowało wartość większą od przyjętej przez autorów \cite{Koponen97} -- problem ten znika. Dla kwadratów o boku $a=10$, wielkość $\ln(\phi_c)\approx -1.0$ i powyższe wyrażenie na $\phi_\mathrm{eff}(\phi)$ przyjmuje uproszczoną postać:  $\phi_\mathrm{eff}=1+\ln(\phi)$. Widać wyraźnie (rysunek \ref{pic:PHIPHIEFF}), że przy uwzględnieniu poprawnej wartości progu perkolacji, zależność ta (dla tego modelu oraz tych wielkości przeszkód) jest logarytmiczna. Ma to bardzo interesujące konsekwencje, bo również logarytmiczne zależności powierzchni swobodnej $S$ i krętości $T$ pozwalają postulować inne związki, z których najciekawszym wydaje się proporcjonalność między krętością a porowatością efektywną:
\begin{equation} \label{eq:TPHIEFF}
T \propto \phi_\mathrm{eff}.
\end{equation}
Porównanie wzoru (\ref{eq:TPHIEFF}) z rezultatami symulacji zostało przedstawione na rysunku \ref{pic:TPHIEFF}.
Widać wyraźnie, że w całym zakresie porowatości efektywnej krętość jest do niej wprost proporcjonalna. Współczynnik nachylenia prostej wynosi $a=-0.68$. Na tym etapie nie ma jednak podstaw do interpretacji tej wartości. Podobnie jak w przypadku powierzchni swobodnej, przedstawioną proporcjonalność należałoby zbadać w szerszym zakresie układów o różnej budowie, co pozwoliłoby potwierdzić bądź odrzucić ją jako prawo uniwersalne.


\section{Omówienie wyników i podsumowanie}

Przedstawiona w tym rozdziale dokładna analiza procedury numerycznej do obliczania krętości oraz wyniki, jakie uzyskaliśmy, pokazują, z jak trudnym problemem mamy do czynienia. Skomplikowana struktura linii prądu, których używaliśmy do obliczeń krętości i duże różnice prędkości minimalnych i maksymalnych wzdłuż pojedynczej linii spowodowały potrzebę użycia wyrafinowanych metod całkowania ze zmiennym krokiem czasowym. Okazało się, że nawet z użyciem takich metod numerycznych, linie prądu nie zawsze znajdują drogę przez cały układ, mogą rozszczepiać się na charakterystyczne ,,wyspy'', a następnie trafić w obszary o bardzo niskich prędkościach lokalnych, gdzie nawet zastosowanie zmiennego kroku czasowego nie umożliwia uzyskania zadowalającego rozwiązania. Oszacowanie tych błędów pokazało jednak, że jest to efekt, który nie wpływa znacząco na obliczaną krętość.

Kolejnym krokiem w stronę usprawnienia procedury numerycznego wyznaczania krętości układu było wprowadzenie rozkładu linii prądu uwzględniającego warunek stałego strumienia cieczy pomiędzy sąsiednimi liniami prądu. Dzięki temu linie prądu w obszarach o większej prędkości są zagęszczone, a średnia z linii prądu jest już zwykłą średnią arytmetyczną. Nie chodzi tu tylko o uproszczenie wzoru na krętość, ale przede wszystkim o zwiększenie liczby linii prądu w miejscach szczególnie ważnych, a takimi są właśnie miejsca przez które następuje przepływ. Miejsca o lokalnie niskich prędkościach nie wnoszą wiele do transportu i nie są brane pod uwagę przy obliczeniach. Realizowana wcześniej w literaturze procedura regularnego rozkładu linii prądu w połączeniu z obliczaniem średniej ważonej z wagami równymi prędkościom lokalnym nie jest w stanie dać dokładności, jaką oferuje algorytm zaproponowany w niniejszej rozprawie.

W trakcie obliczeń okazało się również, że metoda gazu sieciowego LBM jest bardzo dobrym narzędziem dla średnich porowatości. Dla $\phi\rightarrow\phi_c$ oraz $\phi\rightarrow 1$ pojawiały się różne, niepożądane efekty. Przede wszystkim, dla obu skrajnych wartości $\phi$ znacznie rośnie czas relaksacji $t_\mathrm{rel}$ krętości. Oznacza to konieczność znacznego zwiększenia liczby kroków czasowych potrzebnych do uzyskania stanu stacjonarnego oraz -- w niektórych przypadkach -- ekstrapolacji tych wartości za pomocą funkcji wykładniczej. Źródło zwiększenia czasu relaksacji jest jednak w obu przypadkach inne. Dla $\phi\approx 1$ układ zbudowany jest z pojedynczych przeszkód, które nie tworzą skomplikowanej sieci kanałów, a ich liczba jest tak mała, że stan stacjonarny osiągany jest bardzo powoli (czego powodem jest dyfuzyjny charakter relaksacji pola ciśnienia). Jest to znana cecha modeli opartych o gaz sieciowy. W drugim skrajnym obszarze, bliskim progowi perkolacji, wzrost $t_\mathrm{rel}$ można tłumaczyć istnieniem sieci długich, wzajemnie połączonych kanałów o małej średnicy, w których relaksacja ciśnienia następuje również w sposób dyfuzyjny. Okazało się, że konfiguracje o porowatości w okolicach $\phi=0.6$ mają budowę optymalną dla czasu relaksacji. Jest to obszar porowatości w którym układy mają charakter pośredni -- już nie składają się jedynie z odizolowanych wysp, a jeszcze nie tworzą sieci długich, pojedynczych, odizolowanych kanałów.

\vspace{0.5cm}

Do analizowanej w tym rozdziale empirycznej relacji logarytmicznej pomiędzy krętością, a porowatością (\ref{fit:comiti}) należy podejść krytycznie. Dopasowanie, jakie zostało uzyskane jest bardzo dobre, ale nie można zapominać, że uzyskano je tylko dla modelu pokrywających się przeszkód kwadratowych. Na tym etapie nie można wnioskować, że prawo to spełnione jest dla dowolnego układu, ani że ma ono taką samą postać w przestrzeni trójwymiarowej. Dość obiecująco wyglądają natomiast relacje, jakie udało się zaproponować pomiędzy krętością a charakterystyczną powierzchnią mikroporów oraz efektywną porowatością, gdzie uzyskana została bardzo dobra zgodność. Jednym z możliwych dalszych kierunków badań nad omawianymi zagadnieniami jest dalsze poszukiwanie związków między tymi parametrami makroskopowymi, zarówno dla różnych modeli, jak i w badaniach eksperymentalnych.


\chapter{Podsumowanie}
\label{sec:chapterpodsumowanie}


Głównym celem niniejszej rozprawy było zbadanie dwóch szczególnych zagadnień związanych z transportem płynów przez ośrodki porowate: dyfuzji ,,anomalnej'' w \nlb modelu K\"{u}ntza i Lavall\'{e}e'go oraz analiza krętości przepływu przez ośrodek porowaty.


W rozdziałach \ref{sec:gazsieciowy} i \ref{sec:chapterdyfuzja} wprowadzony został model K\"{u}ntza i Lavall\'{e}e'go (KL) ośrodka porowatego złożonego z losowo rozmieszczonych rozpraszaczy. W niniejszej rozprawie udało się pokazać, że obserwowana w tym modelu ewolucja profilu koncentracji, uprzednio interpretowana jako efekt dyfuzji anomalnej, może być wyjaśniona jako złożenie dwóch klasycznych zjawisk: dyfuzji Ficka oraz słabego dryfu hydrodynamicznego \cite{Matyka07}.


Naturalną konsekwencją tych badań było przejście do badań nad zjawiskami hydrodynamicznymi w ośrodkach o bardziej skomplikowanej budowie. Problemem, który łączy oba zagadnienia ze sobą, jest zbadanie korelacji pomiędzy makroskopowymi wielkościami fizycznymi opisującymi transport cieczy przez ośrodek porowaty. Po wprowadzeniu modelu gazu sieciowego Boltzmanna w rozdziale \ref{sec:chapter3}, omówiona została nowa procedura numeryczna do wyliczenia krętości (T) -- parametru makroskopowego opisującego stopień wydłużenia drogi transportowanych przez ośrodek cząsteczek materii (rozdział \ref{sec:chapterkretosc}). Dzięki dokładnej analizie błędów o różnym źródle pochodzenia, udało się przedstawić nowe wyniki dotyczące zależności pomiędzy porowatością a krętością ośrodka \cite{Matyka08}. Nowe, dokładniejsze wyniki dotyczące zależności krętości od porowatości pozwoliły wysunąć hipotezę dotyczącą korelacji pomiędzy takimi parametrami jak porowatość efektywna, powierzchnia charakterystyczna i krętość.

Kolejnym krokiem w kontynuacji przedstawionych w niniejszej rozprawie badań będzie sprawdzenie uniwersalności przedstawionych korelacji. Równie istotne wydaje się być sprawdzenie, jak badane wielkości zachowują się w układach (modelach) trójwymiarowych. Bardzo ważne, szczególnie z punktu widzenia ewentualnych zastosowań, byłoby również opracowanie dokładnych i efektywnych metod pomiaru eksperymentalnego wielkości takich jak $\phi_\mathrm{eff}$, $S$ i $T$. Mogłyby w nich znaleźć zastosowanie narzędzia numeryczne opracowane na potrzeby niniejszej rozprawy.

\bibliographystyle{ieeetr}
{\bibliography{PhD}}


\end{document}